\documentclass[useAMS,usenatbib]{mnras}
\include{journals}
\usepackage{amsmath}
\usepackage{amssymb}
\usepackage{graphicx}
\usepackage{subfig}
\usepackage{float}
\usepackage{hyperref}
\usepackage{fancyvrb}

\usepackage{mathtools}
\newcommand\textlcsc[1]{\textsc{\MakeLowercase{#1}}}
\newcommand{\bff}{\bf}
\renewcommand{\bff}{}   

\usepackage{datatool}
\usepackage{booktabs}

\def\jref@jnl#1{{\rm#1}}

\def\aj{\jref@jnl{AJ}}                   
\def\araa{\jref@jnl{ARA\&A}}             
\def\apj{\jref@jnl{ApJ}}                 
\def\apjl{\jref@jnl{ApJ}}                
\def\apjs{\jref@jnl{ApJS}}               
\def\ao{\jref@jnl{Appl.~Opt.}}           
\def\apss{\jref@jnl{Ap\&SS}}             
\def\aap{\jref@jnl{A\&A}}                
\def\aapr{\jref@jnl{A\&A~Rev.}}          
\def\aaps{\jref@jnl{A\&AS}}              
\def\azh{\jref@jnl{AZh}}                 
\def\baas{\jref@jnl{BAAS}}               
\def\jrasc{\jref@jnl{JRASC}}             
\def\memras{\jref@jnl{MmRAS}}            
\def\mnras{\jref@jnl{MNRAS}}             
\def\pasa{\jref@jnl{PASA}}
\def\pra{\jref@jnl{Phys.~Rev.~A}}        
\def\prb{\jref@jnl{Phys.~Rev.~B}}        
\def\prc{\jref@jnl{Phys.~Rev.~C}}        
\def\prd{\jref@jnl{Phys.~Rev.~D}}        
\def\pre{\jref@jnl{Phys.~Rev.~E}}        
\def\prl{\jref@jnl{Phys.~Rev.~Lett.}}    
\def\pasp{\jref@jnl{PASP}}               
\def\pasj{\jref@jnl{PASJ}}               
\def\qjras{\jref@jnl{QJRAS}}             
\def\skytel{\jref@jnl{S\&T}}             
\def\solphys{\jref@jnl{Sol.~Phys.}}      
\def\sovast{\jref@jnl{Soviet~Ast.}}      
\def\ssr{\jref@jnl{Space~Sci.~Rev.}}     
\def\zap{\jref@jnl{ZAp}}                 
\def\nat{\jref@jnl{Nature}}              
\def\iaucirc{\jref@jnl{IAU~Circ.}}       
\def\aplett{\jref@jnl{Astrophys.~Lett.}} 
\def\apspr{\jref@jnl{Astrophys.~Space~Phys.~Res.}}
\def\bain{\jref@jnl{Bull.~Astron.~Inst.~Netherlands}} 
\def\fcp{\jref@jnl{Fund.~Cosmic~Phys.}}  
\def\gca{\jref@jnl{Geochim.~Cosmochim.~Acta}}   
\def\grl{\jref@jnl{Geophys.~Res.~Lett.}} 
\def\jcp{\jref@jnl{J.~Chem.~Phys.}}      
\def\jgr{\jref@jnl{J.~Geophys.~Res.}}    
\def\jqsrt{\jref@jnl{J.~Quant.~Spec.~Radiat.~Transf.}}
\def\memsai{\jref@jnl{Mem.~Soc.~Astron.~Italiana}}
\def\nphysa{\jref@jnl{Nucl.~Phys.~A}}   
\def\physrep{\jref@jnl{Phys.~Rep.}}   
\def\physscr{\jref@jnl{Phys.~Scr}}   
\def\planss{\jref@jnl{Planet.~Space~Sci.}}   
\def\procspie{\jref@jnl{Proc.~SPIE}}   

\bff

\title[The Auriga Project]{The Auriga Project: the properties and formation mechanisms of disc galaxies across cosmic time}
\author[Grand et al.]{\parbox[t]{\textwidth}{
Robert J. J. Grand$^{12}$\thanks{robert.grand@h-its.org}, Facundo A. G\'{o}mez$^3$, Federico Marinacci$^4$, R\"{u}diger Pakmor$^1$, Volker Springel$^{12}$, David J. R. Campbell$^5$, Carlos S. Frenk$^5$, Adrian Jenkins$^5$ and Simon D. M. White$^3$} \vspace{10pt} \\
$^1$Heidelberger Institut f\"{u}r Theoretische Studien, Schloss-Wolfsbrunnenweg 35, 69118 Heidelberg, Germany\\
$^2$Zentrum f\"{u}r Astronomie der Universit\"{a}t Heidelberg, Astronomisches Recheninstitut, M\"{o}nchhofstr. 12-14, 69120 Heidelberg, Germany\\
$^3$Max-Planck-Institut f\"{u}r Astrophysik, Karl-Schwarzschild-Str. 1, D-85748, Garching, Germany  \\
$^4$Department of Physics, Kavli Institute for Astrophysics and Space Research, MIT, Cambridge, MA 02139, USA \\
$^5$Institute for Computational Cosmology, Department of Physics, Durham University, South Road, Durham, DH1 3LE, UK\\
}

\date{Accepted XXX. Received YYY; in original form ZZZ}

\pubyear{2016}
    
\begin{document}

\label{firstpage}

\pagerange{\pageref{firstpage}--\pageref{lastpage}}
\maketitle

\begin{abstract}
We introduce a suite of thirty cosmological magneto-hydrodynamical zoom simulations of the formation of {\bff galaxies in isolated Milky Way mass dark haloes.} These were carried out with the moving mesh code \textlcsc{AREPO}, together with a comprehensive model for galaxy formation physics, including AGN feedback and magnetic fields, which produces realistic galaxy populations in large cosmological simulations. We demonstrate that our simulations reproduce a wide range of {\bff present-day} observables, in particular, two component disc dominated galaxies with appropriate stellar masses, sizes, rotation curves, star formation rates and metallicities. We investigate the driving mechanisms that set present-day disc sizes/scale lengths, and find that they are related to the angular momentum of halo material. We show that the largest discs are produced by quiescent mergers that inspiral into the galaxy and deposit high angular momentum material into the pre-existing disc, simultaneously increasing the spin of dark matter and gas in the halo. More violent mergers and strong AGN feedback play roles in limiting disc size by destroying pre-existing discs and by suppressing gas accretion onto the outer disc, respectively. The most important factor that leads to compact discs, however, is simply a low angular momentum for the halo. In these cases, AGN feedback plays an important role in limiting central star formation and the formation of a massive bulge.
\end{abstract}

\begin{keywords}
galaxies: evolution - galaxies: kinematics and dynamics - galaxies: spiral - galaxies: structure
\end{keywords}

\section{Introduction}

Much recent effort has been devoted to simulating the formation and evolution of disc galaxies like the Milky Way and M31 in their full cosmological context. For simulations in which baryonic physics is neglected (collisionless dark matter-only simulations), the formation of haloes like that of the Milky Way has been studied in exquisite resolution using the ``zoom-in'' technique \citep[e.g., the Via Lactea and Aquarius simulations, see][respectively]{DKM07,SWV08}. However, it has proven exceedingly difficult to achieve comparably robust results for simulations that aim to follow detailed baryonic physics and build a comprehensive galaxy formation model. Early attempts suffered from the so-called ``overcooling'' problem, in which gas cooled too rapidly and flowed to the central regions of the galaxy \citep[e.g.,][]{KG91}. Owing to the high densities the gas elements achieved, numerical prescriptions for star formation then transformed much of the gas mass into star particles in a short amount of time, and resulted in very compact, bulge dominated galaxies \citep[e.g.,][]{NS00} with too many stars with respect to their total mass \citep{GWB10}.   

Since this time, many studies have reported well formed disc galaxies that match a number of key observables such as rotation curves and abundance matching predictions \citep[e.g.,][]{OEF05,GC11,BSG11,AWN13,SBR13,MPS14,WDS15}. In large part, this success can be attributed to the modelling of stellar feedback, for example, by injecting energy released from supernovae explosions into surrounding gas. The energy injected from supernova may be either thermal \citep[e.g.,][]{KG03,SSK06,Sc08,KWB14}, in which the temperature of surrounding gas cells is raised, or kinetic \citep[e.g.,][]{NW94,K01,SH03,OD06,DS08}, in which gas elements are ``kicked'' with some velocity given by the supernova energy. In the former, many studies have found thermally heated gas quickly radiates and becomes dense once more \citep{K92}. Attempts to mitigate this effect include turning off cooling for a period after the energy injection \citep[e.g.,][]{PJF99,SSK06,SBR13} and directly dumping the thermal energy into a single gas cell, thus allowing the dynamical effects to propagate forward into the Euler equations via the pressure while maintaining a high gas temperature for longer \citep[e.g.,][]{KG03,DVS12}. For the case in which energy is injected at least partially in the form of kinetic energy, excessive dissipation is less of a problem because it cannot be radiated away until it thermalizes. The adoption of any of these schemes ultimately removes low angular momentum gas from the central regions of the galaxy \citep{GBM10,BGR11} and prevents the overcooling of gas and the growth of massive stellar bulges. Currently, supernova feedback is being studied further, in particular their placement and ability to drive winds have been studied in stratified box simulations of the ISM \citep[e.g.,][]{CTB13,WN15,SPM16,SSO16}.

In addition to stellar feedback, there are a variety of other forms of feedback that have been studied, including cosmic rays \citep[e.g.,][]{PPS16,PPS16b,SPM16}, radiation pressure \citep[e.g.,][]{HQM12}, early stellar feedback \citep{SBR13,AWN13} and Active Galactic Nuclei (AGN) feedback from the growth of supermassive black holes at the galaxy centre \citep[e.g.,][]{SMH05,SSD07,DGP13,WSH16}. The last of these is thought to become important for galaxies of total mass above $\sim 10^{12}$$\rm M_{\odot}$ \citep{BBF03}, but it is reasonable to expect that it plays some role in Milky Way-mass galaxies as they lie at the lower end of this mass range. However, it remains much less explored than stellar feedback. Prompt stellar feedback was introduced by \citet{SBR13} and \citet{AWN13} and is reported to be critical in preventing the early onset of star formation and thereby avoiding the growth of a central spheroidal component into a massive bulge. Others claim that very strong radiation pressure feedback from stars is of fundamental importance and regulates star formation in a bursty fashion \citep[e.g.,][]{O13,SHF16}, although it is unclear whether radiation pressure is able to play such an important role in reality given the porous structure of the ambient ISM around star-forming regions \citep{KT13}.

While many studies have highlighted feedback as an essential ingredient for the formation of realistic disc galaxies, the numerical codes employed are also an important factor. A variety of methods have been used to follow the hydrodynamics: e.g., adaptive mesh refinement (AMR) codes \citep[e.g.,][]{T02,BNO14}; smoothed-particle-hydrodynamics (SPH) codes \citep[used in, e.g.,][]{Sc08,GC11,SBR13,SFF15} and moving mesh codes \citep{MPS14}. Even for a given hydrodynamic scheme, the differences in the galaxy formation physics and in their implementation have led to diverse models, each reporting different ``optimal'' values for the parameters that control processes such as feedback. For example, a high density threshold for star formation is often required in high resolution simulations \citep{GC11}, and very different values for star formation efficiency have been advocated in the literature, both low \citep{ATM11} and high \citep{SNS10}. Indeed, the lack of agreement on optimal parameters between different codes is consistent with the variable outcomes of simulations reported in the Aquila comparison project \citep{SWP12}, in which results varied widely between codes even for the same initial conditions \citep[see][for a more recent example]{FDP16}. For many codes this reflects lack of robustness to changes in resolution, and the necessity to re-calibrate free parameters for different resolutions. The final outcome of the simulations appears to be dependent on details of the implementation of feedback and star formation \citep[see][]{OEF05}, a sensitivity which is exacerbated by resolution changes. This is particularly true for the cases in which the effects of non-linear feedback cycles, especially those driven by AGN, affect a wide range of physical scales \citep[e.g.,][]{SMH05,SSD07,DGP13,WSH16}.

Part of the numerical uncertainties discussed above may be accounted for by the accuracy of the method employed to follow the hydrodynamics. As an example, traditional forms of SPH suffer from the formation of spurious dense gas clumps which may then artificially sink to the centres of haloes, and are unable to resolve important mixing processes brought about by the Kelvin-Helmholtz instability. Recent modifications to the standard SPH method, such as pressure-based SPH \citep{RHA10,SM13} and the inclusion of artificial thermal conductivity, have improved the situation somewhat, particularly with regard to mixing and hydrodynamical instabilities \citep{ATM11,SDS15}. Although SPH codes by construction follow fluid elements in the frame of their trajectory and naturally provide high resolution in dense regions, they have difficulty in resolving the effects of shocks, and viscosity and noise effects in low density regions can lead to artificial numerical quenching of the cooling rate in large haloes \citep{BS12}.

In this paper, we present a suite of cosmological zoom simulations of {\bff galaxy formation in isolated haloes similar in mass to that of the} Milky Way with the magneto-hydrodynamical simulation code \textlcsc{AREPO} \citep{Sp10}, which is a quasi-Lagrangian method that {\bff solves the fluid equations on a moving mesh}. We include a comprehensive galaxy formation model capable of producing realistic populations of galaxies in a cosmological context \citep{VGS13,GVS14}. In particular, the models for star formation and feedback processes, which include both stellar and black hole feedback, are physically well motivated and do not require re-tuning of parameters between resolution levels.

We build on the study of \citet{MPS14}, who introduced an earlier version of the galaxy formation model in eight zoom simulations of the Aquarius haloes \citep{SWV08} with \textlcsc{AREPO}. We have now implemented magnetic fields into the galaxy formation model \citep[see][]{PakS13,PMS14}, which until recently was not possible owing to well known difficulties in controlling amplification and divergence errors. Their inclusion represents a major step forward in computational galaxy formation. The suite of Auriga simulations that we present in this study comprises thirty haloes at our standard resolution \citep[level 4 resolution in the nomenclature of the Aquarius project, and equivalent to the highest resolution simulation, Aquarius C-4, presented in the convergence study of][]{MPS14}, and three haloes at level 3 resolution. In addition, we make some modifications to the stellar feedback implementation with respect to \citet{MPS14}, which we outline in Section 2. This paper thus constitutes one of the largest samples of cosmological zoom simulations to date with high resolution and a state-of-the-art galaxy formation model and hydrodynamic method \citep[see also][]{WDS15}. 

In this paper, we introduce the Auriga project: we describe methodology, numerical details and properties of the complete simulation suite. We note that other studies, such as \citet{GSG16,MGG16,GSK16,GWG16} have already made use of a subset of the simulation suite, and that forthcoming studies will focus on topics including the properties of HI gas, magnetic fields, satellites, ex-situ discs and stellar and X-ray haloes. A goal of this paper is to demonstrate that we are able to successfully simulate the formation and evolution of Milky Way-sized late type galaxies, the properties of which match a wide range of observables, and resolve the typical morphological features of late type galaxies such as bars and spiral arms. We further demonstrate that these properties are well converged across three levels of resolution (a factor of 64 in mass and 4 in spatial resolution), without the need to recalibrate model parameters. The relatively large simulation sample allows us to make a quantitative analysis of the formation histories of the simulated galaxies. In particular, we analyse the controlling mechanisms behind disc growth. We show that a high degree of specific angular momentum in the dark matter and gas is required to build radially extended disc galaxies that grow inside-out, whereas haloes with little angular momentum evolve into more compact, but still well defined, discs. We determine that a key mechanism in providing the angular momentum to the gas and dark matter is the quiescent accretion of massive, gas-rich subhaloes, which occurs over a wide range of look back times and imparts significant angular momentum to all components (including pre-existing stars) leading to the largest discs in the sample. In addition, we explore how disc properties are affected by AGN feedback, the importance of which remains uncertain in $L_*$ galaxies. We demonstrate that strong AGN feedback is effective in curtailing star formation, particularly in the central disc regions, and we comment on its impact on the radial extent of the disc.

This paper is organised as follows: in Section \ref{sec2}, we outline the criteria of halo selection from the parent cosmological box, and we describe the simulation code and the galaxy formation model. In Section \ref{sec3}, we introduce the simulation suite and describe their present-day galaxy properties, including morphologies, density structure, mass distribution, kinematics and size. In Section \ref{discsizes}, we investigate the mechanisms responsible for the large variation in disc size, with particular focus on merger histories and feedback. In Section \ref{sec5}, we present star formation histories (SFHs) of all the haloes, and show how they shape galaxy observables such as colour, stellar age, metallicity and stellar mass. In Section \ref{resstudy}, we present a resolution study of three of the simulated haloes across three levels of resolution. Finally, in Section \ref{sec7}, we discuss and summarise the impact of our findings.

\section{Simulations}
\label{sec2}

\subsection{Overview}

In this section, we describe the halo selection criteria, the generation of the initial conditions, the simulation code and the details of the galaxy formation model. The Auriga simulations\footnote{see http://auriga.h-its.org} represent a significant improvement on the Aquarius hydrodynamic simulations presented in \citet{MPS14}: Auriga comprises 30 haloes simulated at high resolution, in comparison to the 8 lower resolution Aquarius haloes. Moreover, magnetic fields are seeded and their evolution followed, and several improvements to the physics model and numerics have been made.

\subsection{Halo selection}

The host dark matter haloes in our set of zoomed galaxy simulations have been drawn from a parent dark matter only counterpart to the Eagle simulation of comoving side length 100 cMpc (L100N1504) introduced in \citet{SCB15} (see Appendix~\ref{appa} for a list of halo ID numbers). This simulation is evolved from redshift 127 to the present day, using particles of fixed mass $1.15 \times 10^7$ $\rm M_{\odot}$. The linear phases for this parent simulation, and for all of the zoom simulations, are taken from the public Gaussian white noise field realisation, \textlcsc{PANPHASIA} \citep{J13}. The adopted cosmological parameters are $\Omega _m = 0.307$, $\Omega _b = 0.048$, $\Omega _{\Lambda} = 0.693$ and a Hubble constant of $H_0 = 100 h$ km s$^{-1}$ Mpc$^{-1}$, where $h = 0.6777$, taken from \citet{PC13}. Haloes were identified at redshift $z=0$ using ``Friends of Friends'' (FOF) with the standard linking length \citep{DEF95}. Our selection procedure made use of the position of the centre of potential of each FOF group, and the corresponding virial mass, $M_{200}$, defined as the mass contained inside the radius at which the mean enclosed mass volume density equals 200 times the critical density for closure. The selection criteria were a cut in halo mass of $1 < M_{200} / 10^{12}$ $\rm M_{\odot} < 2$, and the requirement that each candidate halo be relatively `isolated', at redshift zero.

To quantify their isolation, the 697 haloes within the chosen $M_{200}$ range were ranked in order of their maximum value $\tau_{\rm iso,max}$ of the tidal isolation parameter $\tau_{\mathrm{iso},i}$ with respect to the $i$-th other FOF halo, where

\begin{equation}
\tau_{\mathrm{iso},i} = (M_{200,i}) / (M_{200}) \times (R_{200} / R_i)^3
\end{equation}
where $M_{200}$ and $R_{200}$ are the virial mass and radius of the halo of interest, and $M_{200,i}$ and $R_i$ are the virial mass of, and distance to, the $i$-th halo in the simulation, respectively (neglecting haloes with mass less than three percent of that of the halo under consideration). As the virial mass of a halo is proportional to the cube of its virial radius, applying an upper limit to $\tau_{\rm iso,max}$ is equivalent (for a given candidate halo) to requiring the distance to  other haloes in the simulation to be a least a certain number times their virial radii away.

The most isolated quartile (lowest quartile in $\tau_{\rm iso,max}$) was selected to form a sample of 174 candidates for re-simulation. This sample has a range of a factor of 16 in $\tau_{\rm iso,max}$. To be included in the sample, the centre of a target halo must be located outside of 9 times the $R_{200,i}$ of any other halo that has a mass greater than $3\%$ of the target halo mass. The 30 haloes which have been re-simulated were randomly selected from the relatively isolated sample of 174 (their IDs are randomly ordered).

The initial conditions for the re-simulation follow the \citet{Z70} approximation, the procedure for which is outlined in \citet{J10}. Particles within a sphere of radius $4R_{200}$ of the halo centre at redshift zero were traced to their positions in the initial conditions, and used to identify the amoebae-shaped Lagrangian region from which the halo formed. This region was populated with particles of relatively low mass, thus sampling the target halo and its immediate environment at relatively high resolution. Particles of progressively higher mass were used at larger distances. Such particle sampling increases computational efficiency while the correct representation of relevant external effects such as mass infall and the cosmological tidal field is maintained. It is also ensured that there are no heavy dark matter particles within $R_{200}$ at redshift zero. Once the initial dark matter particle distribution is set, gas is added by splitting each original dark matter particle into a dark matter particle and gas cell pair, with masses determined from the cosmological baryon mass fraction. The dark matter particle and gas cell in each pair are separated by a distance equal to half the mean inter-particle spacing, and the centre of mass and centre of mass velocity is retained. 

The typical mass of a high resolution dark matter particle is $\sim 3 \times 10^{5}$ $\rm M_{\odot}$, and the baryonic mass resolution is $\sim 5 \times 10^{4}$ $\rm M_{\odot}$. The comoving gravitational softening length for the star particles and high resolution dark matter particles is set to $500$ $ h^{-1}$ cpc. The physical gravitational softening length grows with the scale factor until a maximum physical softening length of $369$ pc is reached. This corresponds to $z=1$, after which time the softening is kept constant. The softening length of gas cells is scaled by the mean radius of the cell, which ensures that low density gas cells are softened with higher values relative to high density gas. The minimum comoving softening length of the gas is set to $500$ $h^{-1}$ cpc (minimum physical softening set to $369$ pc at $z=1$, as for star particles) and the maximum physical softening to $1.85$ kpc. We note that the gas cell size is allowed to become less than the softening length in high density regions, because each gas cell is designed to hold a pre-designated target mass (see below). 

The motivation for these choices of softening comes from \citet{PNJ03}, who performed detailed convergence studies of dark matter only simulations, and empirically derived optimal softening lengths for systems of given mass, size and particle number. Although the relaxation timescale depends primarily on the mass resolution (or equivalently, the number of particles employed), the gravitational softening length enters also in the equation, albeit logarithmically, which makes the relaxation timescale sensitive to order of magnitude changes in softening. \citet{PNJ03} found that the optimal softening length followed the scaling: $\epsilon \sim 4 r_{200}  \sqrt{m_p / M_{200}}$, which for our simulations at redshift zero corresponds to a few hundred parsecs. A softening length of an order magnitude larger or smaller than this optimal value was shown to either produce unrealistic central mass profiles or a growing significance of two-body interactions that give rise to excessive local dynamical scatter and violate the collisionless nature of the system, particularly for small progenitor haloes at high redshift.

\addtocounter{footnote}{-2}

\DTLsetseparator{	}
\begin{filecontents*}{./figures/fit_table.txt}
         Run        Time  virialmass virialradius Stellarmass    discmass          Rd   bulgemass        reff           n         D/T		Ropt		facc		fgas
         Au1       0.000      93.376     206.032       2.745       1.065       5.444       1.587       1.346       1.414        0.40		20.0		0.289		0.414
         Au2       0.000     191.466     261.757       7.045       6.377      11.644       2.341       2.056       1.529        0.73		37.0		0.133		0.169
         Au3       0.000     145.777     239.019       7.745       6.288       7.258       2.039       1.488       0.987        0.76		31.0		0.140		0.191
         Au4       0.000     140.885     236.310       7.095       3.662       3.929       2.005       1.740       1.352        0.65		24.5		0.373		0.331
         Au5       0.000     118.553     223.091       6.722       4.509       3.583       1.806       0.839       0.874        0.71		21.0		0.103		0.271
         Au6       0.000     104.385     213.825       4.752       3.315       5.949       1.649       2.851       2.000        0.67		26.0		0.109		0.193
         Au7       0.000     112.043     218.935       4.875       2.458       5.140       2.045       1.731       1.740        0.55		25.0		0.336		0.394
         Au8       0.000     108.062     216.314       2.990       2.457       6.572       0.652       2.147       1.328        0.79		25.0		0.163		0.234
         Au9       0.000     104.971     214.224       6.103       3.597       3.367       2.169       0.999       0.948        0.62		19.0		0.060		0.213
        Au10       0.000     104.710     214.061       5.939       2.378       2.596       3.152       1.080       1.181        0.43		16.0		0.043		0.238
        Au11       0.000     164.935     249.053       6.909       1.133       3.055       2.750       1.046       1.354        0.29		16.0		0.458		0.215
        Au12       0.000     109.275     217.117       6.010       4.315       3.290       1.134       0.892       0.759        0.79		19.0		0.212		0.310
        Au13       0.000     118.904     223.325       6.194       1.675       3.382       3.798       1.403       1.549        0.31		15.5		0.136		0.170
        Au14       0.000     165.721     249.442      10.393       6.359       4.186       3.184       1.138       1.586        0.67		26.0		0.173		0.284
        Au15       0.000     122.247     225.400       3.930       2.772       5.320       1.047       2.216       2.000        0.73		23.0		0.178		0.402
        Au16       0.000     150.332     241.480       5.410       5.059       9.030       1.175       1.825       1.391        0.81		36.0		0.110		0.224
        Au17       0.000     102.835     212.769       7.608       2.563       4.191       4.641       1.208       0.831        0.36		16.0		0.031		0.191
        Au18       0.000     122.074     225.288       8.037       5.205       3.719       2.461       1.225       0.950        0.68		21.0		0.045		0.130
        Au19       0.000     120.897     224.568       5.320       3.532       4.805       1.428       1.416       2.000        0.71		24.0		0.226		0.260
        Au20       0.000     124.922     227.028       4.740       2.248       8.019       2.353       2.174       1.886        0.49		30.0		0.356		0.332
        Au21       0.000     145.090     238.645       7.717       6.005       4.607       1.240       1.188       1.064        0.83		24.0		0.226		0.275
        Au22       0.000      92.621     205.476       6.020       2.851       2.249       2.709       1.014       0.934        0.51		13.5		0.029		0.107
        Au23       0.000     157.539     245.274       9.023       5.547       4.985       3.192       1.708       1.438        0.63		25.0		0.111		0.198
        Au24       0.000     149.178     240.856       6.554       3.756       5.570       2.186       0.946       0.969        0.63		30.0		0.119		0.155
        Au25       0.000     122.109     225.305       3.142       2.475       6.695       0.934       2.951       1.879        0.73		21.0		0.043		0.343
        Au26       0.000     156.384     244.685      10.967       4.697       3.141       5.456       1.116       1.017        0.46		18.0		0.131		0.120
        Au27       0.000     174.545     253.806       9.606       7.229       4.287       1.742       0.947       1.065        0.81		26.0		0.137		0.217
        Au28       0.000     160.538     246.833      10.448       6.761       2.159       2.359       0.948       1.109        0.74		17.5		0.203		0.138
        Au29       0.000     154.243     243.553       9.033       4.084       3.204       2.610       1.190       1.147        0.61		19.0		0.368		0.207
        Au30       0.000     110.827     218.148       4.251       1.446       5.010       2.429       1.901       1.223        0.37		17.0		0.474		0.271
\end{filecontents*}
\DTLloaddb{fittable}{./figures/fit_table.txt}
\begin{filecontents*}{./figures/fit_tablelv3.txt}
         Run        Time  virialmass virialradius Stellarmass    discmass          Rd   bulgemass        reff           n         D/T		Ropt		facc		fgas
       	 Au6        0.000     101.480     211.834       6.077       4.344       3.198       1.203       0.748       1.470        0.78	  	26.0		0.12		0.137
      	 Au16       0.000     150.430     241.530       7.854       6.468       6.367       1.410       0.532       1.601        0.82	  	33.0		0.09		0.137
         Au24       0.000     146.791     239.568       7.766       5.698       6.521       1.872       0.623       1.914        0.75	  	32.0		0.12		0.137
\end{filecontents*}
\DTLloaddb{fittablelv3}{./figures/fit_tablelv3.txt}
\begin{filecontents*}{./figures/fit_tablelv5.txt}
         Run        Time  virialmass virialradius Stellarmass    discmass          Rd   bulgemass        reff           n         D/T		Ropt		facc		fgas
     	Au6       0.000     102.649     212.640       3.835       3.548       4.585       0.271       1.085       0.605        0.93		26.0		0.06		0.272
    	Au16       0.000     149.469     241.021       3.748       3.090       7.692       0.824       1.468       0.872        0.79		33.0		0.08		0.272
    	Au24       0.000     148.153     240.308       6.259       4.509       4.445       1.334       1.437       0.814        0.77		32.0		0.07		0.268
\end{filecontents*}
\DTLloaddb{fittablelv5}{./figures/fit_tablelv5.txt}

\begin{table*}
\caption{Table of simulation parameters at $z=0$. The columns are 1) Model name; 2) Halo virial mass \protect\footnotemark; 3) Halo virial radius\protect\footnotemark; 4) Stellar mass; 5) Inferred stellar disc mass; 6) Radial scale length; 7) Inferred stellar bulge mass; 8) Bulge effective radius; 9) Sersic index of the bulge; 10) Disc to total mass ratio; 11) Optical radius; 12) accreted stellar fraction {\bff within $0.1R_{200}$ and 13) gas fraction within $0.1 R_{200}$.}}
\centering
\begin{tabular}{c c c c c c c c c c c c c}
\toprule
Run &       
$\frac{M_{200}} {[\rm 10^{10} M_{\odot}]}$ &
$\frac{R_{200}} {[\rm kpc]}$ &
$\frac{M_{*}} {[\rm 10^{10} M_{\odot}]}$  &
$\frac{M_{\rm d}} {[\rm 10^{10} M_{\odot}]}$  &
$\frac{R_{\rm d}} {[\rm kpc]}$  &
$\frac{M_{\rm b}} {[\rm 10^{10} M_{\odot}]}$ &
$\frac{R_{\rm eff}} {[\rm kpc]}$ &
 $n$ &
 $D/T$ &
 $\frac{R_{\rm opt}} {[\rm kpc]}$ &
 $f_{\rm acc}$ &
 $f_{\rm gas}$ \\ %
 \hline
 \multicolumn{12}{c}{Level 4 resolution} \\
 \hline
  \vspace{-0.3cm}
\DTLforeach{fittable}{%
\run=Run,\time=Time,\mv=virialmass,\rv=virialradius,\mstar=Stellarmass,\mdisc=discmass,\rd=Rd,\mbulge=bulgemass,\reff=reff,\n=n,\dt=D/T,\ropt=Ropt,\facc=facc, \fgas=fgas}{%
\DTLground{\mv}{\mv}{2}
\DTLground{\rv}{\rv}{2}
\DTLground{\mstar}{\mstar}{2}
\DTLground{\mdisc}{\mdisc}{2}
\DTLground{\rd}{\rd}{2}
\DTLground{\mbulge}{\mbulge}{2}
\DTLground{\reff}{\reff}{2}
\DTLground{\n}{\n}{2}
\DTLground{\ropt}{\ropt}{2}
\DTLground{\facc}{\facc}{2}
\DTLground{\fgas}{\fgas}{2}
\DTLiffirstrow{\\}{\\}%
\run & \mv & \rv & \mstar & \mdisc & \rd & \mbulge & \reff & \n & \dt & \ropt & \facc & \fgas \\ \vspace{-0.3cm} }

\\\hline
 \multicolumn{12}{c}{Level 5 resolution} \\
 \hline
 \vspace{-0.3cm}
\DTLforeach{fittablelv5}{%
\run=Run,\time=Time,\mv=virialmass,\rv=virialradius,\mstar=Stellarmass,\mdisc=discmass,\rd=Rd,\mbulge=bulgemass,\reff=reff,\n=n,\dt=D/T,\ropt=Ropt,\facc=facc, \fgas=fgas}{%
\DTLground{\mv}{\mv}{2}
\DTLground{\rv}{\rv}{2}
\DTLground{\mstar}{\mstar}{2}
\DTLground{\mdisc}{\mdisc}{2}
\DTLground{\rd}{\rd}{2}
\DTLground{\mbulge}{\mbulge}{2}
\DTLground{\reff}{\reff}{2}
\DTLground{\n}{\n}{2}
\DTLground{\ropt}{\ropt}{2}
\DTLground{\facc}{\facc}{2}
\DTLground{\fgas}{\fgas}{2}
\DTLiffirstrow{\\}{\\}%
\run & \mv & \rv & \mstar & \mdisc & \rd & \mbulge & \reff & \n & \dt & \ropt & \facc & \fgas \\ \vspace{-0.3cm} }

\\\hline
 \multicolumn{12}{c}{Level 3 resolution} \\
 \hline
  \vspace{-0.3cm}
\DTLforeach{fittablelv3}{%
\run=Run,\time=Time,\mv=virialmass,\rv=virialradius,\mstar=Stellarmass,\mdisc=discmass,\rd=Rd,\mbulge=bulgemass,\reff=reff,\n=n,\dt=D/T,\ropt=Ropt,\facc=facc, \fgas=fgas}{%
\DTLground{\mv}{\mv}{2}
\DTLground{\rv}{\rv}{2}
\DTLground{\mstar}{\mstar}{2}
\DTLground{\mdisc}{\mdisc}{2}
\DTLground{\rd}{\rd}{2}
\DTLground{\mbulge}{\mbulge}{2}
\DTLground{\reff}{\reff}{2}
\DTLground{\n}{\n}{2}
\DTLground{\ropt}{\ropt}{2}
\DTLground{\facc}{\facc}{2}
\DTLground{\fgas}{\fgas}{2}
\DTLiffirstrow{\\}{\\}%
\run & \mv & \rv & \mstar & \mdisc & \rd & \mbulge & \reff & \n & \dt & \ropt & \facc & \fgas \\ \vspace{-0.3cm} }

\\\bottomrule
\end{tabular}
\end{table*}

\subsection{Simulation code}

The zoom re-simulations are performed with the $N$-body, magnetohydrodynamics code \textlcsc{AREPO} \citep{Sp10}, which we describe briefly here. For further detailed description, we refer the reader to \citet{Sp10}. \textlcsc{AREPO} is a moving-mesh code that follows magnetohydrodynamics and collisionless dynamics in a cosmological context. Gravitational forces are calculated by a standard TreePM method \citep[e.g.][]{SP05}, which itself employs a Fast Fourier Transform method for long range forces, and a hierarchical oct-tree algorithm \citep{BH86} for short range forces, together with adaptive time-stepping. 

To follow the magnetohydrodynamics, \textlcsc{AREPO} utilises a dynamic unstructured mesh constructed form a Voronoi tessellation of a set of mesh-generating points (the so-called Voronoi mesh), that allows for a finite-volume discretisation of the magneto-hydrodynamic (\textlcsc{MHD}) equations. The \textlcsc{MHD} equations are solved with a second order Runge-Kutta integration scheme with high accuracy least square spatial gradient estimators of primitive variables  \citep{PSB15}, that improve on the estimators in the original version of \textlcsc{AREPO} \citep{Sp10}. 

A unique feature of \textlcsc{AREPO} is that the mesh can be transformed through a mesh reconstruction at any time-step, which is not the case for standard grid-based methods. The mesh construction ensures that each cell contains a given target mass (specified to some tolerance), such that regions of high density are resolved with more cells than regions of low density. Furthermore, the mesh generating points are able to move with the fluid flow velocity, such that each cell of the newly constructed mesh moves approximately with the fluid. In this way, \textlcsc{AREPO} overcomes the Galilean non-invariance problem of standard Eulerian mesh codes and significantly reduces the advection errors that plague fixed mesh codes when applied to complex supersonic flows. The quasi-Lagrangian characteristic of the method makes it relatable to other Lagrangian methods such as Smoothed Particle Hydrodynamics (SPH), although several limitations of the SPH method are eliminated, for example, there is no artificial viscosity, and the hydrodynamics of under-dense regions are treated with higher accuracy.

\addtocounter{footnote}{-2}

\stepcounter{footnote}\footnotetext{Defined to be the mass inside a sphere in which the mean matter density is 200 times the critical density, $\rho _{\rm crit} = 3H^2(z)/(8 \pi G)$.} 

\stepcounter{footnote}\footnotetext{Defined as the stellar mass within $0.1$ times the virial radius.}

\subsection{Physics model}


The interstellar medium (ISM) is described by the subgrid model first presented in \citet{SH03}, in which star-forming gas is treated as a two phase medium: a phase of cold, dense clouds embedded in a hot, ambient medium. The gas is assumed to be star-forming and thermally unstable for densities higher than a threshold density, which we derive from the parameters of the two gas phases and the desired star formation timescale to be $n=0.13$ $\rm cm^{-3}$. The motivation for the ISM model comes from the assumption that, at the onset of thermal instability, the many small scale processes that govern the mass fractions of molecular clouds and ambient gas, such as radiative cooling, thermal conduction, star formation and feedback, quickly establish a pressure equilibrium between the hot and cold phases. In this regime, the gas pressure is a function of density only. Owing to the occurrence of star formation and stellar feedback, the equilibrium temperature of the two-phase medium ought to be higher than that of a pure isothermal gas, as a result of energy injection from supernovae, and is therefore governed by an effective equation of state (eEOS) at densities higher than the threshold density for star formation, which is `stiffer' than an isothermal equation of state. The model therefore provides a representation of the ISM and star formation on scales below the resolution limit, which is physically motivated by and consistent with the important small scale processes that operate in the ISM.


Once a gas cell enters the thermally unstable star-forming regime, star particles are formed stochastically according to a probability that scales exponentially with time in units of the star formation timescale, which we set to $t_{\rm sf} = 2.2$ Gyr. The amount of gas mass converted to stellar mass depends on the mass of the gas cell selected for star formation: all mass is converted and the cell removed if the gas cell holds less than twice the target cell mass, defined as $ m_{\rm target} =  \bar{m}_b$, where $\bar{m}_b \sim 5 \times 10^4$ $\rm M_{\odot}$ is the mean cell mass (as quoted above). Otherwise a mass of only $m_{\rm target}$ is converted and the cell retained with reduced mass.

We assume that each star particle represents a single stellar population (SSP), which is characterised by a given age and metallicity. The distribution of stellar masses contained in each SSP is given by a \citep{C03} Initial Mass Function (IMF), which allows the calculation of mass that leaves the main sequence at a given time-step. The mass loss and metal return is calculated for SNII, SNIa and AGB stars each time-step as the mass moving off the main sequence for each SSP. The yields used for this calculation are those reported in \citet{K10} for AGB stars, and \citet{PCB98} for core collapse SN. The mass and metals are then distributed among nearby gas cells with a top-hat kernel. 

For a given SSP, the number of SNIa events per time-step is calculated by integrating the delay time distribution function (DTD):

\begin{equation}
N_{\rm Ia} (t, \Delta t) = \int _t ^{t+\Delta t} g(t' - t_0) \mathrm{d}t',
\end{equation} 
where $t_0$ is the birth time of the SSP and $\Delta t$ is an interval of time thereafter. The DTD takes the form

\begin{equation}
\label{frsol}
g (t) = 
\begin{cases}
0 & \quad \text{if $t < \tau _{8 \rm M_{\odot}}$}, \\
N_0 \left( \frac{t}{\tau _{8 \rm M_{\odot}}} \right) ^{-s} \frac{s-1}{\tau _{8\rm M_{\odot}}} & \quad \text{if $t > \tau _{8 \rm M_{\odot}}$},
\end{cases}
\end{equation}
where the normalisation $N_0$ is calculated to be $1.99 N_{\rm SN} / M$ assuming the upper limit of a Hubble time for the DTD \citep[as in][]{MPS14b}, where the supernova rate $N_{\rm SN} / M$ is taken to be $1.3 \times 10^{-3}$ $\rm SN / M_{\odot}$ \citep{MMB12}. The power law index is $s=1.12$ and $\tau _{8 \rm M_{\odot}} = 40$ $\rm Myr$, which corresponds to the main sequence lifetime of a $8\rm M_{\odot}$ star, and is the upper mass limit for SNIa. The amount of mass and metals returned from SNIa to the ISM is then calculated from SNIa yield tables \citep{TAB03,THR04}, and distributed among neighbouring gas cells.

The number of SNII is calculated to be the number of stars in an SSP that lie in the mass range $8$-$100$ $\rm M_{\odot}$. SNII events are assumed to occur instantaneously, and are implemented such that an active gas cell (star-forming) is probabilistically chosen to either form a star or become a site for SNII \citep[see][]{VGS13}. SNII feedback is modelled by converting the gas cell into a wind particle, which is launched in an isotropically random direction {\bff \citep[as opposed to the bipolar wind model of][see also \citealt{WSH16}]{MPS14}. We set the wind velocity equal to $3.46 \sigma_{DM}^{1D}$, where $\sigma_{DM}^{1D}$ is the local one-dimensional dark matter velocity dispersion \citep[see e.g.,][]{OFJ10}, which is calculated from the nearest 64 dark matter particles}. Upon launch, a wind particle is loaded with $1-\eta _w$ times the metal mass of the gas cell from which it is created, where $\eta _w =0.6$ is the metal loading parameter. The wind particle travels away from its launch site and interacts only gravitationally until either a gas cell with a density below 0.05 times the physical density threshold for star formation is reached, or the maximum travel time is exceeded. Once either of these criteria is met the wind particle deposits its mass, metals, momentum and energy into the gas cell in which it is located. {\bff Following \citet{MPS14}, the energy deposited is split into equal parts kinetic and thermal, which produces smooth, regular winds in contrast to cold, purely kinetic winds.} This parametrisation of wind outflows is required in order to reproduce both the stellar mass and oxygen abundances of low mass haloes \citep{PS13}.

{\bff To facilitate comparison between our models and observations, we derive photometric properties for the star particles. Because we treat each star particle as a single stellar population of a given age, mass and metallicity, it is possible to use stellar population synthesis models to estimate their luminosity in a series of broad bands. Our model currently tabulates luminosities for U-, V-, B-, K-, g-, r-, i- and z- bands from the \citet{BC03} catalogues. We caution that we do not model the effects of dust attenuation.}

\begin{figure*} 
\centering
\includegraphics[scale=.3]{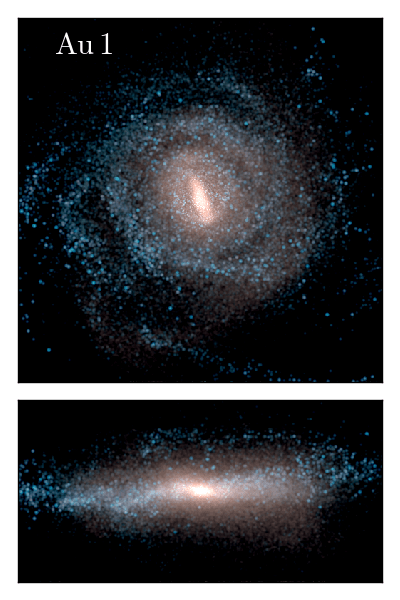} 
\includegraphics[scale=.3]{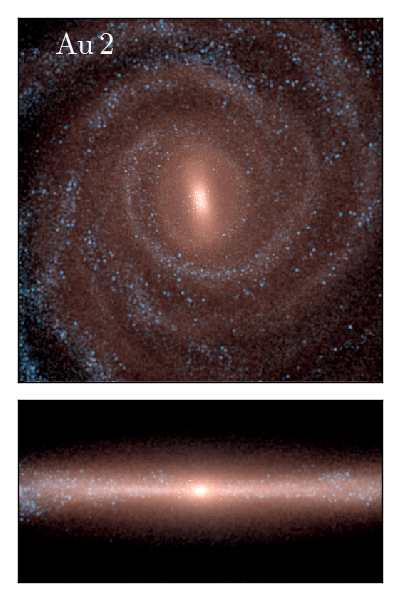} 
\includegraphics[scale=.3]{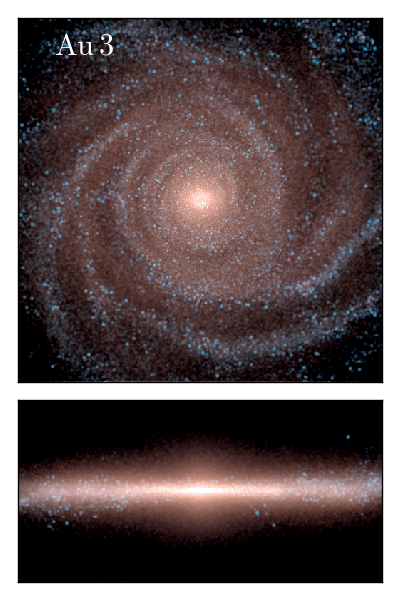} 
\includegraphics[scale=.3]{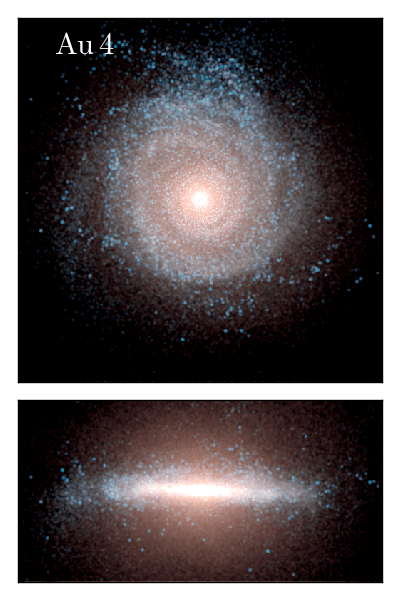} \\
\includegraphics[scale=.3]{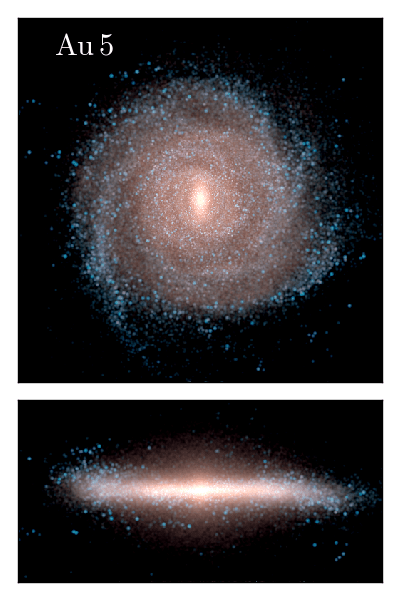} 
\includegraphics[scale=.3]{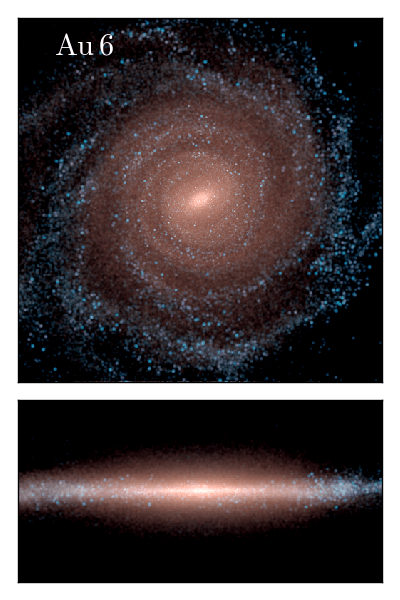} 
\includegraphics[scale=.3]{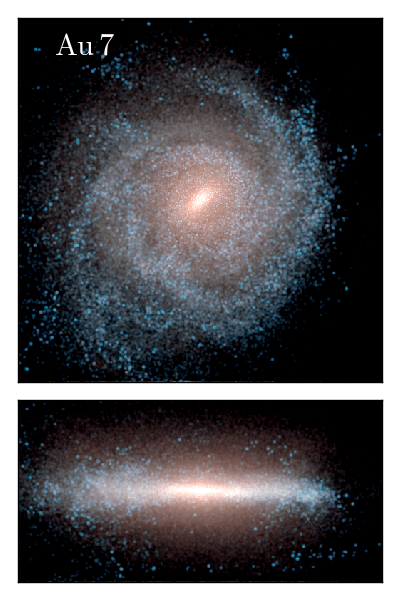} 
\includegraphics[scale=.3]{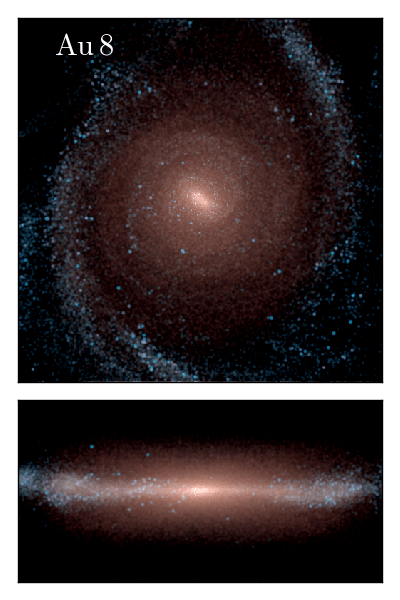} \\
\includegraphics[scale=.3]{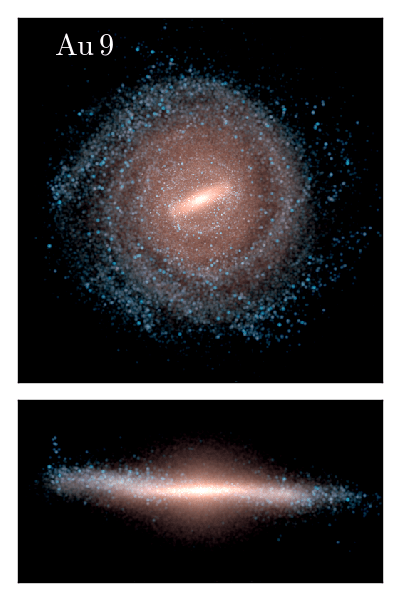} 
\includegraphics[scale=.3]{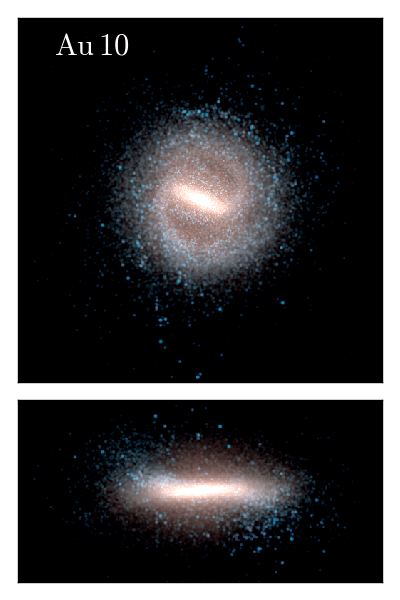} 
\includegraphics[scale=.3]{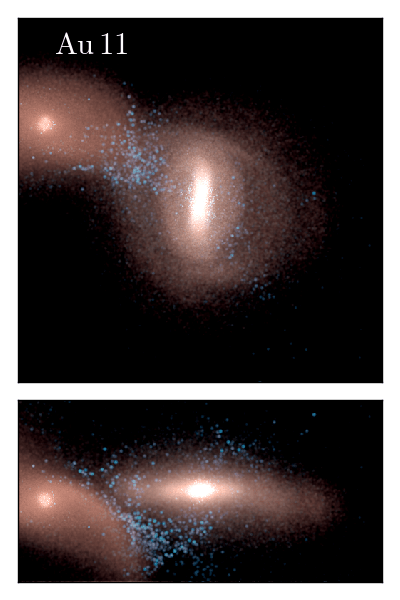} 
\includegraphics[scale=.3]{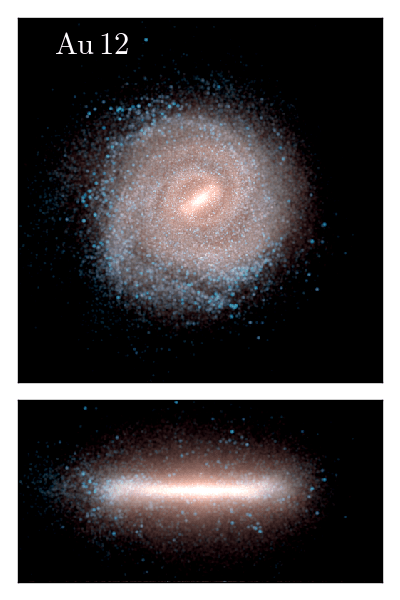} \\
\caption{The face-on and edge-on projected stellar density at $z=0$ for simulations Au 1-12. The images are synthesised from a projection of the $K$-, $B$- and $U$-band luminosity of stars, which are shown by the red, green and blue colour channels, in logarithmic intervals, respectively. Younger (older) star particles are therefore represented by bluer (redder) colours. The plot dimensions are $50\times 50 \times 25$ kpc. For movies and images, go to \url{http://auriga.h-its.org}.}
\label{proj1}
\end{figure*}

\begin{figure*} 
\centering
\includegraphics[scale=.3]{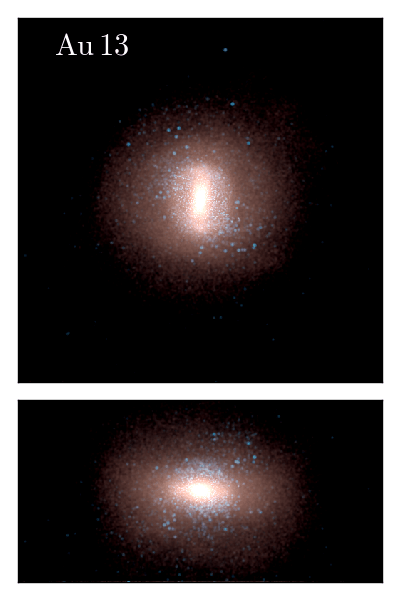} 
\includegraphics[scale=.3]{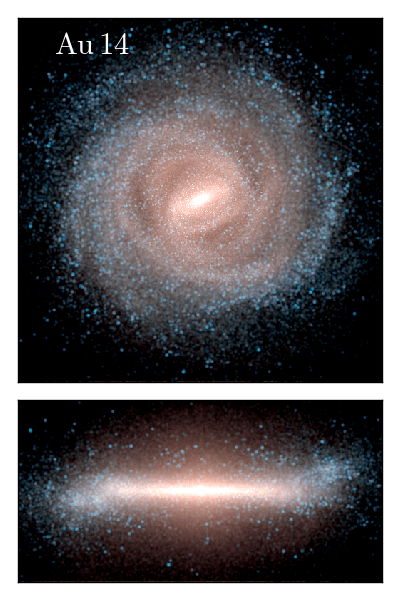} 
\includegraphics[scale=.3]{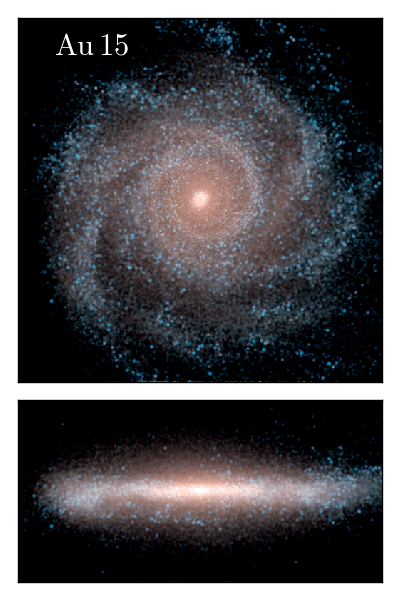} 
\includegraphics[scale=.3]{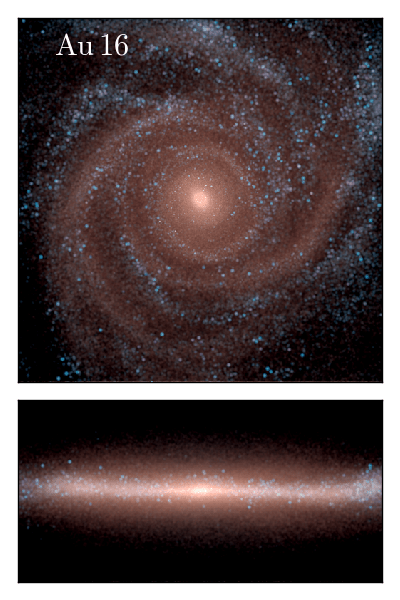} \\
\includegraphics[scale=.3]{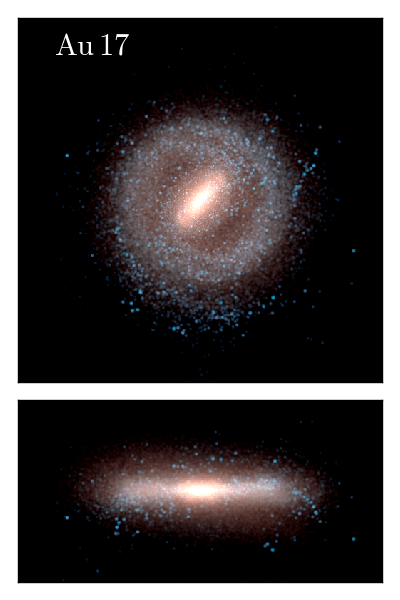} 
\includegraphics[scale=.3]{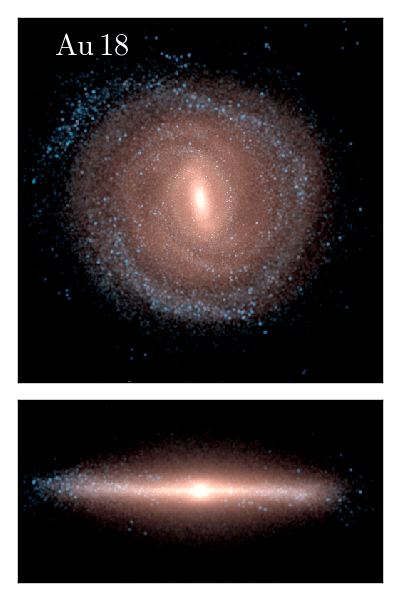} 
\includegraphics[scale=.3]{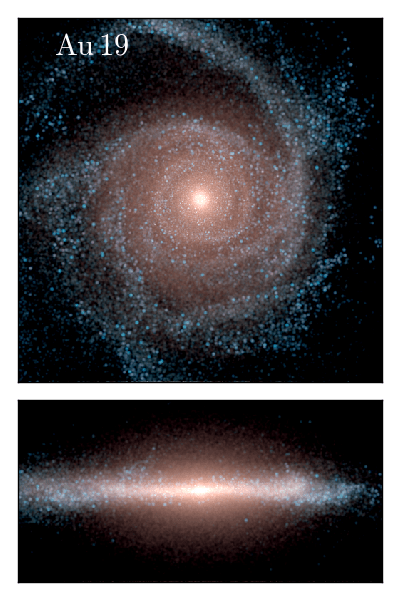} 
\includegraphics[scale=.3]{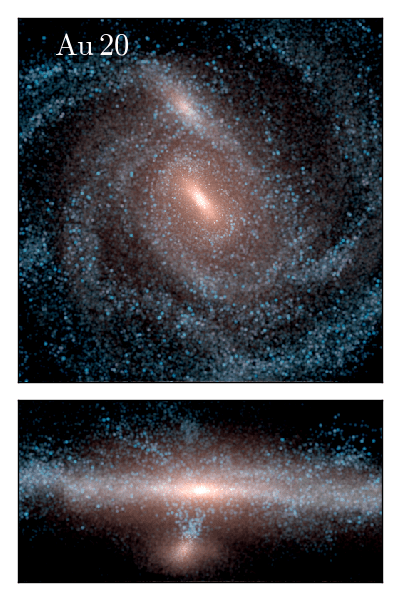} \\
\includegraphics[scale=.3]{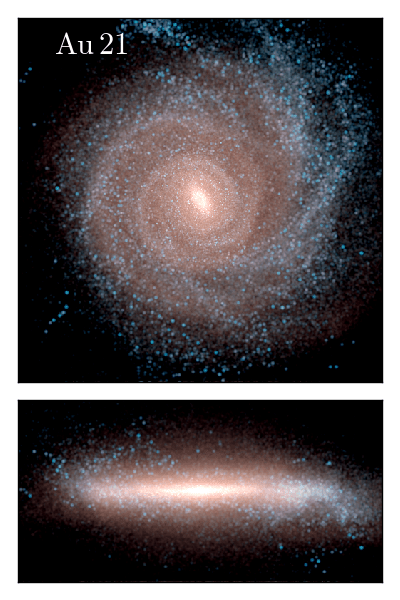} 
\includegraphics[scale=.3]{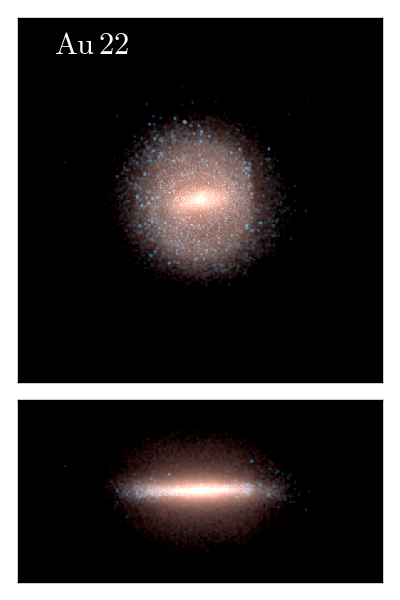} 
\includegraphics[scale=.3]{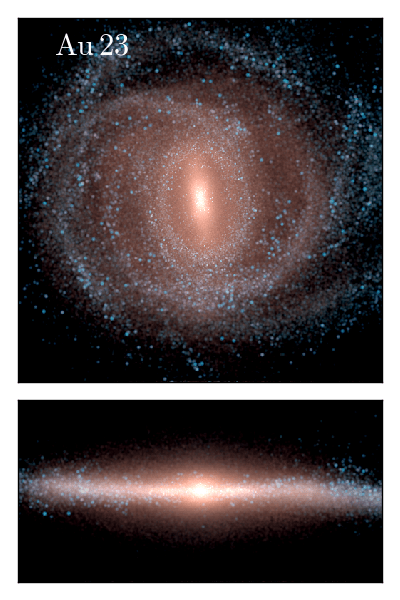} 
\includegraphics[scale=.3]{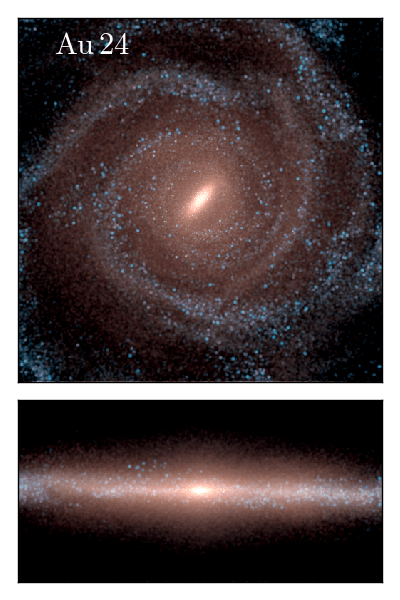} \\
\caption{As Fig.~\ref{proj1} for simulations Au 13-24.}
\label{proj2}
\end{figure*}

\begin{figure*} 
\centering
\includegraphics[scale=.3]{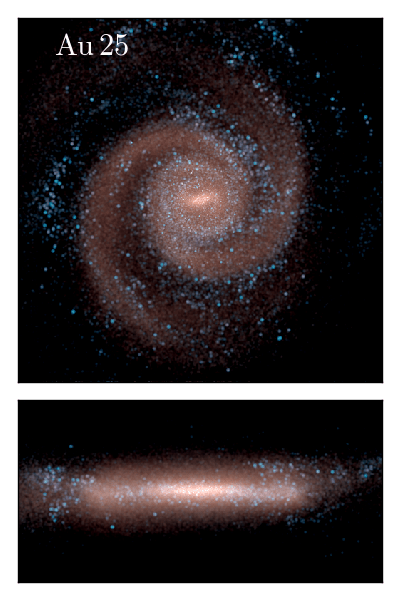} 
\includegraphics[scale=.3]{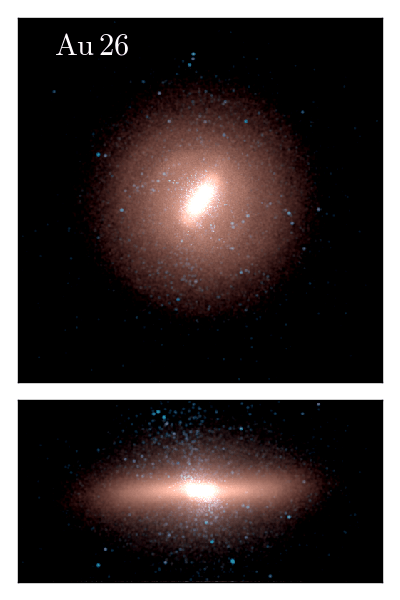} 
\includegraphics[scale=.3]{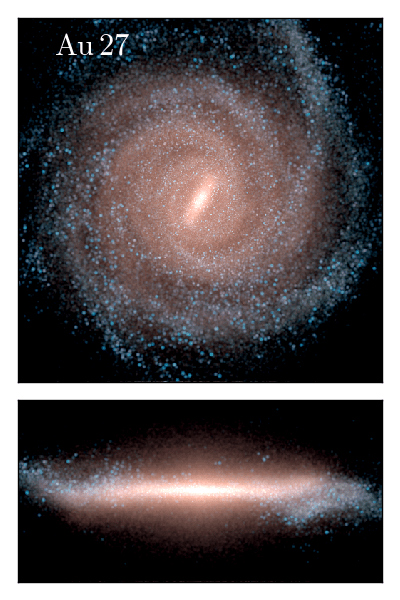} 
\includegraphics[scale=.3]{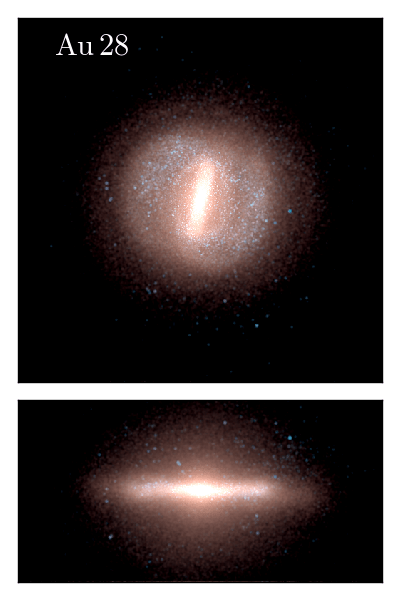} \\
\includegraphics[scale=.3]{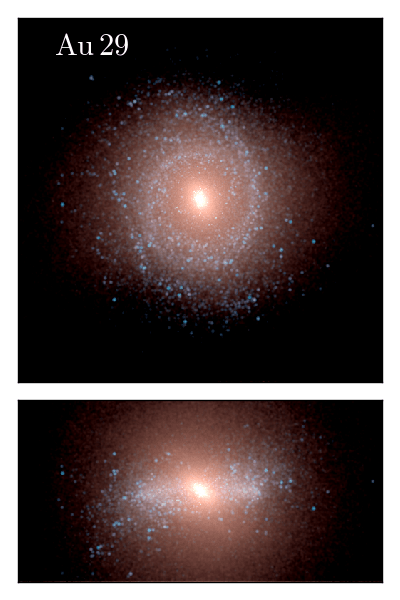} 
\includegraphics[scale=.3]{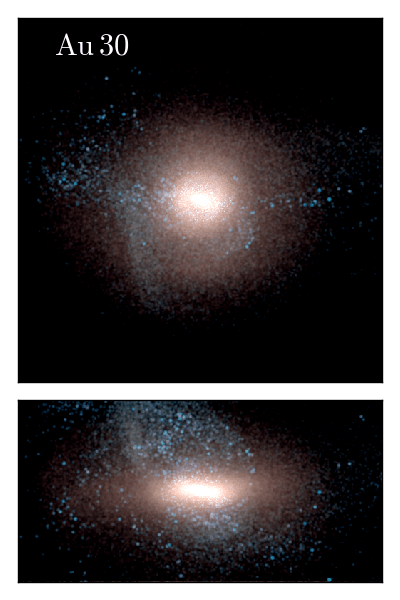} 

\caption{As Fig.~\ref{proj1} for simulations Au 25-30.}
\label{proj3}
\end{figure*}


Primordial and metal-line cooling with self-shielding corrections is enabled. Following the model of \citet{FG09}, a spatially uniform UV background field is included, which completes reionization at redshift $z\sim6$ \citep[for more details see][]{VGS13}. 


Black Holes are seeded with a seed mass of $10^{5}$ $\rm M_{\odot}$ $h^{-1}$ in halo FOF groups of masses greater than $M_{\rm FOF} = 5 \times 10^{10}$ $\rm M_{\odot}$ $h^{-1}$, and are placed at the position of the most dense gas cell. These black hole sink particles acquire mass from nearby gas cells or by merging with other black hole particles. The black hole mass growth from accretion and merger processes follows \citet{SMH05}. The mass growth from gas accretion is described by Eddington-limited Bondi-Hoyle-Lyttleton accretion \citep{BH44,B52} in addition to a term that models the radio mode accretion. This radio mode term comes from the assumption that the ICM of the halo is in thermodynamic equilibrium, and therefore the thermal energy added to the ICM from the radio mode AGN feedback,

\begin{equation}
L_{\rm radio} = \epsilon _f \epsilon _r c^2 \dot{M}_{\rm radio},
\end{equation}
is radiated away in X-ray losses, and is set equal to $R(T,z) L_{\rm X}$, where $L_{\rm X}$ is the X-ray luminosity and $R(T,z)$ is a scaling factor. The scaling factor is calculated from relations presented in \citet{NF00}, in which the accretion rate, $\dot{M'}_{\rm BH}$, is inferred from the state of the gas at large distances from the black hole, and the X-ray luminosity, $L'_{\rm X}$, is assumed to depend on the observational scaling relation given by \citet{PCA09}:

\begin{equation}
R(T,z) = \frac{\epsilon _f \epsilon _r c^2 \dot{M'}_{\rm BH}}{L'_{\rm X}},
\end{equation}
where  
\begin{equation}
\dot{M'}_{\rm BH} = \frac{2\pi Q (\gamma -1) k_B, T_{\rm vir}}{\mu m_H \Lambda (T_{\rm vir})} G M'_{\rm BH} \left[ \frac{\rho ^2}{n_e n_H} \right] \mathcal{M} ^{3/2}, \nonumber
\end{equation}

\begin{equation}
M'_{\rm BH} = M_0 \left( \frac{\sigma}{\sigma _0} \right)^4 ,
\end{equation}
and

\begin{equation}
L'_{\rm X} = \frac{H(z)}{H_0} C \left(\frac{T}{T_0} \right)^{2.7}.
\end{equation}
Here, $\gamma = 5/3$ is the adiabatic index of the gas, $T_{\rm vir}$ is the virial temperature, $\mathcal{M}$ is the Mach number of gas far from the black hole, $\sigma$ is the velocity dispersion of stars in the galactic bulge, $Q=2.5$, $C=6\times 10^{44}$ $\rm ergs$ $\rm s^{-1}$, and $T_0 = 5$ keV. The Mach number is set to $\mathcal{M} =0.0075$ because the black hole-stellar mass relation is reproduced for this value in cosmological simulations. We adopt this parametrisation because the accretion rate inferred in \citet{NF00} depends on the virial temperature, and therefore is insensitive to gas temperature variations that are required to create a self-regulated feedback process.

The total black hole accretion rate is given by

\begin{equation}
\begin{split}
\dot{M}_{\rm BH} &= \mathrm{min} \left[ \frac{4 \pi G^2 M_{\rm BH}^2 \rho}{(c_s^2 + v_{\rm BH}^2)^{3/2}} + \frac{R(T,z) L_{\rm X}}{\epsilon _f \epsilon _r c^2}, \dot{M}_{\rm Edd} \right],
\end{split}
\end{equation}
where $\rho$ and $c_s$ are the density and sound speed of the surrounding gas, $v_{\rm BH}$ is the velocity of the black hole relative to the gas, $\dot{M}_{\rm Edd}$ is the Eddington accretion rate and $L_{\rm X}$ is calculated from the thermal state and cooling time of the non-star-forming gas cells. We define the black hole radiative efficiency parameter to be $\epsilon _r = 0.2$ and the fraction of released energy that couples thermally to the gas to be $\epsilon _f=$ 0.07, respectively.

Feedback from black holes is implemented in two continually active phases: a radio mode and a quasar mode. In both phases, the energy injected into the gas is given by

\begin{equation}
\dot{E} = \epsilon _f \epsilon _r \dot{M}_{\rm BH} c^2.
\label{edot}
\end{equation}
For the quasar mode,  the thermal energy given by Equation (\ref{edot}) is injected isotropically into neighbouring gas cells. For the radio mode, bubbles of gas are gently heated at randomly placed locations {\bff following an inverse square distance profile around the black hole, up to a maximum radius of $0.8R_{\rm vir}$.} The bubbles are chosen to be 0.1 times the virial radius in size, from which the thermal energy budget for each bubble is estimated as

\begin{equation}
E_{\rm b} = \sum _i 0.05 m_i \overline{u}_{\rm ICM},
\end{equation}
where $\overline{u}_{\rm ICM}$ is the mean thermal energy of gas inside the halo. The number of bubbles is then calculated from the total energy available for radio mode feedback based on the mass accretion rate onto the black hole. In our simulations, both the quasar and radio feedback modes are active at all times. 


Magnetic fields are implemented following the method described in \citet{PakS13}. A homogeneous magnetic field of $10^{-14}$ (comoving) Gauss is seeded at $z=127$ in the simulation box $Z'$-coordinate direction. The choice of direction and strength has been shown to have little effect on the evolution \citep{PMS14,MVM15}. The divergence cleaning scheme of \citet{PR99} is implemented to ensure that $\nabla\cdot\boldsymbol{\rm B}\sim 0$. {\bff We note that the presence of magnetic fields has little effect on the global stellar disc properties discussed in this paper. A study of the magnetic field evolution will be presented in Pakmor et al. in prep.}



\section{Disc structure}
\label{sec3}

\begin{figure*} 
\centering
\includegraphics[scale=1.,trim={1.cm 3.cm 1.5cm 0.5cm},clip]{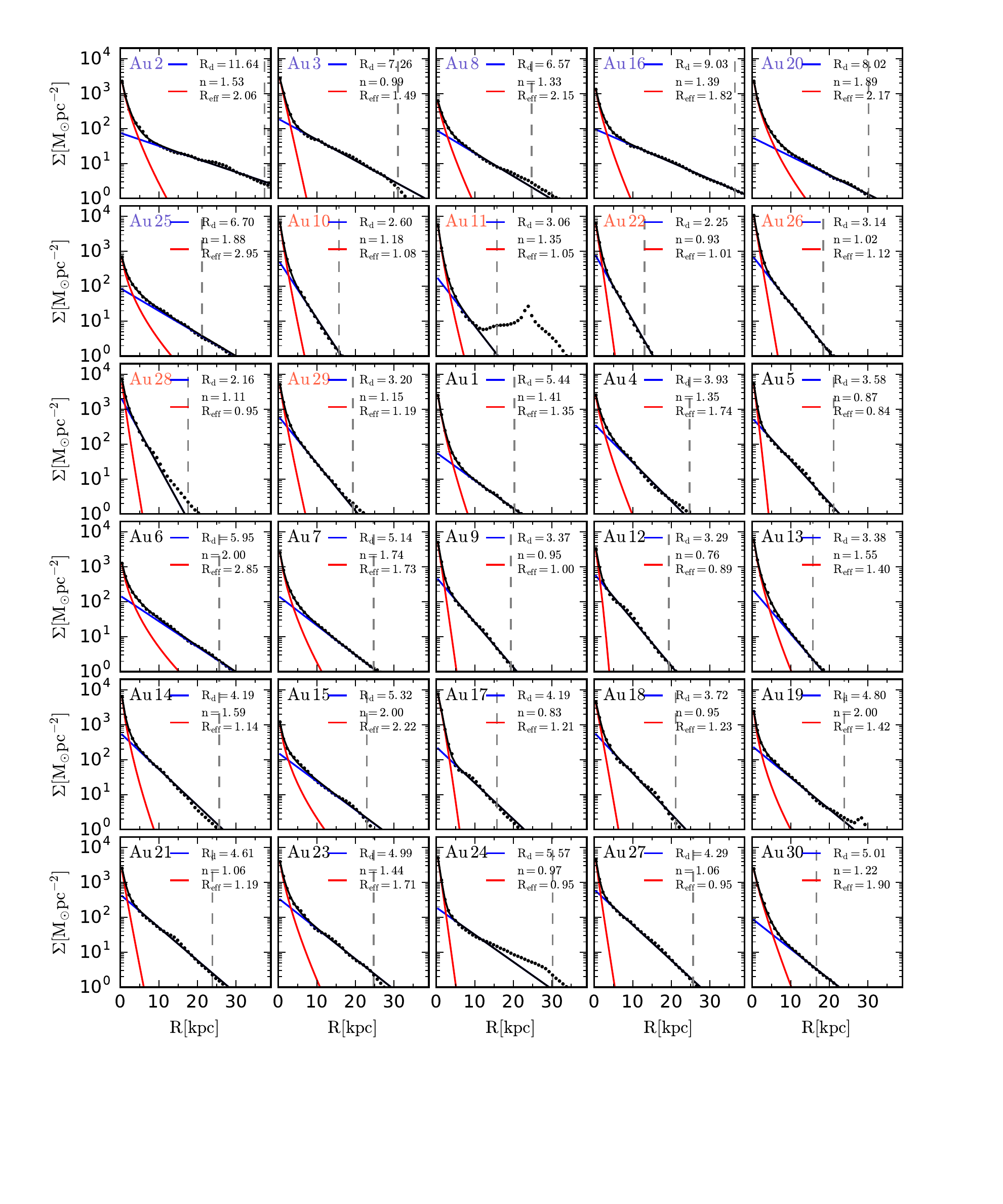}
\caption{Face-on stellar surface density profiles for stellar mass within $\pm 5$ kpc of the mid plane in the vertical direction for all simulations at $z=0$. The profiles are simultaneously fit with a \citet{S63} (red curve) and exponential (blue curve) profile using a non-linear least squares method. The total fitted profile is indicated by the black curve. The fit is carried out to the optical radius (vertical dashed line), defined to be the radius at which the $B$-band surface brightness drops below $\mu _B = 25$ $\rm mag \, arcsec^{-2}$. The first (second) six panels with blue (red) labels represent the discs with the largest (smallest) scale lengths, and are grouped together to facilitate reference. The sample has a wide range of disc scale lengths, from 2.16 kpc to 11.64 kpc.}
\label{sden}
\end{figure*}

\begin{figure*} 
\centering
\includegraphics[scale=0.62,trim={0 0 0.2cm 0},clip]{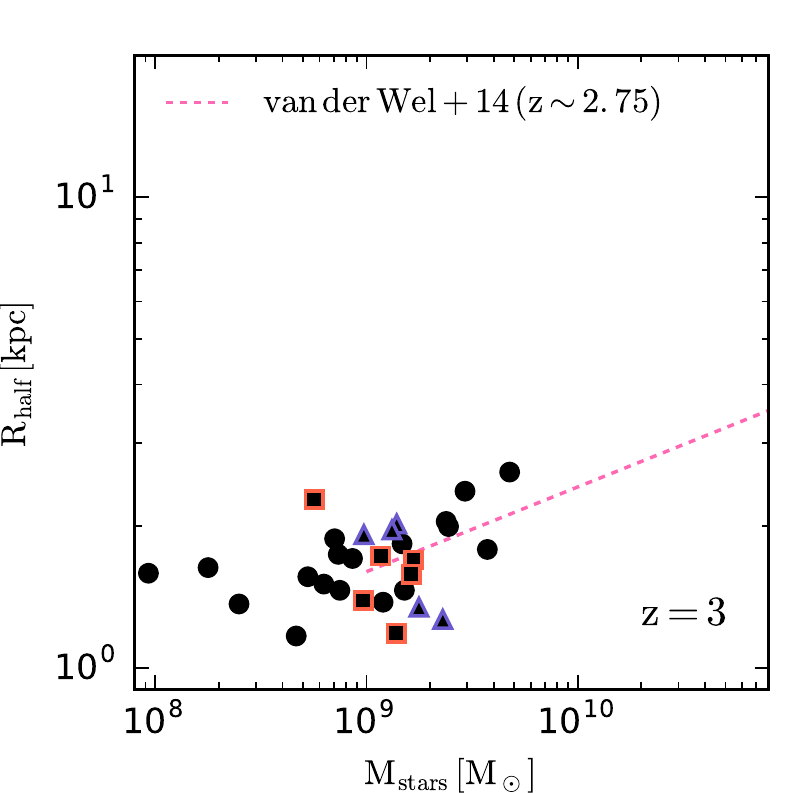}
\includegraphics[scale=0.62,trim={1.2cm 0 0.2cm 0},clip]{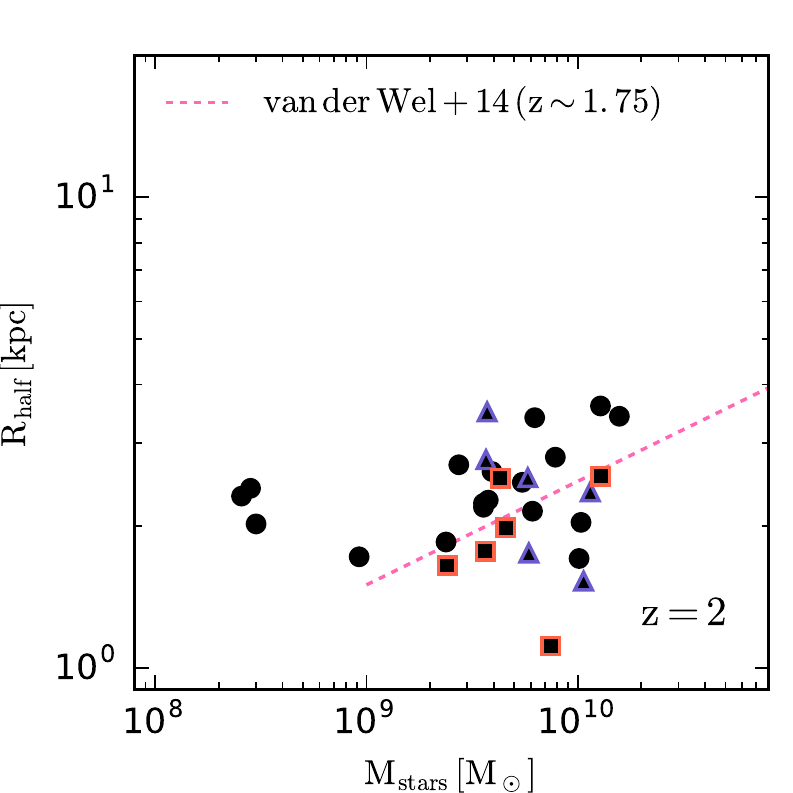}
\includegraphics[scale=0.62,trim={1.2cm 0 0.2cm 0},clip]{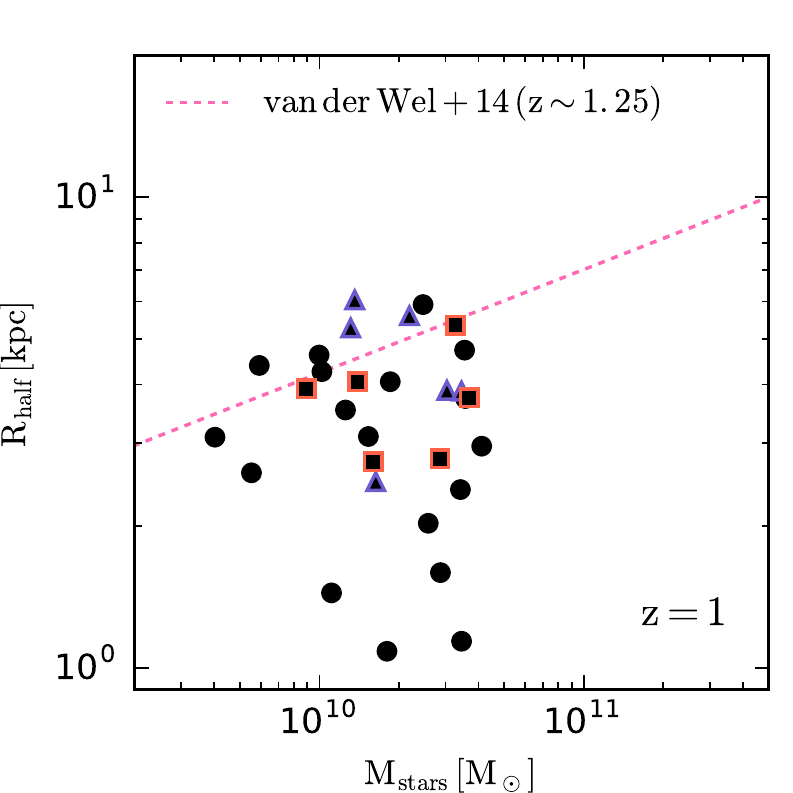}
\includegraphics[scale=0.62,trim={1.2cm 0 0.2cm 0},clip]{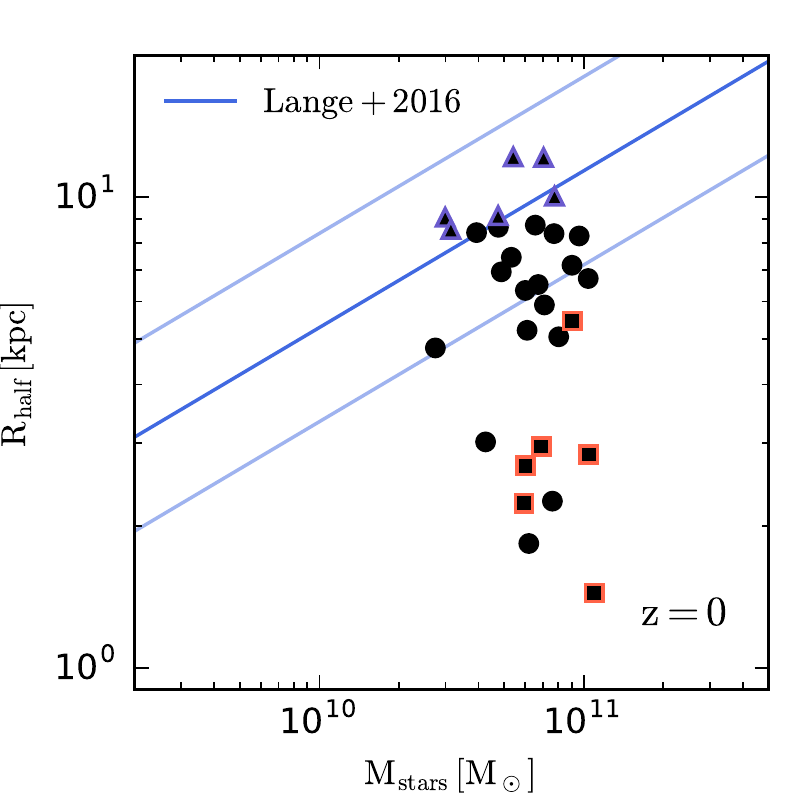}
\caption{The stellar half-luminosity radius measured in the $r$-band is plotted against stellar mass for all simulations{\bff, at a series of redshifts. The best-fit curves presented in \citet{vdW14} for CANDELS survey late-type galaxies are shown for $z = 1$-$3$, and the best-fit (0.2 dex scatter) curve presented in} \citet{LMD16} for GAMA survey late-type galaxies are shown by the (light) blue line, {\bff for $z=0$}. {\bff The discs with the six largest (smallest) scale lengths (see Fig.\ref{rscale})} are indicated by the triangles with blue outline (squares with red outline). }
\label{masssize}
\end{figure*}

\begin{figure}
\centering
\includegraphics[scale=1.,trim={0 0 0 0},clip]{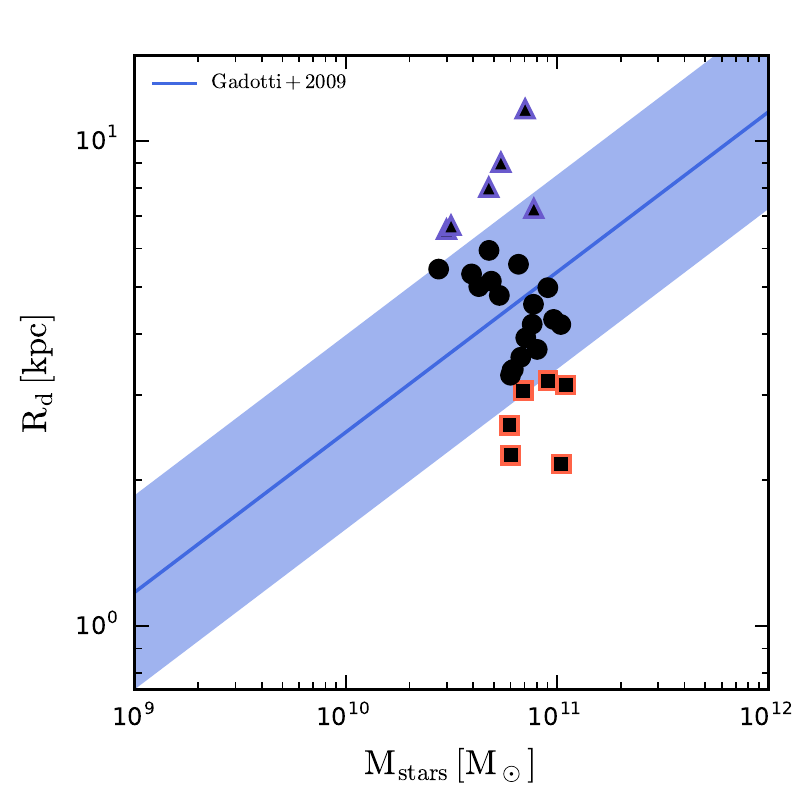} 
\caption{Disc radial scale length plotted against stellar mass for all simulations. The best-fit Sloan Digitial Sky Survey (SDSS) Data Release 2 relation of \citet{G09} is shown by the blue line with a scatter of 0.2 dex indicated by the blue shaded region. }
\label{rscale}
\end{figure}

In Figs.~\ref{proj1}, \ref{proj2} and \ref{proj3}, we present face-on and edge-on projections of all thirty simulated haloes at $z=0$. The spin- (or $Z$-)axis of star particles is found by calculating the dot product of the eigenvectors of the moment of inertia tensor of star particles within $0.1R_{200}$ with the angular momentum vectors of the same star particles in the $X'$, $Y'$, $Z'$ simulation box coordinate reference frame. The eigenvector of the inertia tensor that is most closely aligned with the principal angular momentum axis is chosen as the $Z$-axis. The images are a superposition of the $K$-, $B$- and $U$-band luminosities (mapped to the red, green and blue colour channels, respectively), which indicate the distribution of older (redder colours) and younger (bluer colours) star particles, respectively. In the majority of cases, a radially extended, young star-forming disc component typical of late-type disc galaxies is visible, in addition to clear non-axisymmetric structures such as bars and spiral arms. A red spheroidal bulge is a common feature of the simulations also, as can be clearly seen in the edge-on projections. The bulge component appears to be subdominant in many of the haloes; however, there are some notable exceptions. In particular, haloes Au 13 and Au 30 do not show extended discs, but rather a spheroidal morphology. Au 11 appears to be a special case owing to an imminent major merger.

\begin{figure*} 
\centering
\includegraphics[scale=1.,trim={1.cm 3.cm 1.5cm 0.5cm},clip]{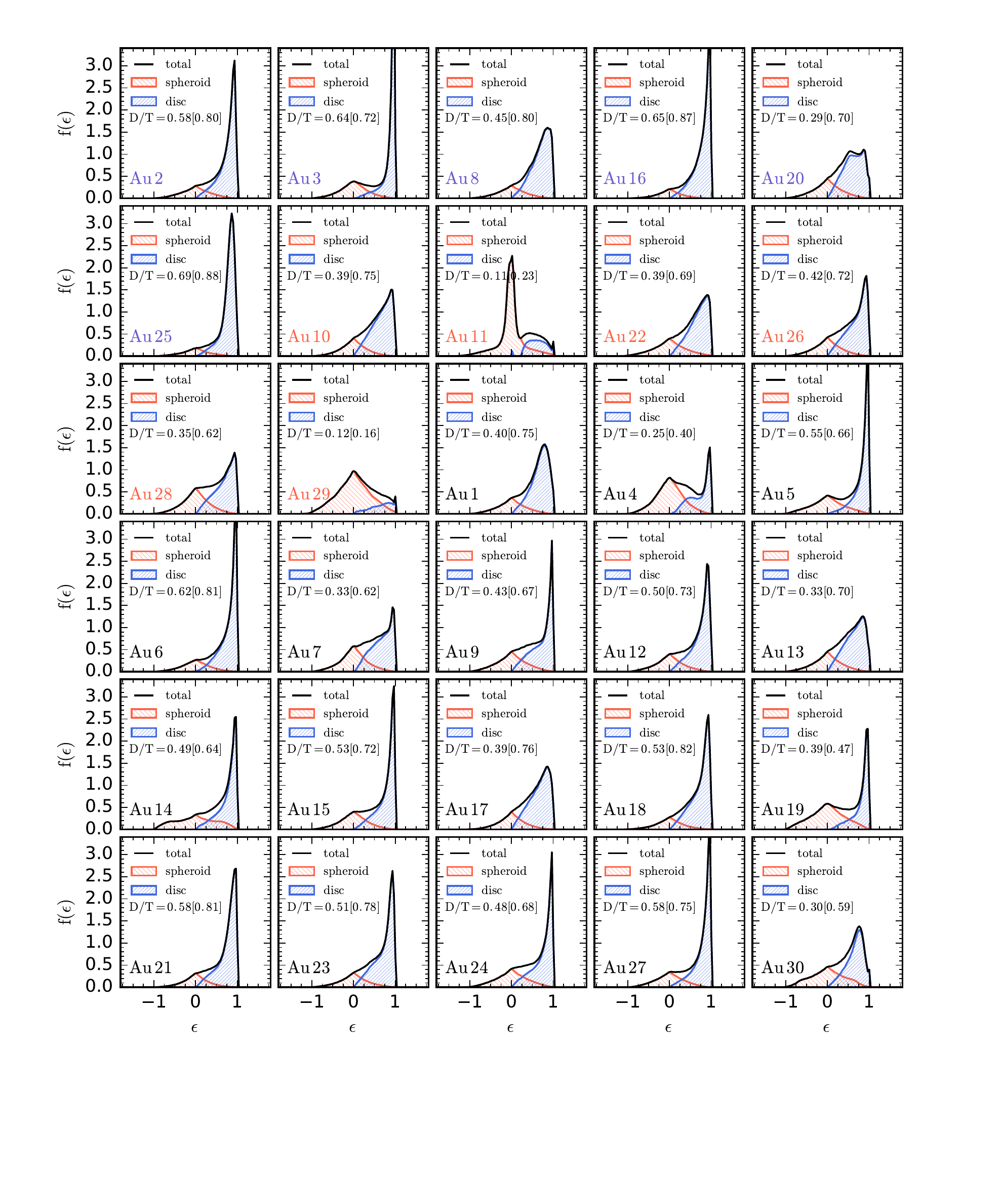}
\caption{The orbital circularity distribution for all simulations at $z=0$. The spheroidal component (red hatches) is calculated assuming the bulge has no net rotation, and is therefore taken to be the material with negative $\epsilon$ mirrored about the $\epsilon = 0$ value (or twice the mass of the counter-rotating material). The disc (blue hatches) is then assumed to make up the remaining part of the distribution. The disc-to-total mass ratio ($D/T$) that results from this decomposition is indicated by the bracketed value in each panel. The unbracketed value represents the decomposition that follows from assuming the disc to be composed of those star particles satisfying $\epsilon > 0.7$, which is generally a smaller fraction than the bracketed value.}
\label{eps}
\end{figure*}

Fig.~\ref{sden} shows stellar surface density profiles, for stellar mass within $\pm 5$ kpc of the mid plane in the vertical direction, for all simulations at $z=0$. The profiles are simultaneously fit to the sum of a \citet{S63} and exponential profile using a non-linear least squares method \citep[as in][]{MPS14}. The total fitted profile is indicated by the black curve. The fit is carried out to the optical radius, defined to be the radius at which the $B$-band surface brightness drops below $\mu _B = 25$ $\rm mag \, arcsec^{-2}$. In general, the stellar profiles are well fit by this bulge-disc decomposition, and exhibit a wide range of disc scale lengths, from 2.16 kpc to 11.64 kpc, whereas the bulge effective radii, $R_{\rm eff}$, are typically between 1 and 2 kpc. The surface density at a radius of 8 kpc (similar to the solar radius) is in many cases a few tens of $\rm M_{\odot}$ pc$^{-2}$, which is similar to the value measured in the Milky Way \citep{FHP06}. We note that we applied this fitting procedure to the $U$-, $B$- and $V$- surface brightness profiles, and found similar results as for the stellar surface density. The best fitting parameters are listed in Table 1.

In Fig.~\ref{masssize}, we calculate the stellar half-luminosity radius measured in the $r$-band, and plot it as a function of stellar mass {\bff at a series of redshifts}. {\bff For $z>0$, we compare our galaxy sizes to the corresponding observational relation for late-type disc galaxies derived from the CANDELS survey by \citet{vdW14}. {\bff We note that all of the Auriga galaxies qualify as late-type galaxies according to the criteria followed in \citet{vdW14}.} On average, our simulated galaxies grow in size over time, and match the observations best at $z>1$, with a clear trend of increasing disc size with increasing stellar mass at $z=3$ (first panel) and $z=2$ (second panel) close to the locus of the observations. At $z=1$ (third panel), this trend weakens, and the scatter in half-luminosity radius increases as several galaxies become very compact. At $z=0$ (right panel),} we compare our simulated galaxies to the observational relation for late-type disc galaxies derived from the Galaxy And Mass Assembly (GAMA) survey by \citet{LMD16}. {\bff According to the classification criteria followed in \citet{LMD16}, 22 of the Auriga galaxies qualify as late-type galaxies, most of which lie within 0.2 dex scatter of the \citet{LMD16}} relation{\bff ; thirteen galaxies lie below the scatter. We note that those with the shortest scale lengths (see Fig.~\ref{rscale}) that belong to this compact group of galaxies are not the most compact at earlier times.} Several (but not all, see Section \ref{discsizes}) of the {\bff these galaxies} have experienced a major merger at $z < 1$, and some even at around $z\sim 0.5$, which creates a massive bulge component {\bff and explains their transition to compact mass distributions}. Bearing this in mind, the majority of simulated galaxies agree well with the observed sizes of late type galaxies, as shown in Fig.~\ref{masssize}.


In Fig.~\ref{rscale}, we show the disc scale lengths of the simulated galaxies as a function of their stellar masses. We compare with the observations of disc galaxies presented in \citet{G09}, who used dust-corrected stellar masses of galaxies selected from the SDSS Data Release Two derived from \citet{KHW03}. The simulation mean appears to lie close to the best-fit line found from observations, and the significant variation of scale length in the simulation suite is evident. In Section \ref{discsizes} we examine the physical processes that drive this scatter, therefore we highlight the six largest (smallest) discs in blue triangles (red squares) here and in subsequent plots in order to emphasize how the disc size relates to other variables.

In addition to the surface density decomposition shown in Fig.~\ref{sden}, we use a kinematic decomposition of the star particles to infer the disc and bulge mass. The kinematic decomposition entails the calculation of the circularity parameter, $\epsilon$, for all star particles according to the procedure outlined in \citet{ANS03},

\begin{equation}
\epsilon = \frac{L_z}{L_{z,\rm max}(E)},
\end{equation}
where $L_z$ is the $z$-component of angular momentum of a given star particle and $L_{z,\rm max}(E)$ is the maximum angular momentum allowed for the orbital energy, $E$, of the same star particle. This ensures that a prograde circular orbit in the disc plane takes the value $\epsilon=1$, retrograde circular orbits correspond to $\epsilon = -1$, and $\epsilon \sim 0$ indicates orbits with a very low $z$-component of angular momentum, which may be highly inclined to the disc spin axis and/or be highly eccentric. 

From the distributions shown in Fig.~\ref{eps} we calculate the $D/T$ values from two different methods: the first follows the assumption made in \citet{ANS03} that the bulge component has zero net rotation and is therefore centred around $\epsilon=0$. The bulge mass can therefore be inferred by doubling the mass of the counter-rotating material. The disc is then assumed to make up the remaining part of the distribution. The $D/T$ values inferred from this method are indicated in each panel (the bracketed values), and are typically around 0.7 to 0.8. {\bff However, \citet{OSD16} showed that $D/T$ values inferred from this method overestimate the $D/T$ values inferred from Gaussian mixture models of dynamical variables by about 5-20$\%$.} The second method assigns all star particles that satisfy $\epsilon > 0.7$ to the disc. This $D/T$ value is typically lower than that of the first method because only the kinematically coldest star particles are considered to be part of the disc, which means there is a large swathe of particles in the range $0 < \epsilon < 0.7$ that are assigned to a spheroidal component. The true value of $D/T$ probably lies between the two values. From Fig.~\ref{eps} we can infer that the majority of the simulated haloes host prominent disc components with relatively small bulges. However, there are notable exceptions that contain a large bulge, for example, Au 29 is almost completely dominated by a spheroidal stellar component, which is caused by a recent merger.

\begin{figure*} 
\centering
\includegraphics[scale=1.,trim={1.cm 3.cm 1.5cm 0.5cm},clip]{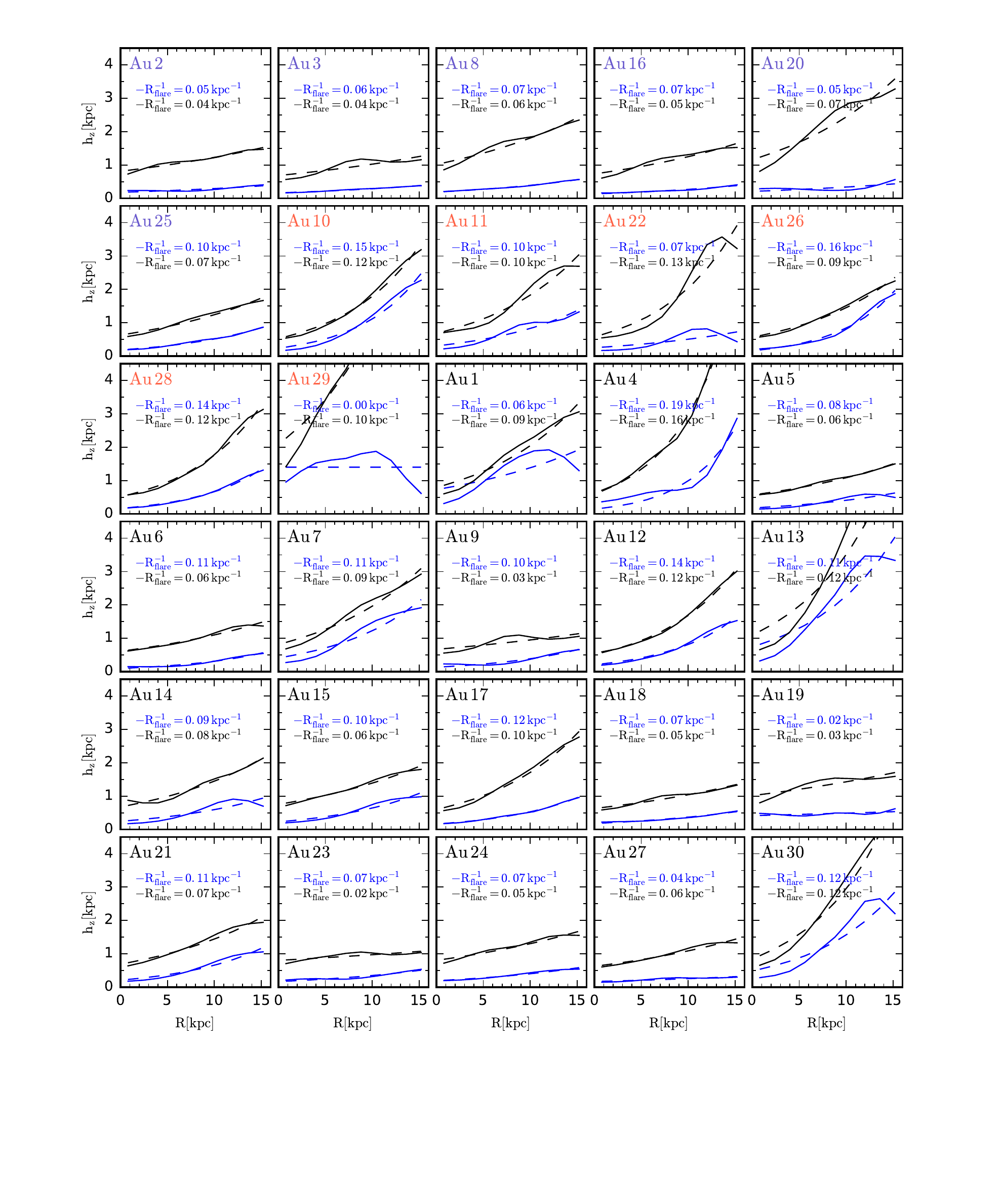}
\caption{The vertical scale height of {\bff all star particles (black) and star particles younger than 3 Gyr (blue) as a function of radius. The scale height at a given radius is calculated by fitting a function composed of two exponentials and a break radius (see text for details).} Each radial profile of vertical scale heights is fit with an exponential function: $A\exp{(-R/R_{\rm flare})}$,  where a large value of $-R_{\rm flare}^{-1}$ indicates a high degree of flaring of the vertical density distribution. These fits are indicated by the dashed lines.}
\label{vstruct}
\end{figure*}

\begin{figure} 
\centering
\includegraphics[scale=1.13,trim={0 0 0 0.7cm},clip]{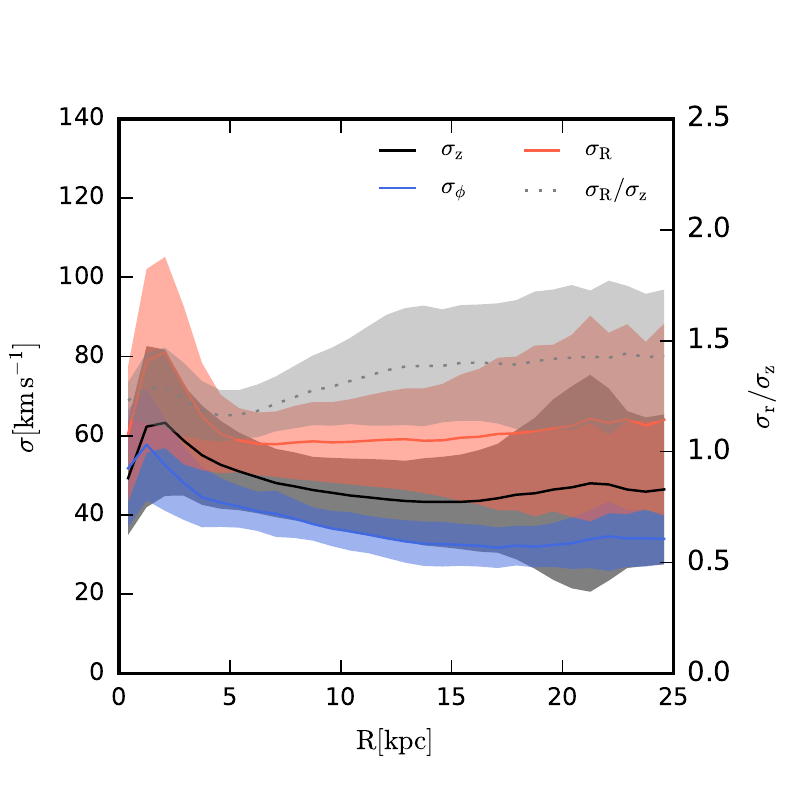}
\caption{The radial (red), tangential (blue) and vertical (black) velocity dispersion profiles of all simulations at $z=0$. The ratio between the radial and vertical velocity dispersions is indicated by the dashed grey curve, and is generally found to be between 1 and 1.5, close to the observational value of $\sim \sqrt{2}$ for the Milky Way in the Solar Neighbourhood. All dispersions are, however, larger than observed near the Sun.}
\label{vdisp}
\end{figure}

We examine the vertical structure of the discs by {\bff fitting their vertical density distribution of the disc at a series of radii. The fitting function we choose is defined by two exponentials and a break radius, which has the flexibility to fit density profiles that are well described by either a single or double exponential\footnote{We note that fitting a single exponential produces similar scale height values in many cases.}. To remove the effect of warps, we bin star particles in a 2D cartesian grid in the disc plane with cells of side length $\sim 2$ kpc, and calculate the mean vertical height in each cell. We then subtract the corresponding mean height from each particle such that the peak density is located in the disc midplane everywhere\footnote{We note that some of the Auriga discs are strongly warped \citep{GWG16}, which can give the false impression of flaring when the vertical density distribution is averaged over azimuth.}.} The radial profiles of the disc scale height for all galaxies are shown in Fig.~\ref{vstruct}. The star particles considered in this figure are those that are kinematically defined to be part of the disc, specifically those that satisfy $\epsilon > 0.7$. This sample is further subdivided into a young stellar age group, defined as star particles younger than 3 Gyr. Fig.~\ref{vstruct} highlights that a common feature in the simulation sample is the presence of discs with scale heights of about 1 kpc at radii less than $\sim 8$ kpc, {\bff which is thicker than the accepted value of $\sim 300$ pc for the thin disc of the Milky Way, and similar to the observed value of the Milky Way's thick disc \citep[e.g,][]{JIB08}}. The younger star population, as expected, is thinner than older star populations, {\bff and in many cases has scale heights of $\sim 200$ - $300$ pc within $R \sim 8$ kpc, which is similar to the thin disc of the Milky Way. \citet{GSG16} showed that this trend is a combination of dynamical scatter of star particles and upside-down formation of the disc through the cooling and accretion of star-forming gas. The thinning of the gas disc (and therefore newborn star particles) with time is likely caused by the decreasing merger activity and star formation rate \citep[see Fig. 8 of][]{MGP16}. Another common feature is a flaring scale height profile for both old and young star particles. The latter is very similar to the flaring star forming gas distribution at $z=0$, which suggests that a significant amount of disc flaring of a stellar population is set at birth. }



In Fig.~\ref{vdisp} we show the radial profiles of the radial, tangential and vertical velocity dispersions of \emph{all} the star particles (not only those kinematically assigned to the disc). In most cases the profiles rise then drop too steeply in the innermost few kpc. Outside of this region, the flaring disc dominates the stellar distribution and the profiles flatten out at radii larger than 10 kpc. The values of the velocity dispersion in all directions at the solar radius {\bff (mean values of $\sigma _R \sim$ 60 $\rm km \, s^{-1}$, $\sigma _Z \sim$ 50 $\rm km \, s^{-1}$)} are too high compared to the {\bff thin disc of the Milky Way: $\sigma _R \sim 35$ $\rm km \, s^{-1}$; $\sigma _Z \sim 25$ $\rm km \, s^{-1}$ \citep[see][and references therein]{BHG16}, although are similar to those values inferred for the thick disc: $\sigma _R \sim \sigma _Z \sim 50$ $\rm km \, s^{-1}$.} The ratio between the radial and vertical velocity dispersions is in nearly all cases similar to the value $\sqrt{2}$ observed in the Milky Way.

\begin{figure} 
\centering
\includegraphics[scale=1.,trim={0 0 0 0},clip]{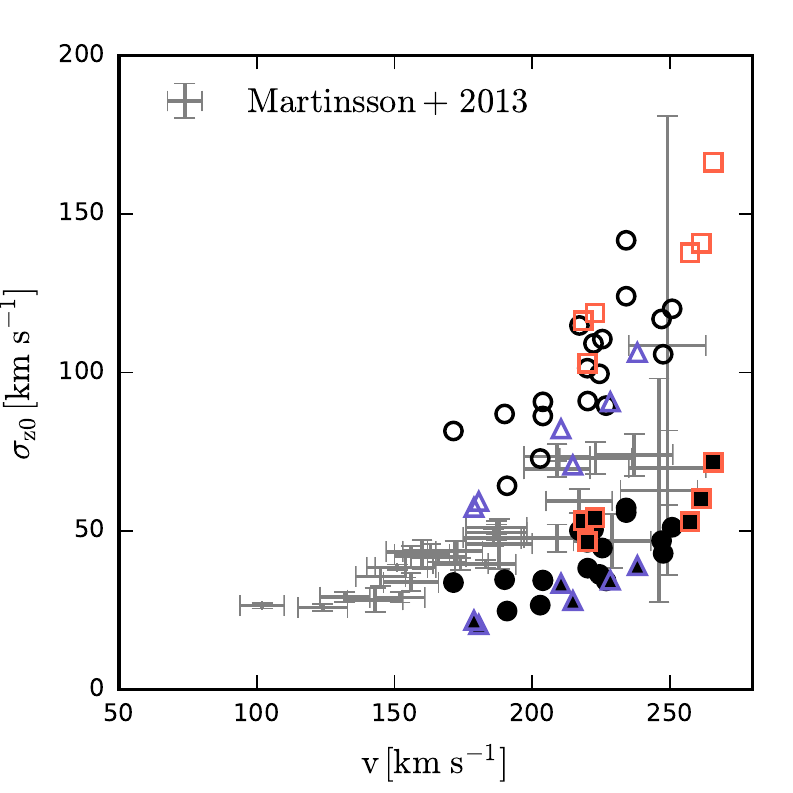}
\caption{The central vertical velocity dispersion of star particles as a function of the circular velocity at the optical radius for all simulated galaxies. The filled (empty) symbols indicate the velocity dispersion of all (disc) star particles within the central kiloparsec of the galaxies. Observational data and error bars (grey points) derived by \citet{MVW13} represent a kinematically cold subsample of disc galaxies from the DiskMass survey.}
\label{sigvc}
\end{figure}

{\bff Following \citet{OSD16}, we exploit data from the DiskMass Integral Field Unit spectroscopic survey \citep{BVS10} in order to compare the vertical structure of our simulated galaxies to external disc galaxies. Specifically, we make use of results obtained by \citet{MVW13} who analysed a subsample of disc dominated galaxies from the DiskMass Survey and presented a correlation between their central vertical velocity dispersion, $\sigma _{z0}$, and the flat part of their rotation curves. In Fig.~\ref{sigvc}, we show $\sigma _{z0}$ calculated for all star particles as a function of the circular velocity at the optical radius, which in most cases coincides with the flatter, outer rotation curve. The simulations have systematically higher velocity dispersions than the observations. However, we note that the galaxy sample analysed in \citet{MVW13} make up the kinematically coldest discs in the DiskMass survey, with no significant bulges or bars. We therefore calculate $\sigma _{z0}$ for all disc star particles ($\epsilon > 0.7$), which brings most simulations with $v > 220$  $\rm km \, s^{-1}$ in line with observations and most with $v < 220$  $\rm km \, s^{-1}$ below the relation. We note that the largest (smallest) discs tend to have lower (higher) $\sigma _{z0}$. This conservative selection of disc star particles likely underestimates the true value of $\sigma _{z0}$ for the disc, which probably lies between the two sets of simulated points in Fig.~\ref{sigvc}. We conclude that the vertical thickness of the inner parts of our simulated galaxies is similar to observations, although on average the outer parts seem to be thicker compared to the Milky Way. }

\begin{figure} 
\centering
\includegraphics[scale=1.,trim={0 0 0 0},clip]{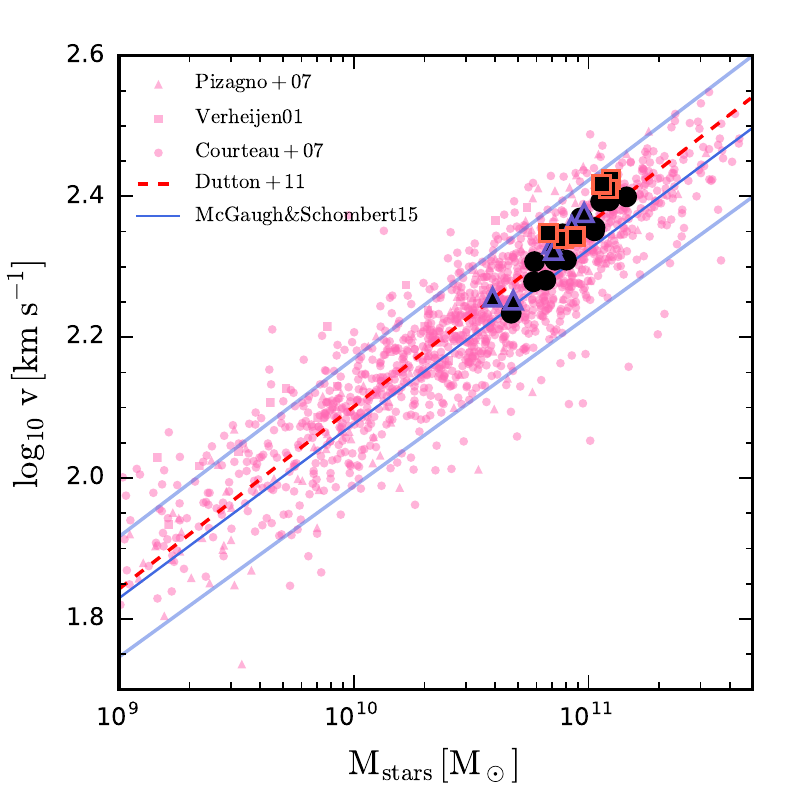} 
\caption{The Baryonic Tully-Fisher relation of all simulated galaxies. The mass includes all the stars and gas within an optical radius, $R_{opt}$, and the velocity is equal to $\sqrt{(GM(<R_{opt})/R_{opt})}$. The optical radius is large enough in most cases to probe the flat part of the rotation curve (in haloes that have flat rotation curves). The results of several observational studies: \citet{V01,CDV07} and \citet{PPW07}, are included for comparison. All the simulated galaxies fall comfortably within the scatter of the observational relations shown.}
\label{tf}
\end{figure}

\begin{figure*} 
\centering
\includegraphics[scale=1.,trim={1.cm 3.cm 1.5cm 0.5cm},clip]{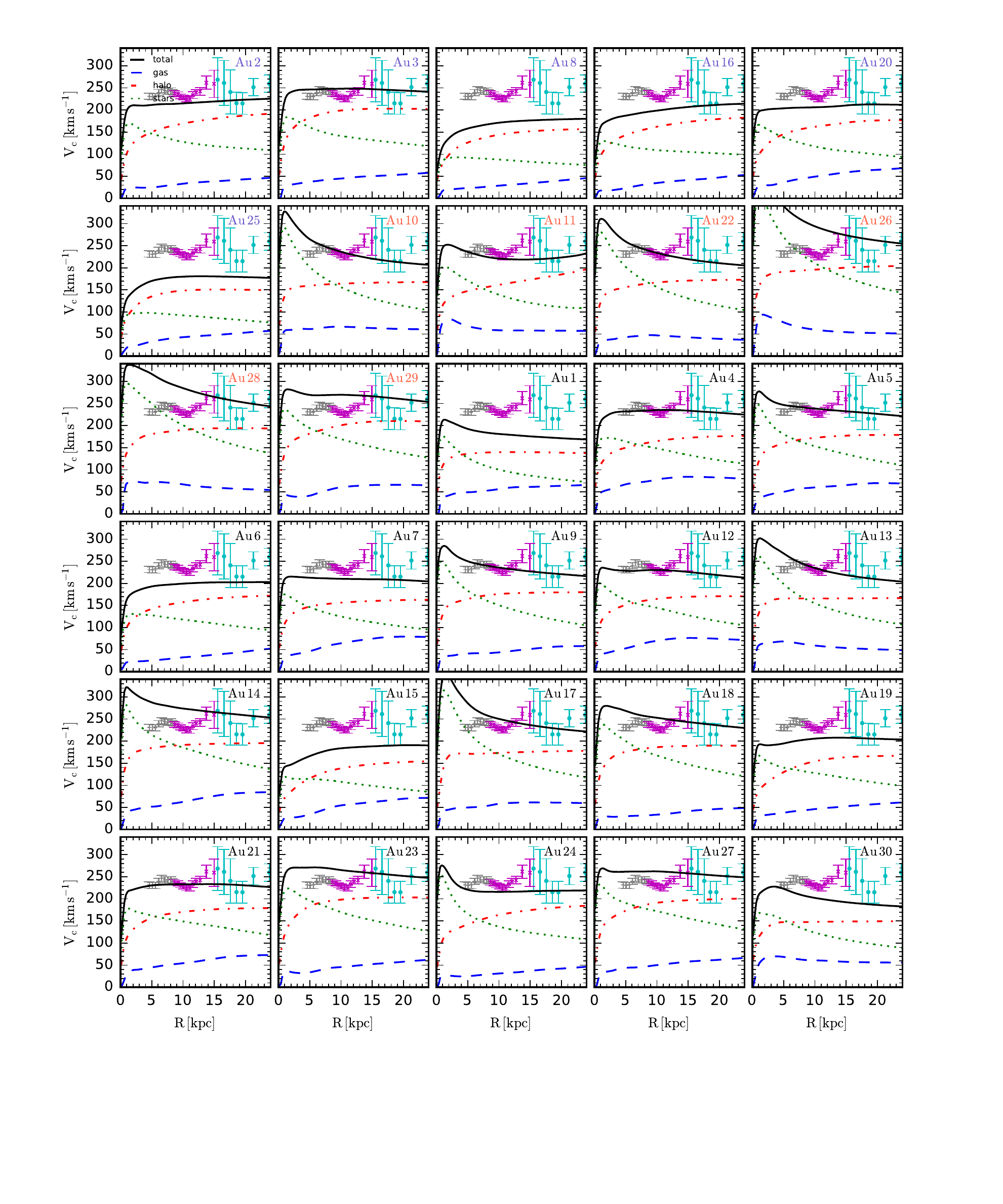}
\caption{Total circular velocity curves for all simulations (black lines), and for the constituent components: stars (green dotted), dark matter (red dot-dashed) and gas (blue dashed). For comparison, a compilation of observational data for the Milky Way from \citet{HLY16} is indicated. Different symbols refer to different sources; HI (grey pluses), red clump giant stars selected from {\bff LSS-GAC and} APOGEE (magenta crosses) and halo K giant stars selected from SEGUE (cyan circles). }
\label{vrot}
\end{figure*}

In Fig.~\ref{tf} we show the Baryonic Tully-Fisher (BTF) relation for all the simulations. The baryonic mass is taken as the sum of all star particles and gas elements within an optical radius, which contains the majority of stellar and cold gas material in the haloes. The circular velocity at the optical radius is measured, which is a large enough radius (in most cases) to be at the flatter, outer part of the rotation curve. For comparison, we include the observational BTF-relation from \citet{MS15}, who determine stellar masses using two methods: the first via spectral energy distribution fitting using population synthesis (consistent with a Chabrier IMF, which we use in this paper) constrained by 3.6 $\mu$m emission, and the second from the DiskMass data, which is independent of the IMF. We retain also the data presented in \citet{MPS14}, which includes the best fit of the dataset determined in \citet{DBF11}. The slope for the simulated galaxies is very similar to that of the set of observations shown in Fig.~\ref{tf}, and they lie well within the scatter. We tested different measures of the rotation velocity including taking the circular velocity at various observationally motivated radii, and verified that there is no significant variation in the trends. We conclude that the consistency of the simulated haloes with the BTFR indicates that the Auriga haloes host rotationally dominated disc galaxies with rotational velocities in agreement with observed disc galaxies.

In Fig.~\ref{vrot} we show the circular velocity curves of the simulations decomposed into stellar, gas and dark matter components. In many cases, the stellar component reaches a peak in the inner regions and steadily decreases with increasing radius, and the dark matter component rises with increasing radius. In most cases, the stellar and dark matter contribution to the profile balance in the outer disc regions, producing a near-flat rotation curve that persists beyond 20 kpc. The gas component comprises star-forming and non-star-forming gas, of which the former is rotationally supported, whereas the latter is warm or hot and rotates slowly. Together the gas components contribute a few tens of $\rm km \, s^{-1}$ to the circular velocity at large radii. Overall, the majority of cases exhibit circular velocity curves that compare well with the current observational data for the Milky Way compiled from HI surveys and stellar kinematics. There are, however, exceptions, such as Au 26 and Au 17, which exhibit high velocity peaks in the inner region as a result of very centrally concentrated stellar distributions in the form of either a nuclear bulge that formed as a result of a late-time merger (e.g., Au 28), or a strong stellar bar (e.g., Au 17).

\section{Disc sizes}
\label{discsizes}

\begin{figure} 
\centering
\includegraphics[scale=1.]{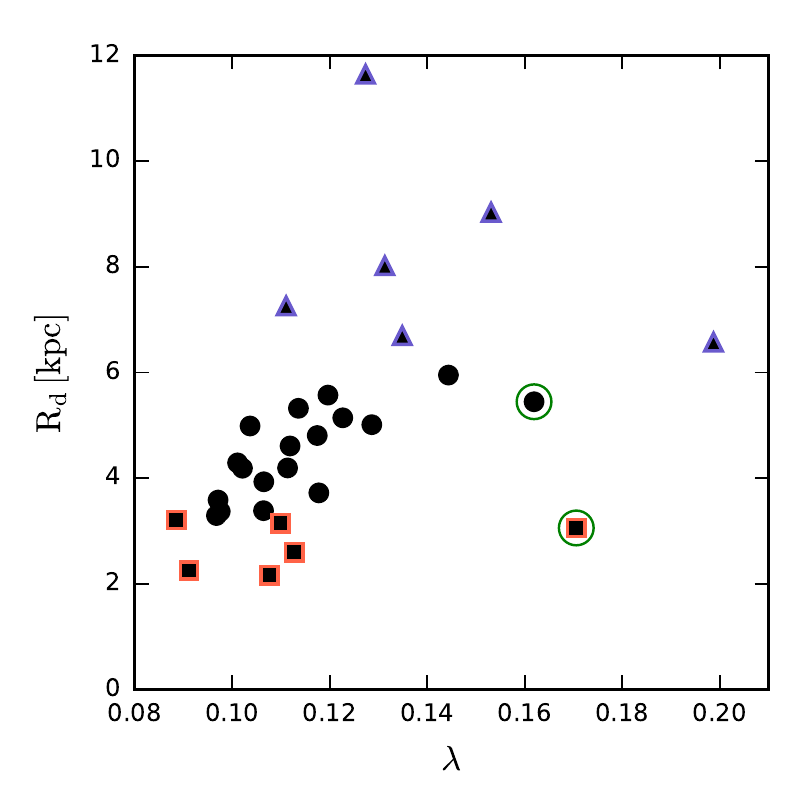}\\
\includegraphics[scale=1.]{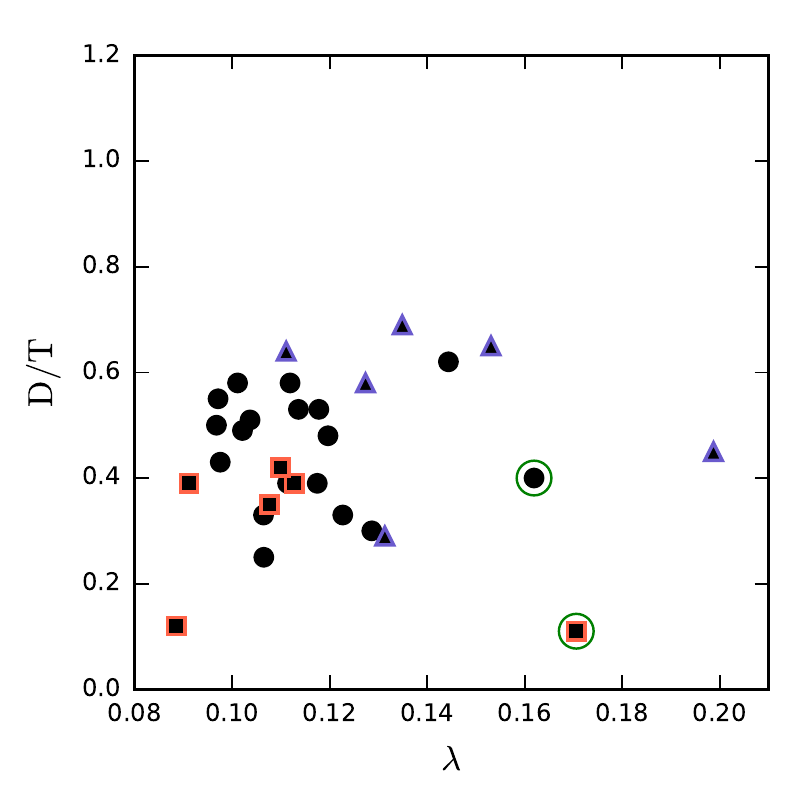}\\
\caption{The disc radial scale length (top panel) and the kinematically defined disc mass-total stellar mass ratio, $D/T$ (bottom panel), as a function of the halo spin parameter (see text for definition), for each simulation. Square (triangle) symbols of red (blue) outline represent the 6 smallest (largest) discs in terms of scale length. The symbols encircled in green highlight Au 1 and Au 11, both of which have massive companions that contribute to large spin values.}
\label{spin}
\end{figure}

As we have shown above in Fig.~\ref{sden}, there is a substantial spread in radial scale length among the simulated galaxies: from 2.16 kpc to 11.64 kpc. This is evident in the present-day stellar projections shown in Fig.~\ref{proj1}. In this section, we aim to determine the factors that control disc size by establishing the principal mechanism(s) that stimulate and suppress disc growth. 

\subsection{Angular momentum}

First, we investigate how discs are affected by the angular momentum of the halo and its various constituents, a dependence which is theoretically expected to be correlated \citep{P69,FE80,MMW98,BDK01}. In the top panel of Fig.~\ref{spin}, we show, for all simulations, the disc scale length as a function of halo spin, defined as

\begin{equation}
\lambda = \frac{l_z}{\sqrt{2}V_{\rm 200}R_{\rm 200}},
\end{equation}
where $l_z$ is the specific angular momentum of the all the material inside the virial radius, $R_{\rm 200}$, and $V_{\rm 200}$ is the circular velocity at $R_{200}$. A positive correlation is present: discs of larger radial scale lengths tend to be hosted by haloes of greater spin\footnote{{\bff \citet{HSD15} found a qualitatively similar correlation for spin values between 0.02 and 0.1, using a simulation suite of isolated haloes that undergo spherical collapse.}}. There is a significant scatter, some of which can be understood by consideration of some outlying haloes. For example, the encircled symbols represent Au 1 and Au 11, both of which have nearby massive companions (in the case of Au 11, it is a massive merger) at $z=0$, which act to increase their spin parameters beyond their normal values. The radially extended disc galaxy with $\lambda \sim 0.2$ has a particularly large spin value owing to a late quiescent merger (see below), and a disc scale length lower than the inferred correlation in the top panel of Fig.~\ref{spin} owing to a low star formation rate at late times and at large radii, which prevents it from building up an extended disc. 

The positive correlation between halo spin and disc scale length at face value seems to contradict the conclusion reached in \citet{SWS09}, who found no correlation between the presence of discs and halo spin parameter in the eight Aquarius simulations re-simulated with a modified version of the SPH code Gadget 2 \citep{SP05}. However, the conclusions of \citet{SWS09} were based on a lack of correlation between disc mass-total mass ratio, $D/T$, and the halo spin. As we show in the bottom panel of Fig.~\ref{spin}, this is also the case for the Auriga simulations. The reason is that $D/T$ is sensitive to various factors, in particular to the interaction with subhaloes and mergers (even quiescent mergers if they occur at late times) that can easily drive down the ratio by perturbing disc star particles, and hence wash out a correlation. We conclude that the discs of the Auriga simulations do correlate with halo spin as expected from theories of inside-out formation, and are at the same time not inconsistent with the results of \citet{SWS09}.

\subsection{Quiescent mergers}
\label{qm}

\begin{figure*} 
\centering
 \hspace{-1.2cm}
\includegraphics[scale=0.59,trim={0 0.9cm 0.18cm 0},clip]{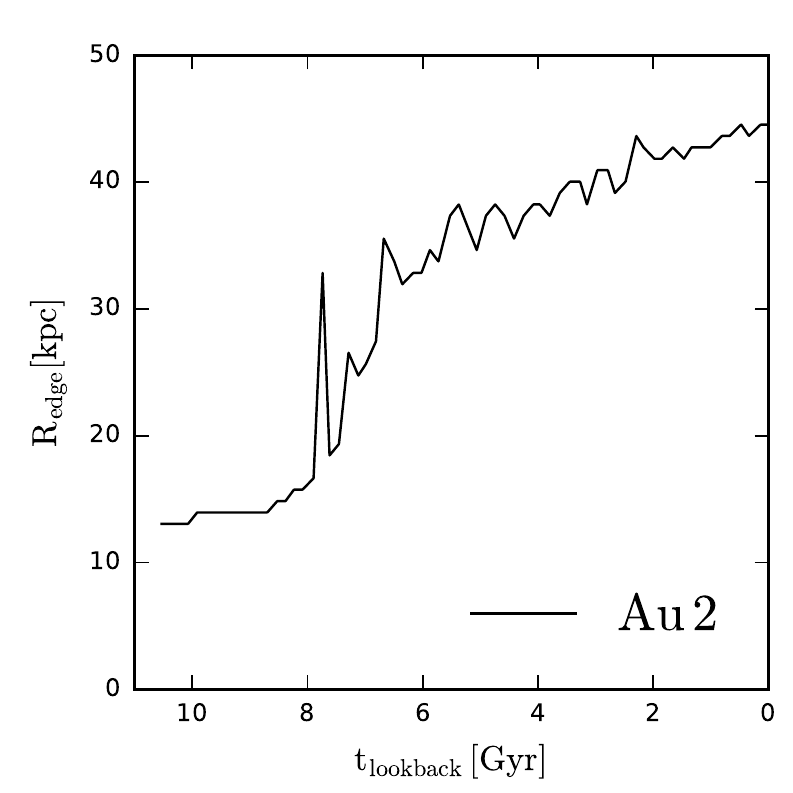} 
\includegraphics[scale=0.59,trim={1.23cm 0.9cm 0.18cm 0},clip]{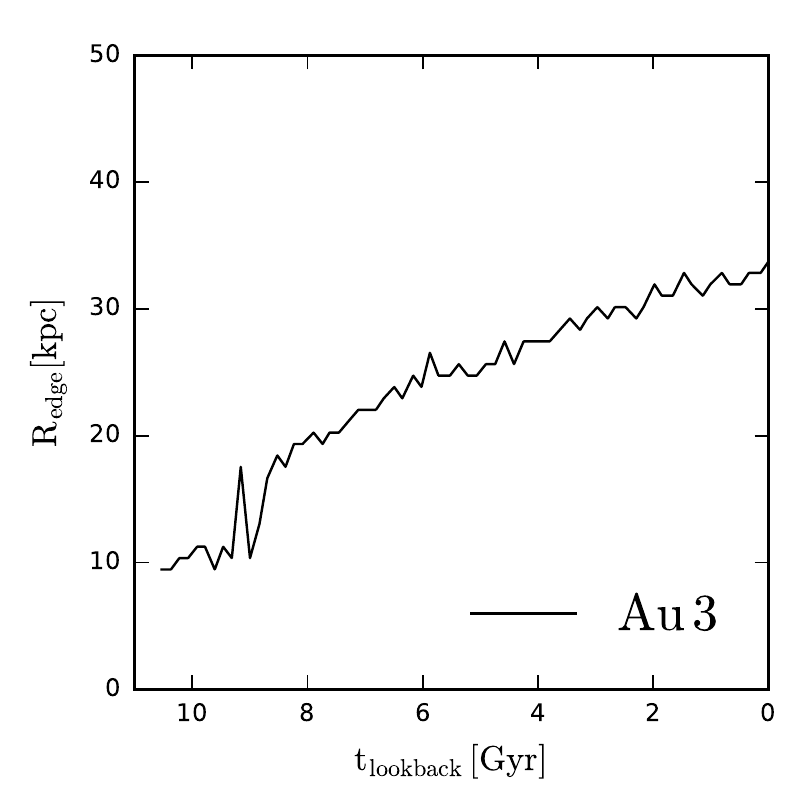}
\includegraphics[scale=0.59,trim={1.23cm 0.9cm 0.18cm 0},clip]{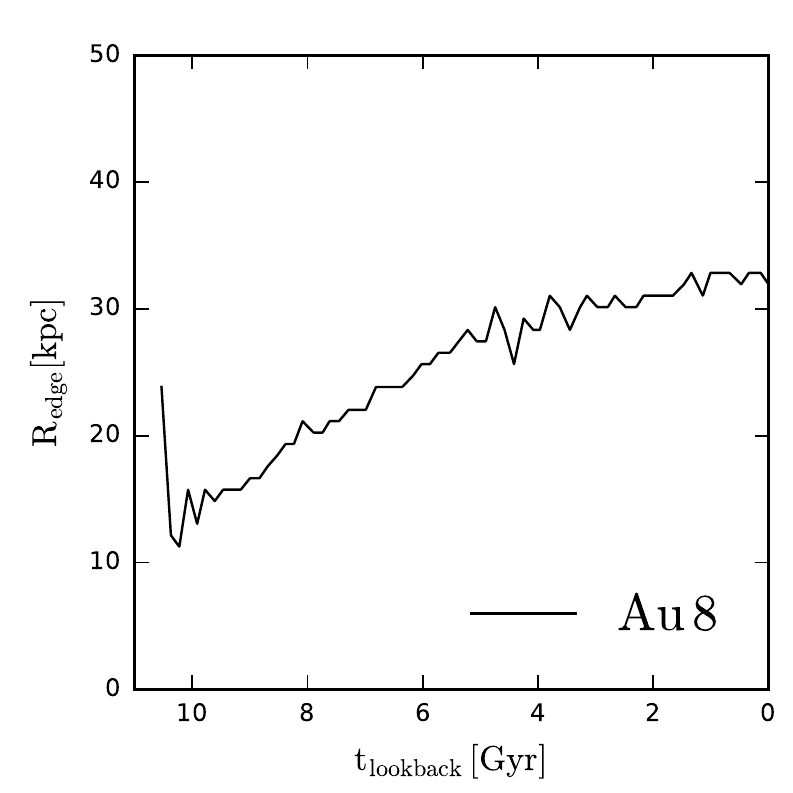} 
\includegraphics[scale=0.59,trim={1.23cm 0.9cm 0 0},clip]{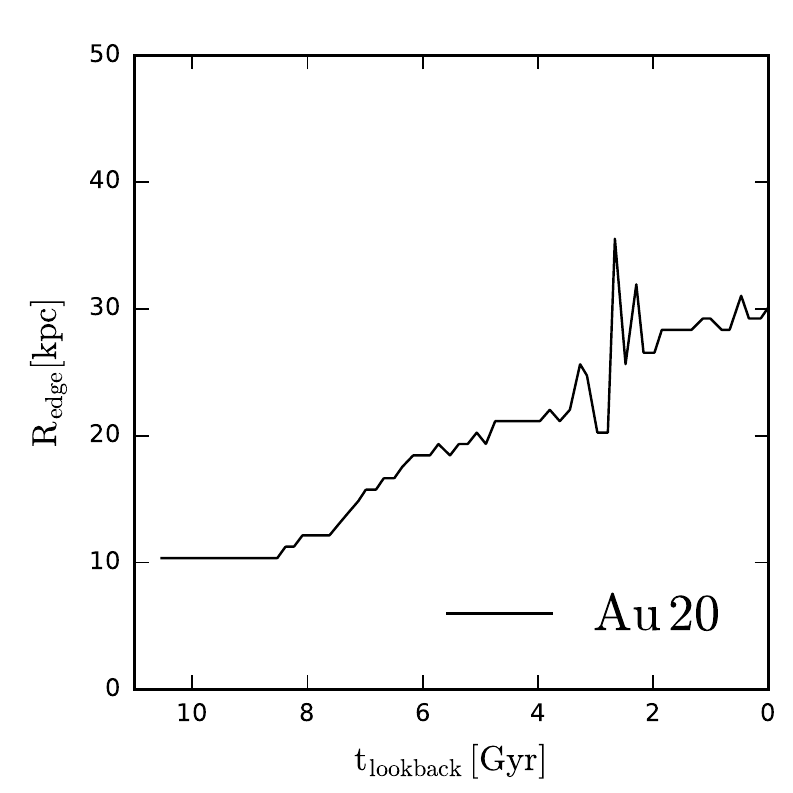} \\ \vspace{0.5cm} 
 
 \includegraphics[scale=0.28,trim={0.1cm 1.4cm 3.8cm 1.cm},clip]{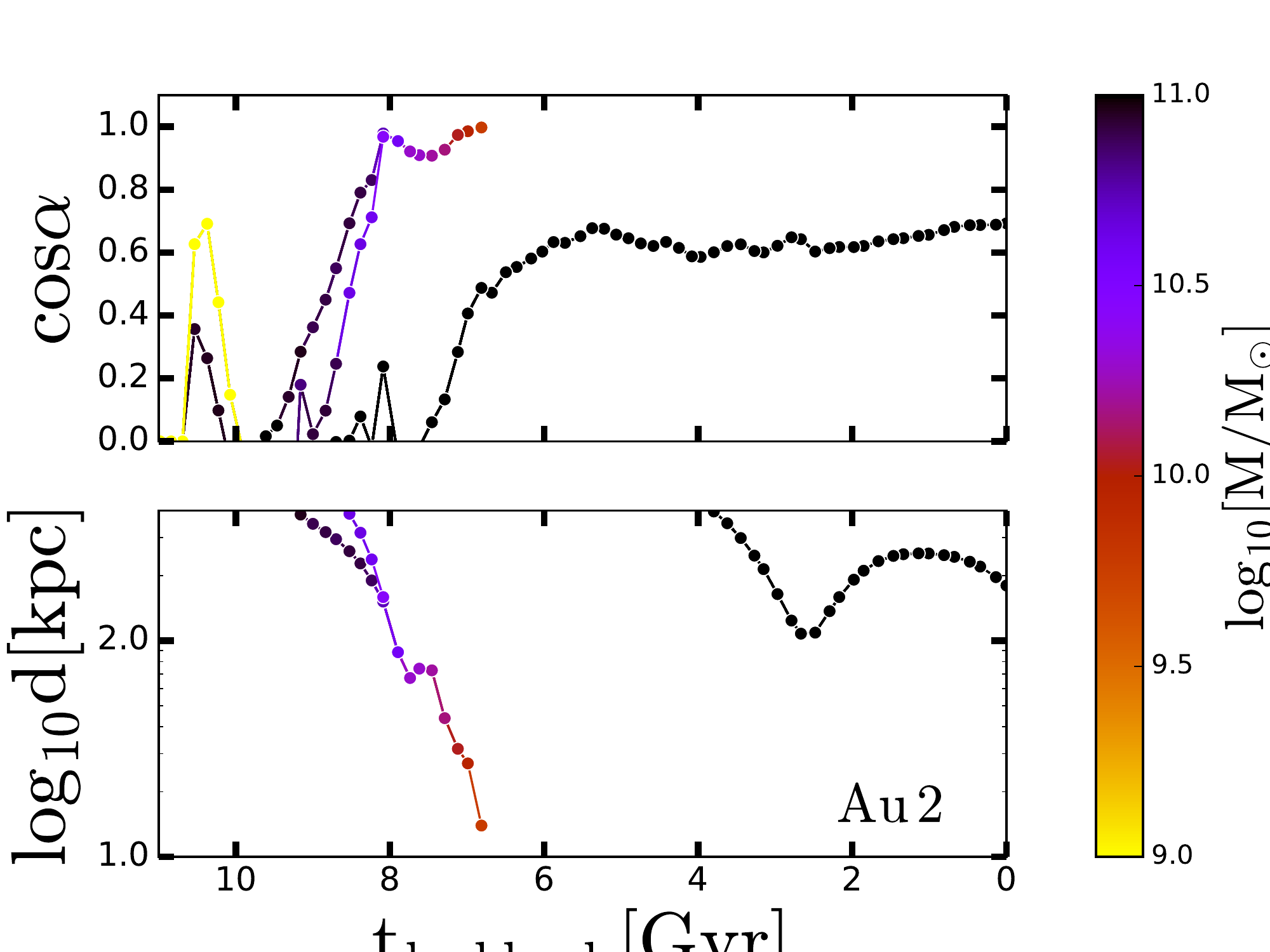}
\includegraphics[scale=0.28,trim={2.5cm 1.4cm 3.8cm 1.cm},clip]{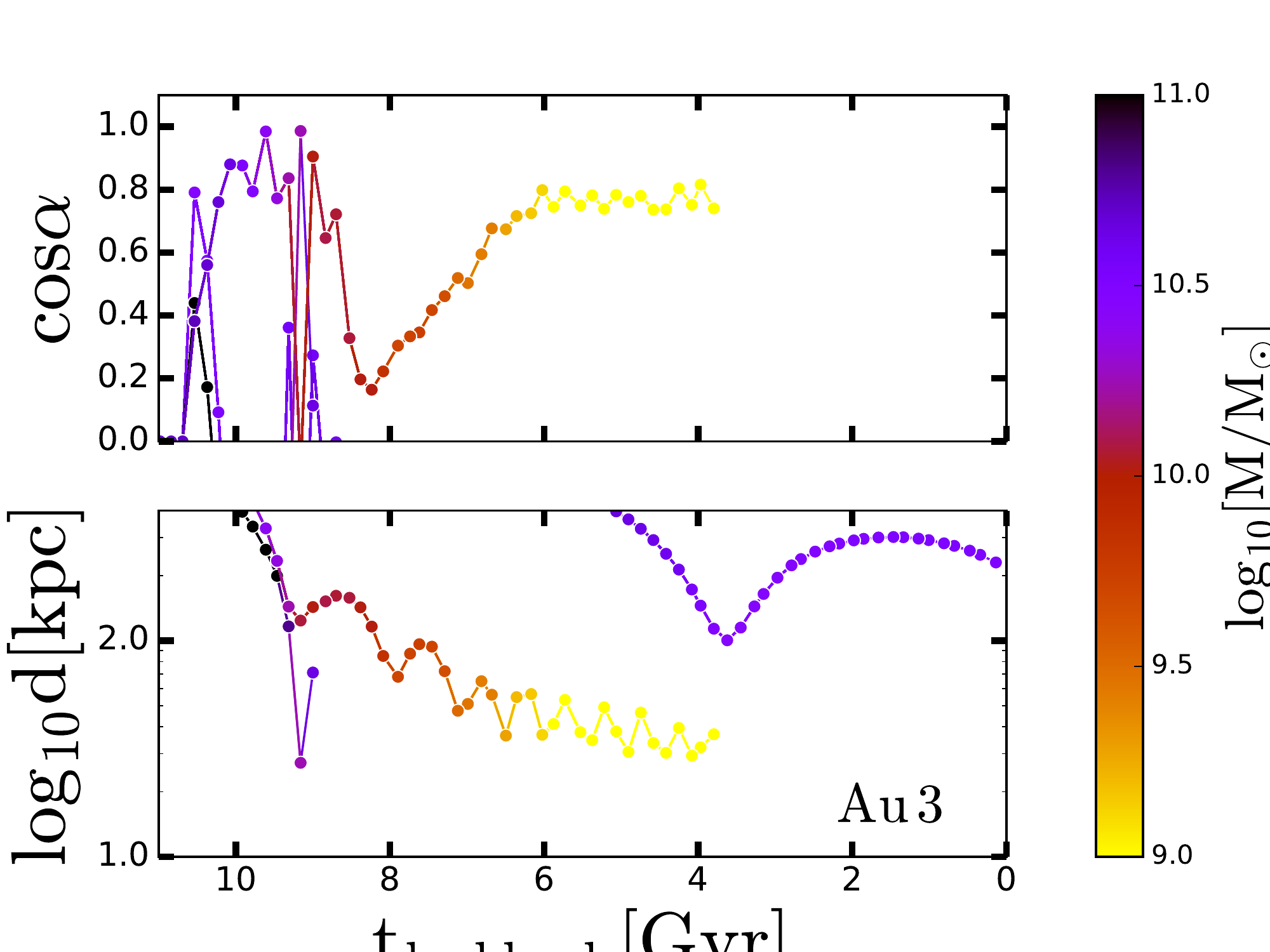}
\includegraphics[scale=0.28,trim={2.5cm 1.4cm 3.8cm 1.cm},clip]{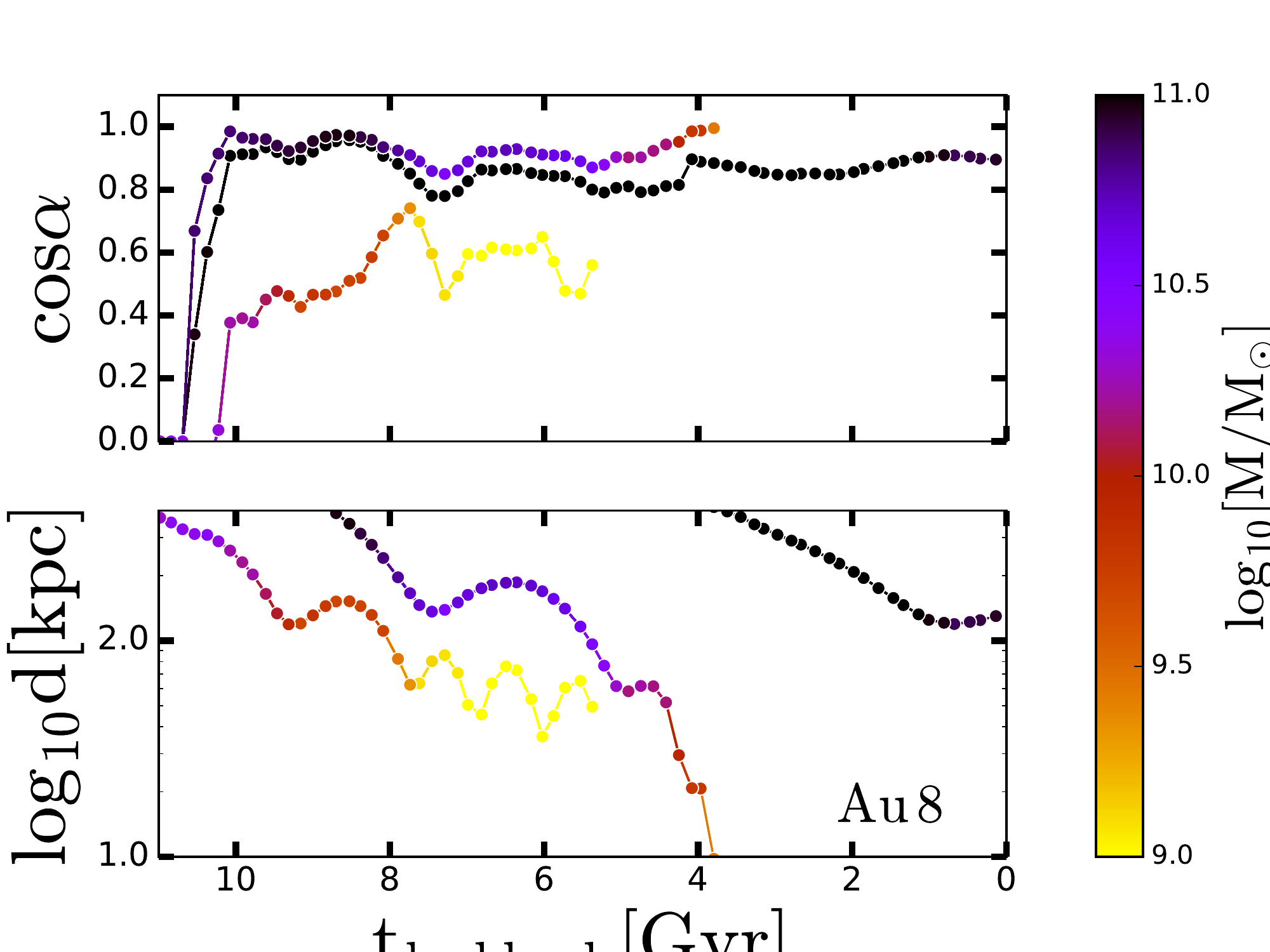}
\includegraphics[scale=0.28,trim={2.5cm 1.4cm 0 1.cm},clip]{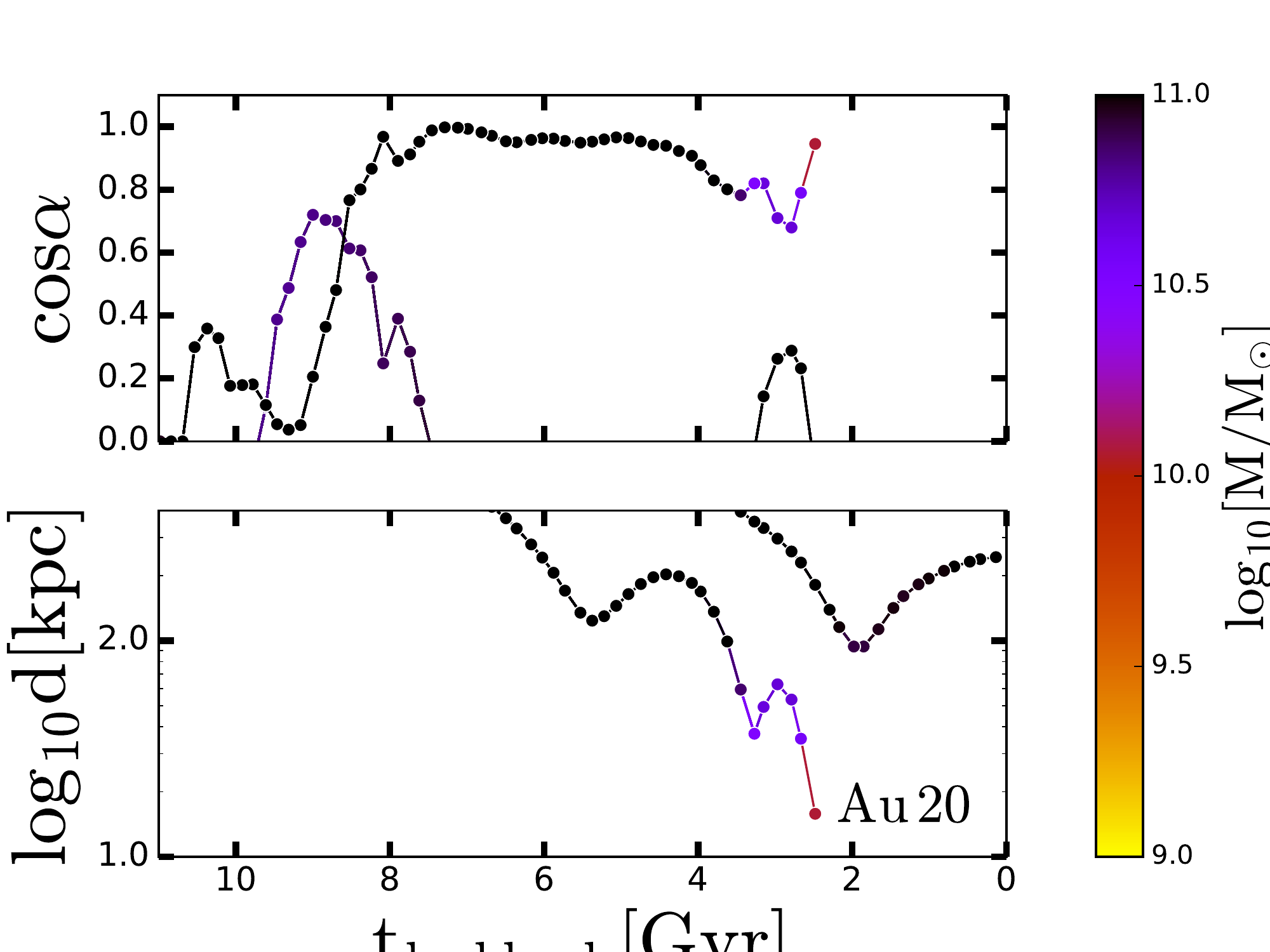} \\ \vspace{0.5cm} 
 
  \hspace{-1.2cm}
\includegraphics[scale=0.59,trim={0 0.9cm 0.18cm 0},clip]{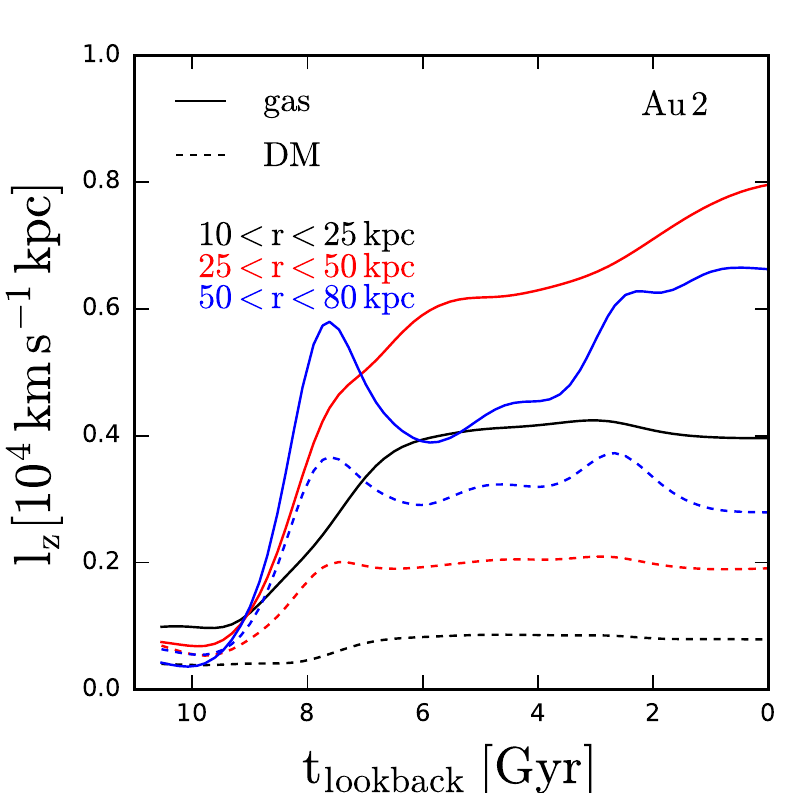} 
\includegraphics[scale=0.59,trim={1.23cm 0.9cm 0.18cm 0},clip]{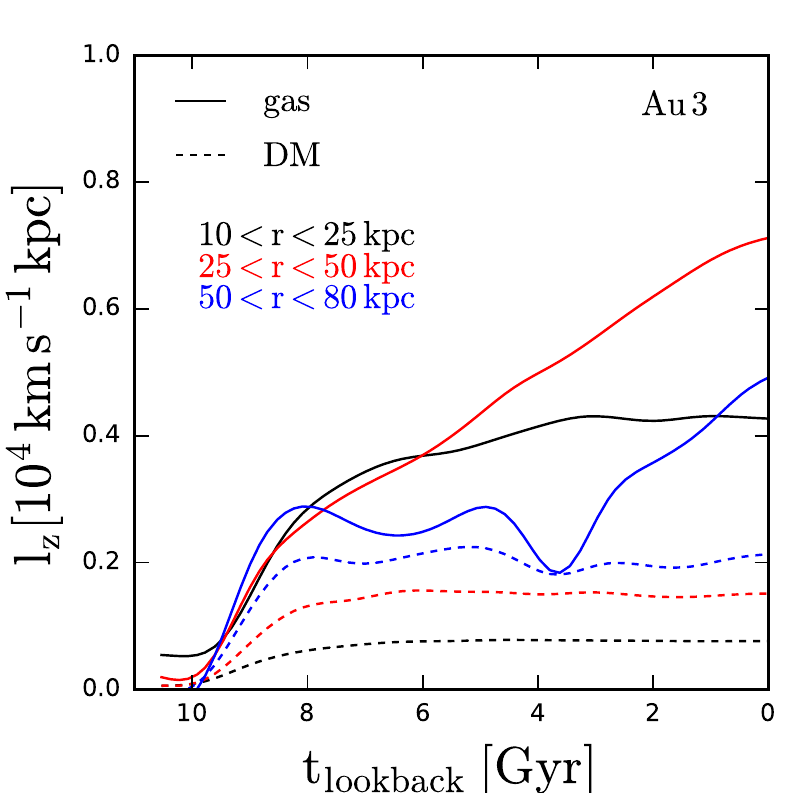}
\includegraphics[scale=0.59,trim={1.23cm 0.9cm 0.18cm 0},clip]{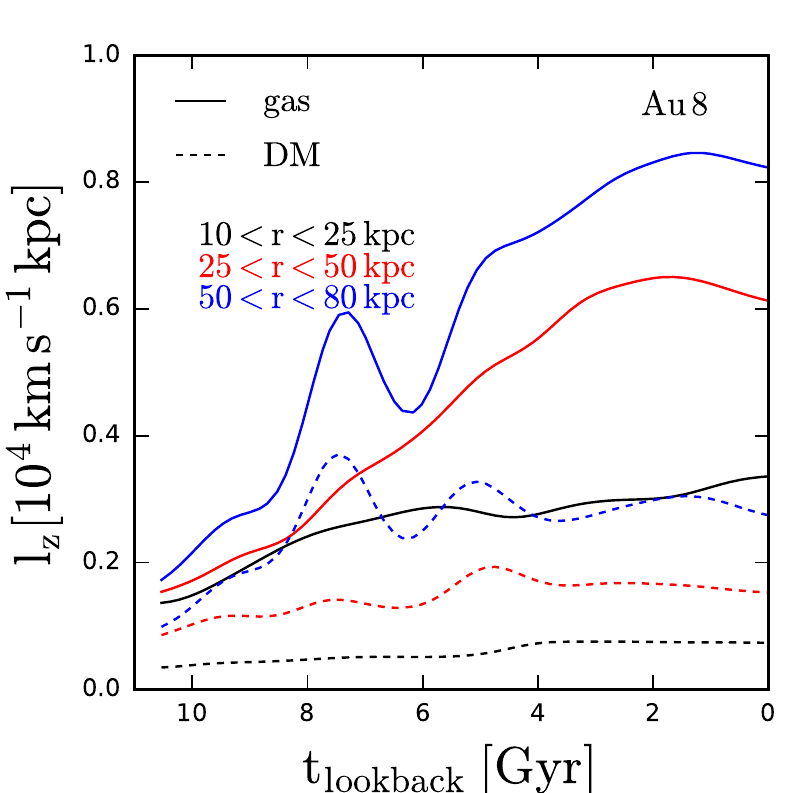} 
\includegraphics[scale=0.59,trim={1.23cm 0.9cm 0 0},clip]{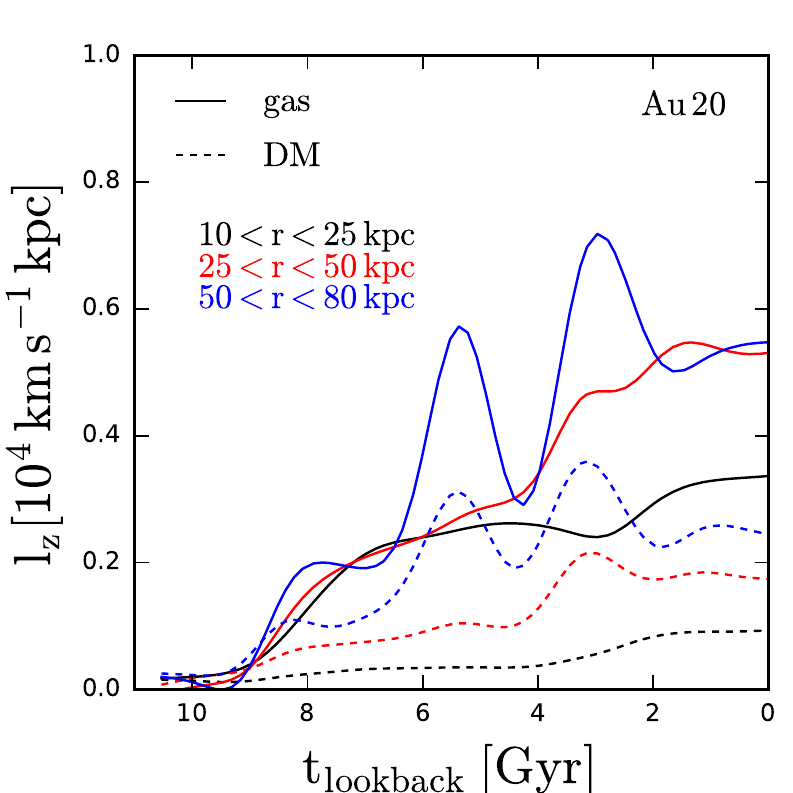} \\ \vspace{0.5cm}
 \hspace{-1.2cm}
\includegraphics[scale=0.59,trim={0 0 0.18cm 0.4cm},clip]{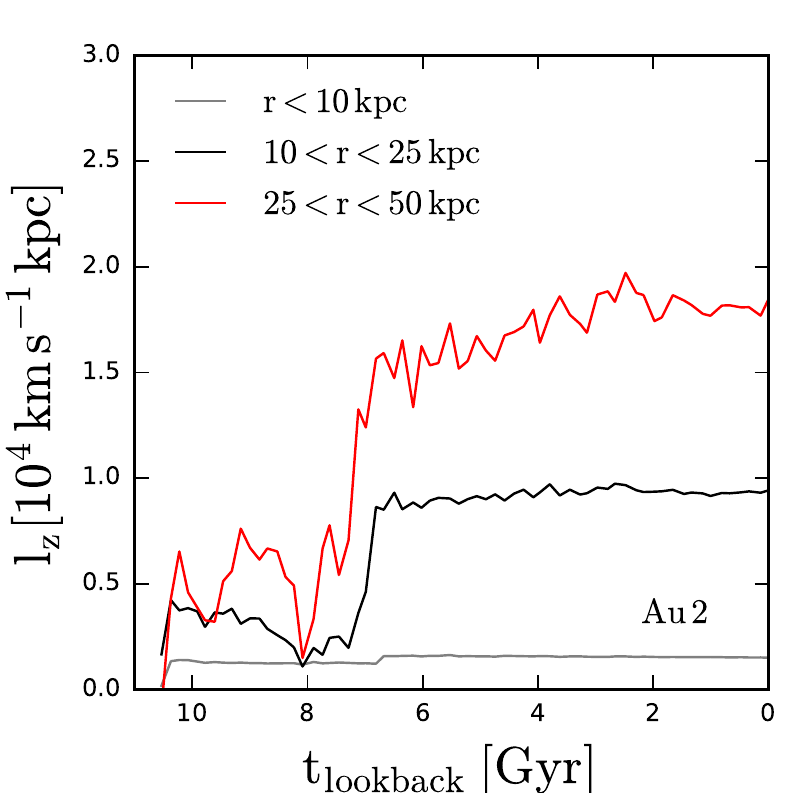}
\includegraphics[scale=0.59,trim={1.23cm 0 0.18cm 0.4cm},clip]{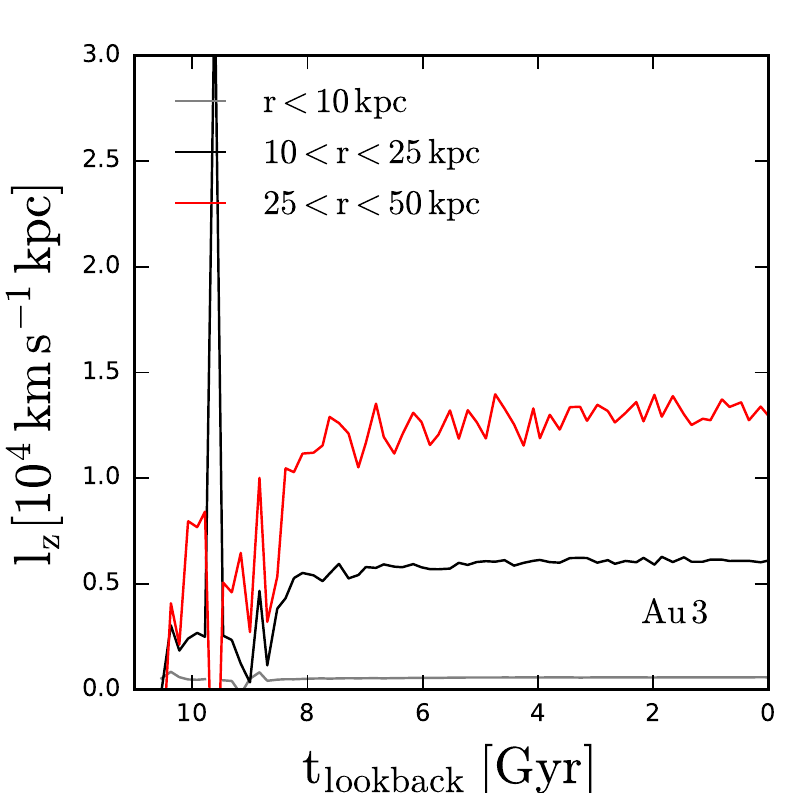}
\includegraphics[scale=0.59,trim={1.23cm 0 0.18cm 0.4cm},clip]{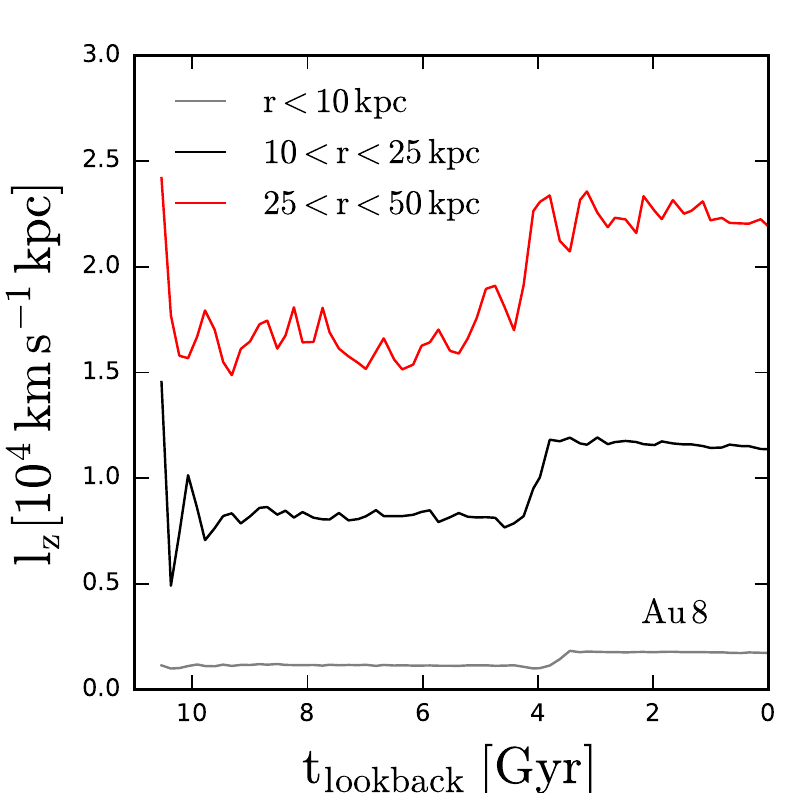}
\includegraphics[scale=0.59,trim={1.23cm 0 0. 0.4cm},clip]{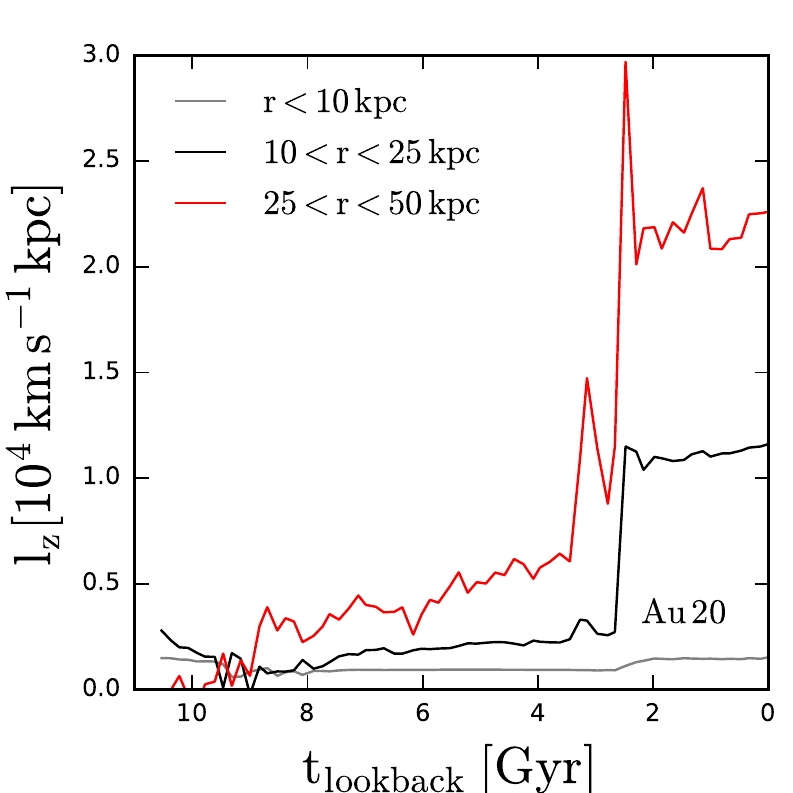} \\ \vspace{0.5cm}

\caption{The evolutionary history of a selection of subhaloes and their impact on the disc size and specific angular momentum of halo components of four of the largest discs in the simulation suite. \emph{Top row}: the evolution of the disc edge, $R_{\rm edge}$, defined as the radius at which the surface density of the stellar component drops below 1 $\rm M_{\odot} pc^{-2}$; \emph{second and third row}: the evolution of the dot product of the angular momentum of the main halo stellar disc and that of the orbital angular momentum of subhaloes (top) and the distance of the subhaloes to the centre of the main halo (bottom). Points are colour coded according to the total mass of the subhaloes at a given time; \emph{fourth row}: the evolution of the specific angular momentum of gas (solid lines) and dark matter (dashed lines) in different radial shells; \emph{fifth row}: the same as the fourth row but for star particles born between $t_{\rm lookback} = 10$ and 12 Gyr (note the radial shells are different from those shown in the fourth row). These four haloes are clear cases in which a massive ($\rm log _{10} [M/M_{\odot}] > 10$) subhalo inspirals into the main halo and quiescently merges in the disc plane. In particular, it is clear that subhaloes may not be completely aligned with the disc spin axis at distances larger than 100 kpc; however they gradually torque the disc as their orbit shrinks, which increases the alignment between their orbital angular momentum vector with the spin axis prior to merging. Their effects on the specific angular momentum of the halo and disc are clearly evident in the fourth and fifth row (see text for details).}
\label{lzvec1}
\end{figure*}

\begin{figure*} 
\centering
\includegraphics[scale=0.47,trim={1.5cm 0.7cm 1.5cm 0.5cm},clip]{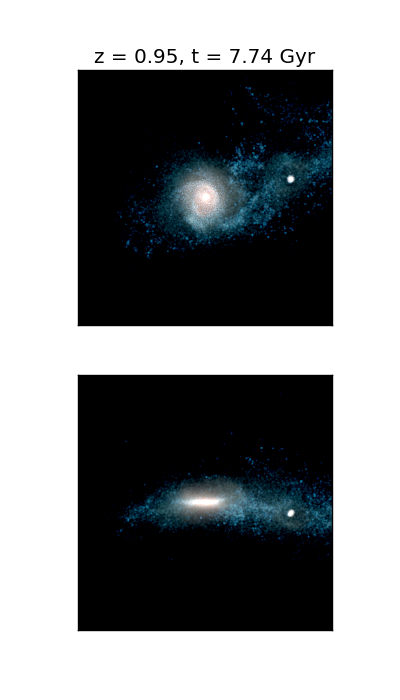} 
\includegraphics[scale=0.47,trim={1.5cm 0.7cm 1.5cm 0.5cm},clip]{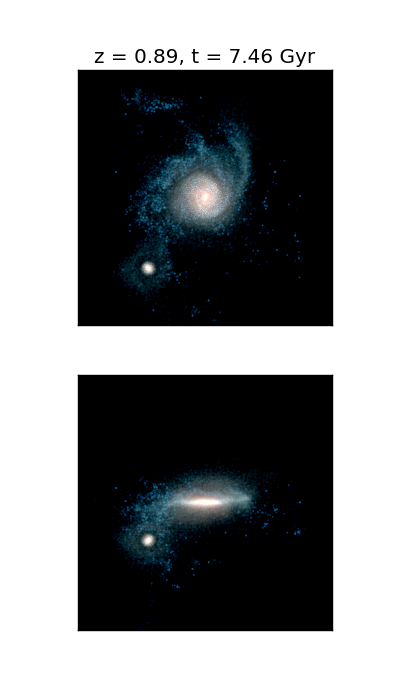} 
\includegraphics[scale=0.47,trim={1.5cm 0.7cm 1.5cm 0.5cm},clip]{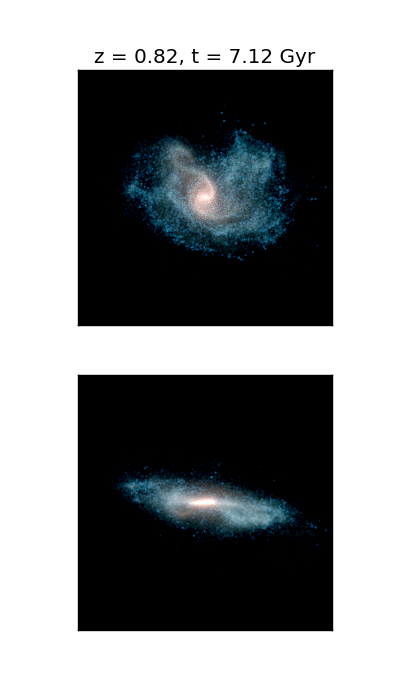} 
\includegraphics[scale=0.47,trim={1.5cm 0.7cm 1.5cm 0.5cm},clip]{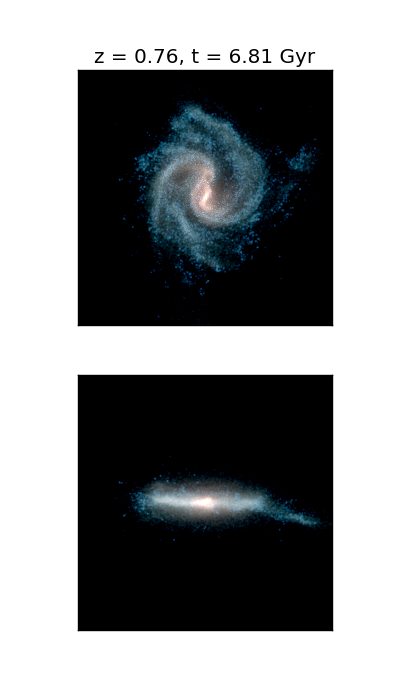}
\includegraphics[scale=0.47,trim={1.5cm 0.7cm 1.5cm 0.5cm},clip]{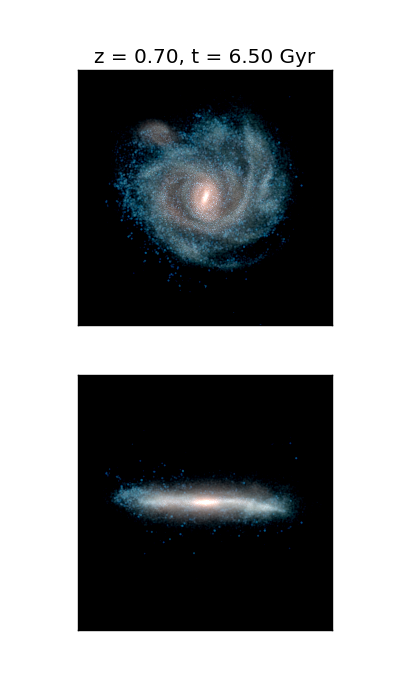} \\
\includegraphics[scale=0.47,trim={1.5cm 0.7cm 1.5cm 1.75cm},clip]{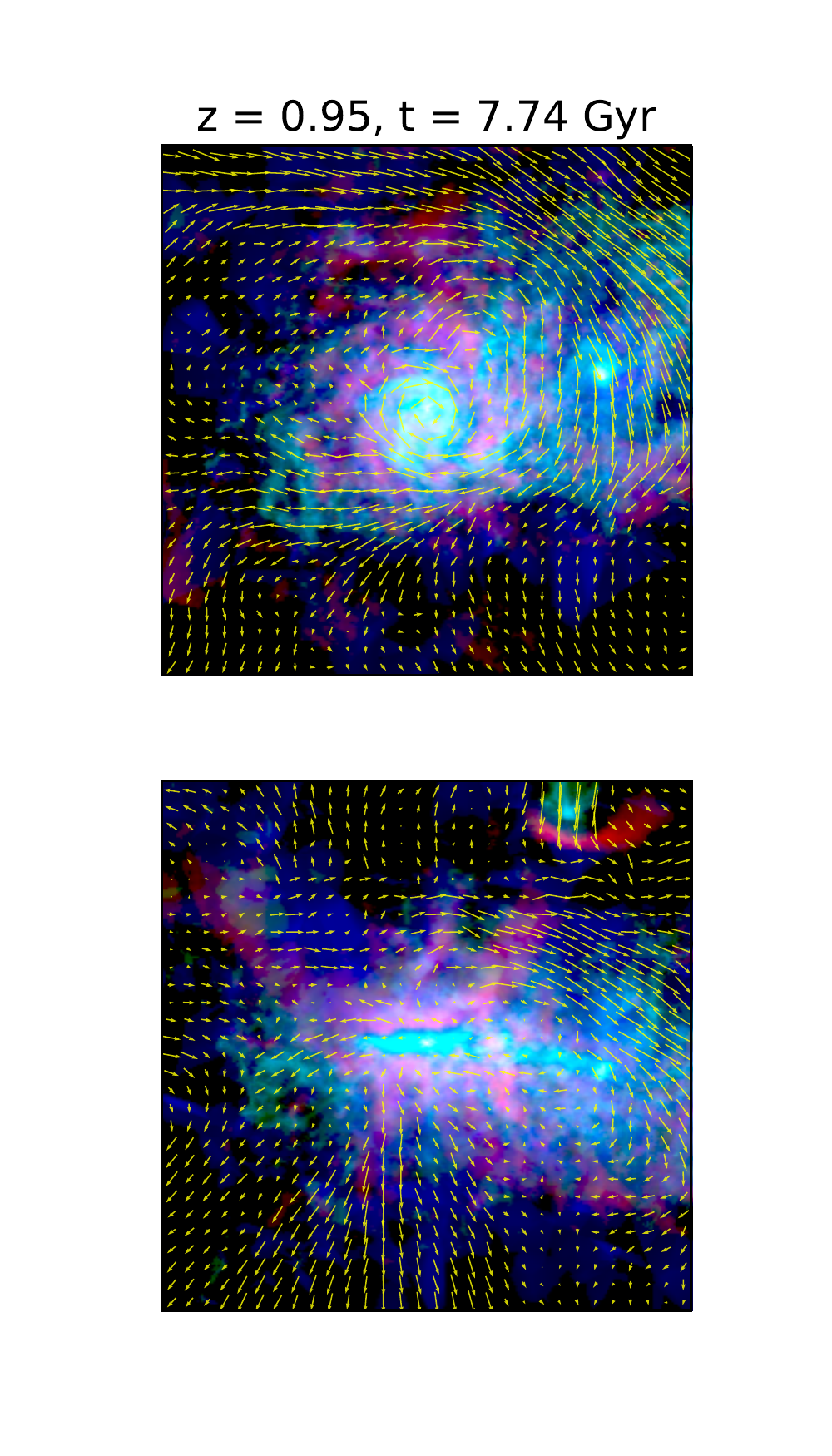} 
\includegraphics[scale=0.47,trim={1.5cm 0.7cm 1.5cm 1.75cm},clip]{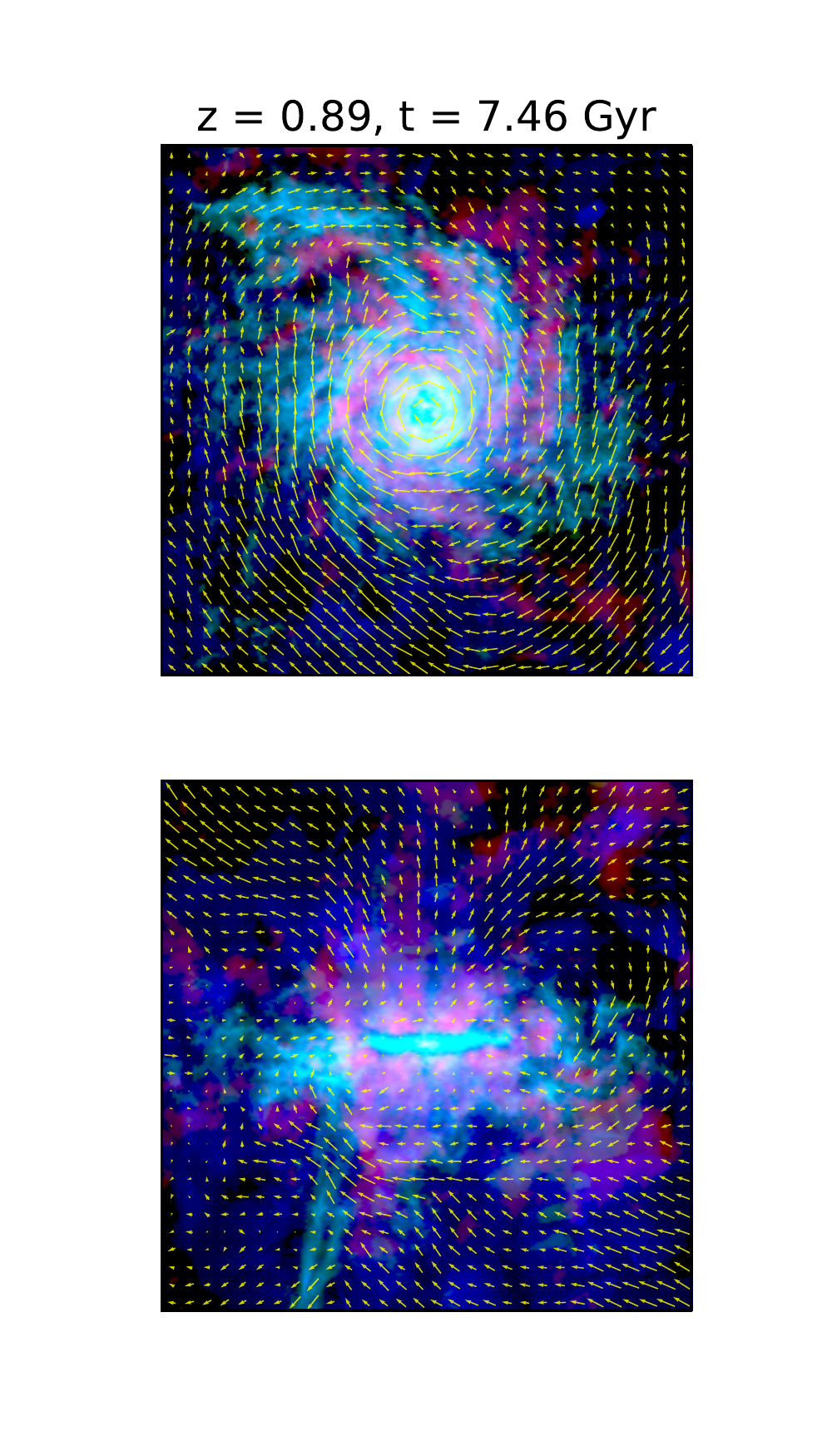} 
\includegraphics[scale=0.47,trim={1.5cm 0.7cm 1.5cm 1.75cm},clip]{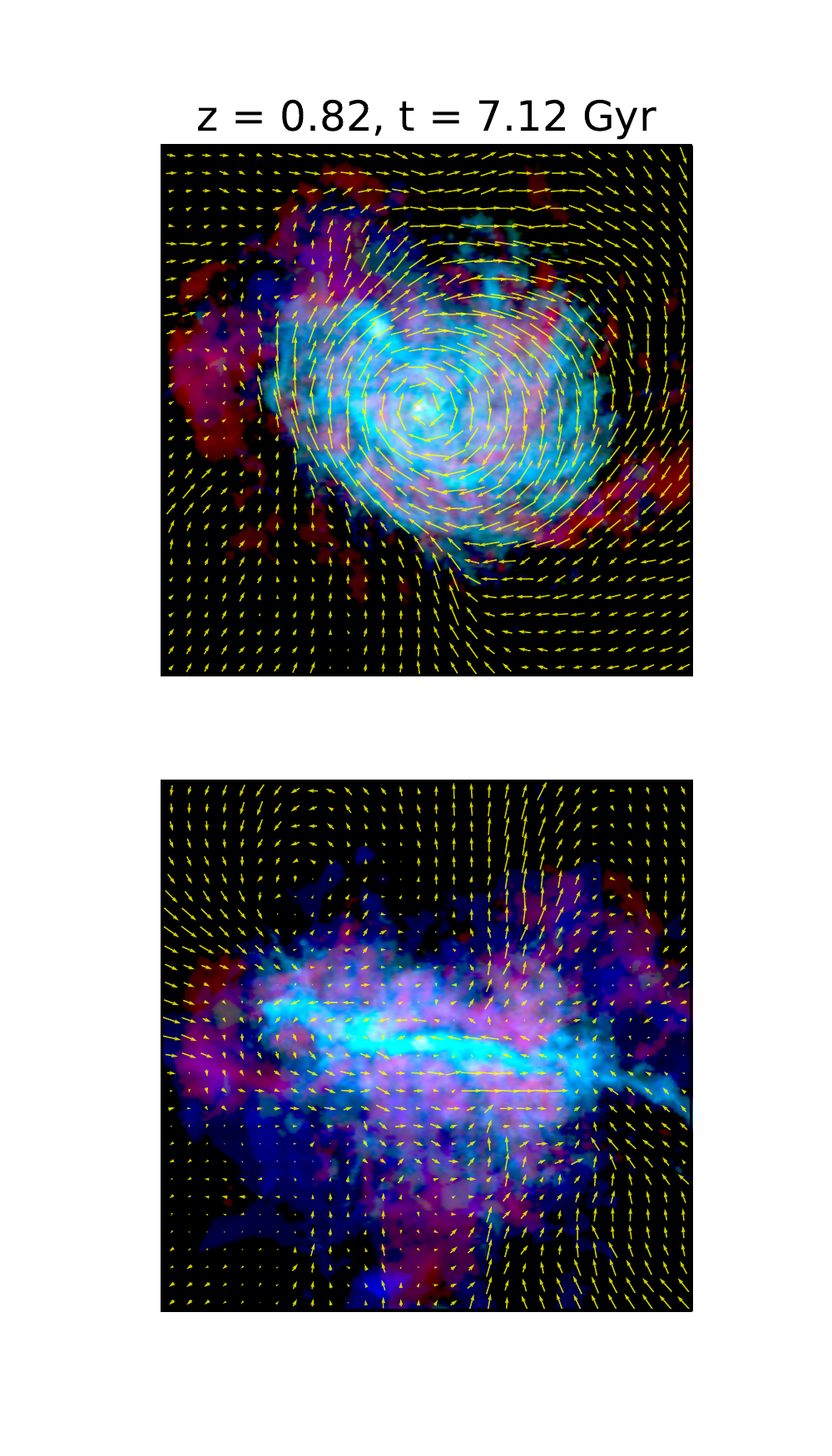} 
\includegraphics[scale=0.47,trim={1.5cm 0.7cm 1.5cm 1.75cm},clip]{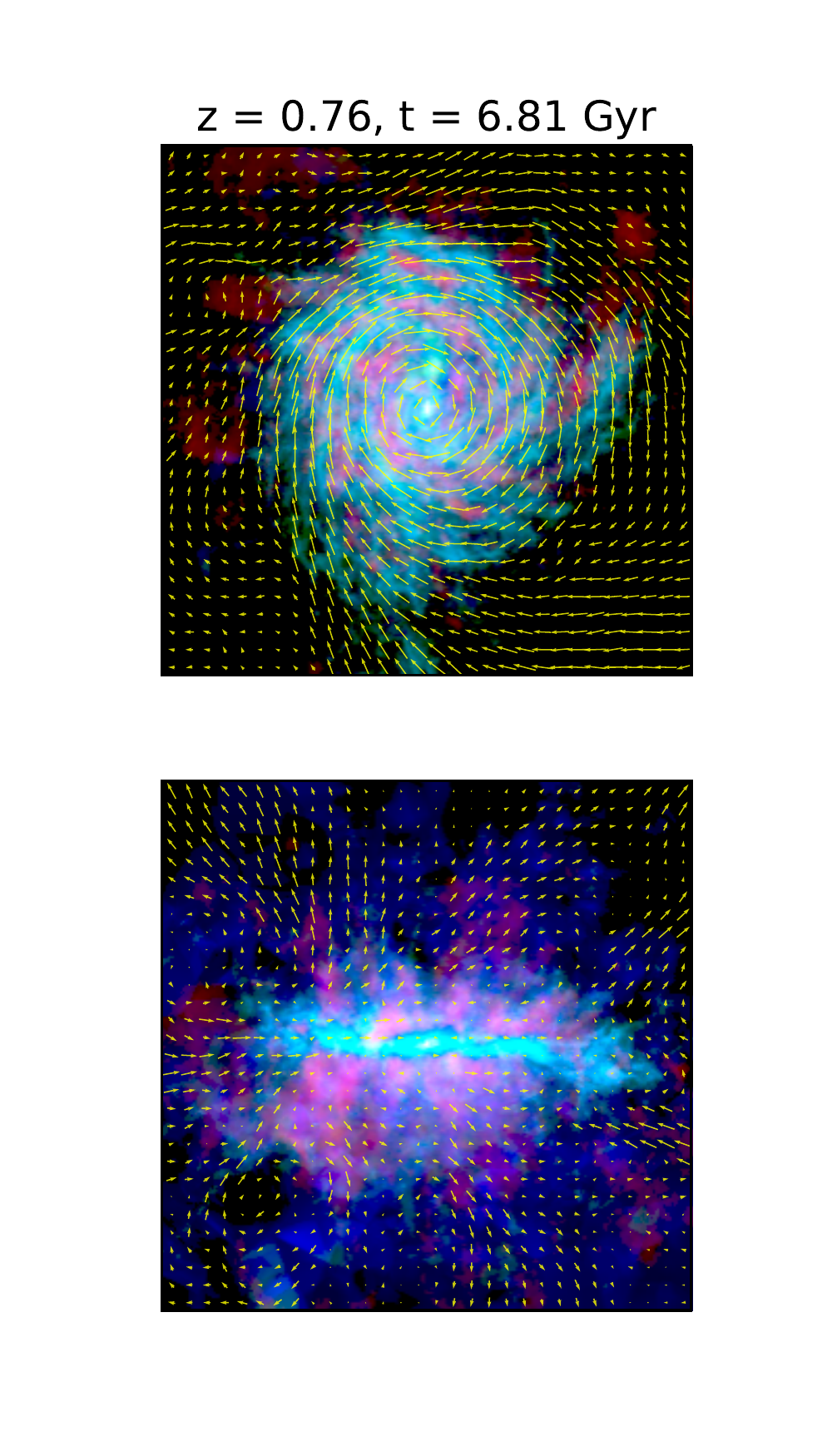}
\includegraphics[scale=0.47,trim={1.5cm 0.7cm 1.5cm 1.75cm},clip]{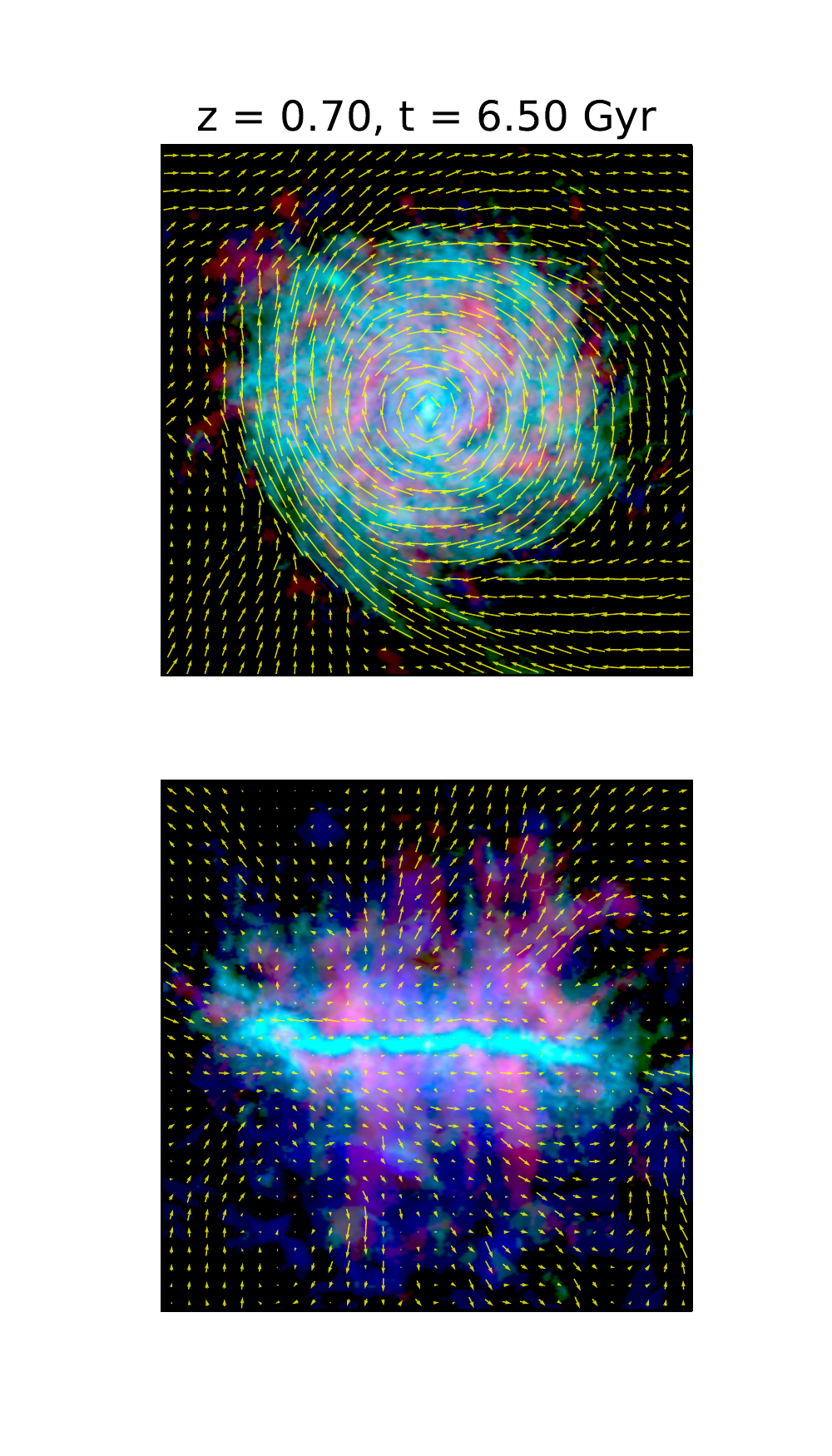} 
\caption{Face-on and edge-on images of the star (top two rows) and gas (bottom two rows) distributions in Au 2 between $z=0.95$ and $z=0.7$, during which time a massive subhalo quiescently merges in the disc plane. In the stellar images, colour indicates stellar age, whereas in the gas images, it indicates gas temperature (blue:cold star-forming gas, $T<2\times 10^4$ K; green: warm gas, $2\times 10^4 < T < 5\times 10^5$ K and red: hot gas, $T>5\times 10^5$ K). Velocity vectors are shown in the gas projections. The in-plane trajectory of the subhalo orbit is evident in both the stars and the gas. The radial size of the disc grows noticeably during this time period, and is well correlated with the increase in specific angular momentum of the gas and dark matter in this halo. Note that the warm gas entrained in the wake of the subhalo (e.g., at $t_{\rm lookback} = 7.74$ Gyr) has significant rotation. This indicates that the subhalo has deposited its gas with substantial residual angular momentum around the central galaxy, in addition to (perhaps) spinning up surrounding gas (alternatively, after stripping it may mix with halo gas, thus increasing its specific angular momentum). Note also the strong non-axisymetries developed in the stellar component, which may allow angular momentum to be transferred from the subhalo to the pre-existing stars (see Fig.~\ref{lzvec1}).}
\label{au2evo}
\end{figure*}


In this section, we investigate {\bff the impact of subhalo accretion on the spin \citep[see also][]{VKK02,DN07,SSB12} and size of our simulated discs}. We first define a measure for the edge of the stellar disc, $R_{\rm edge}$, as the radius at which the stellar surface density drops below $1$ $\rm M_{\odot} \, pc^{-2}$, which is a good proxy for the optical radius of the disc (see Fig.~\ref{sden}). In the top panels of Fig.~\ref{lzvec1}, we show the time evolution of $R_{\rm edge}$ for four of the most radially extended discs in the sample. Focusing on Au 2, we note two features: a sudden increase in $R_{\rm edge}$ at $t_{\rm lookback}\sim8$ Gyr after a period of almost no growth, followed by steady growth of $R_{\rm edge}$ until $z=0$. 

The cause of this increase of $R_{\rm edge}$ appears to be related to the accretion of gas-rich satellites. For example, in the third row of Fig.~\ref{lzvec1}, we show the distance evolution of a subhalo that first comes within $200$ kpc of the main halo at around $t_{\rm lookback}\sim8$ Gyr, and gradually approaches the latter until they merge at around $t_{\rm lookback}\sim6.5$ Gyr. The second row of Fig.~\ref{lzvec1} shows the evolution of the cosine of the angle between the orbital angular momentum vector of subhaloes and the spin axis of the stellar disc of the main halo. It is clear that the subhalo at $d\sim 200$ kpc is not well aligned with the disc of the main halo ($\cos \alpha \sim 0.4$); however, as the subhalo sinks into the main halo, its orbit becomes increasingly more aligned with the disc. At $t_{\rm lookback}\sim8$ Gyr, when the subhalo is $d\sim 80$ kpc from the galaxy centre, the subhalo and disc are almost perfectly aligned \citep[see also Fig. 4 of][]{MGG16}. 

The influence of these well aligned, quiescent mergers on the specific angular momentum, $l_z$, of the dark matter and gas in different radial shells is depicted in the fourth row of Fig.~\ref{lzvec1}. For Au 2, $l_z$ abruptly increases at around the time the subhalo falls into the main halo and crosses the radial shells; first, $l_z$ of the dark matter and gas in the outermost region considered (50 - 80 kpc) increases, followed by the intermediate region (25 - 50 kpc) and finally the inner region (10 - 25 kpc). It is interesting to note that once the subhalo has been accreted, the dark matter retains a constant $l_z$ after its pre-merger increase, whereas the $l_z$ of the gas continues to rise in the intermediate radial region long after the merger has occurred. The explanation of this behaviour can be found from consideration of the angular momentum transfer from the dark matter to the gas component \citep[recently studied in detail in][in the Illustris simulation]{ZS16}. As the subhalo falls into the main halo, the gas component can be affected by the ram pressure in such a way as to offset the centre of mass of the gas and dark matter of the subhalo, and thus allow a torque to come into play and transfer orbital angular momentum from its dark matter to the gas. This is seen in Fig.~\ref{au2evo} as the warm gas (which may also contain halo gas that has been spun up by the subhalo) with a high amount of rotation in the wake of the subhalo, and the slight depression in the dark matter $l_z$ after the initial rise in Fig.~\ref{lzvec1}. The high angular momentum gas then cools and enters the inner regions to increase the $l_z$ there (seen as the dip in the $l_z$ of the gas in the outer region and increase in the $l_z$ of the intermediate region)\footnote{The increasing gas specific angular momentum is consistent with the re-distribution of angular momentum of gas to the inner regions and dark matter to the outer regions reported in \citet{SSB12}}. 

In addition to the increase in $l_z$ of the gas and dark matter, the stellar component is also affected; in the bottom row of Fig.~\ref{lzvec1}, we show the evolution of $l_z$ in a set of radial regions of the \emph{in situ} star particles older than $10$ Gyr, i.e., before the merger events in question occur. The clarity of the effect of the merging subhaloes is striking: the $l_z$ of the old stars is almost constant at all times until a merger event occurs and increases very sharply their specific angular momentum. After this time, their $l_z$ remains constant at a comparatively larger value. The trends described above for Au 2 are applicable also to Au 3 (second column of Fig.~\ref{lzvec1}), Au 8 (third column of Fig.~\ref{lzvec1}), and Au 20 (fourth column of Fig.~\ref{lzvec1}), which confirm that these qualitative effects are not unique to one special case.

The dynamical effects of the merger on the main halo described above are well illustrated in Fig.~\ref{au2evo}, which shows the face-on and edge-on projections of the stars and gas in a 100 kpc side-length box centred on the main halo, at the time of the merger event. The subsequent cooling of the high angular momentum gas leads to the build up of fresh star-forming gas around the disc edges, which is consistent with the inside-out star formation that takes place in the large-disc simulations shown (see Fig.~\ref{sfr}). Indeed, the disc growth even in this small period shown in Fig.~\ref{au2evo} is striking. We note that this mechanism for disc growth is not uncommon among the Auriga simulations: 5 out of the 6 largest discs have experienced quiescent mergers and subsequent disc growth. The exception is Au 25, which already has high angular momentum at $z\sim 3$. In addition, 6 out of the 10 discs that have intermediate scale lengths (between $4$ and $6$ kpc) have experienced quiescent mergers at some point of their evolution, either from a less massive subhalo or in combination with retrograde or violent mergers, which act to reduce overall disc growth at late times. 

In contrast, the smallest discs have no such quiescent gas-rich mergers, and in some cases instead undergo violent major mergers that act to destroy discs \citep[see][]{SS16}. The comparatively small amount of angular momentum in the gas and dark matter that these haloes possess reinforces the idea that angular momentum{\bff , whether acquired from these quiescent mergers or otherwise,} is critical to the build up of large, radially extended discs.

\subsection{AGN feedback}
\label{agn}

\begin{figure} 
\centering
\includegraphics[scale=1.]{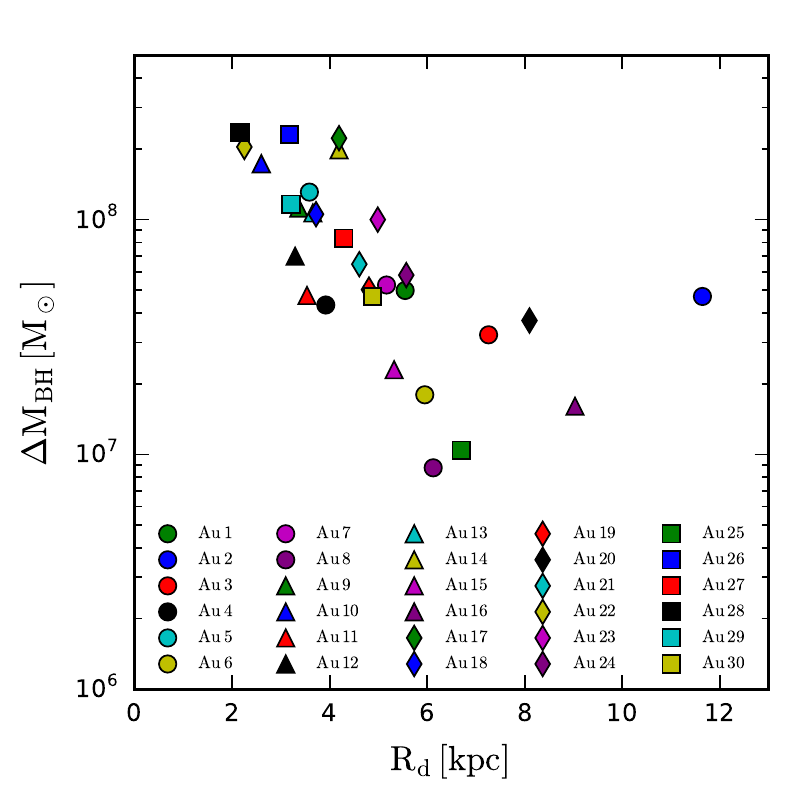}
\caption{Growth in black hole mass growth between $z=1$ and $z=0$ as a function of the disc radial scale length at $z=0$.}
\label{dmbh}
\end{figure}

\begin{figure} 
\centering
\includegraphics[scale=1.1,trim={0 0 0 0},clip]{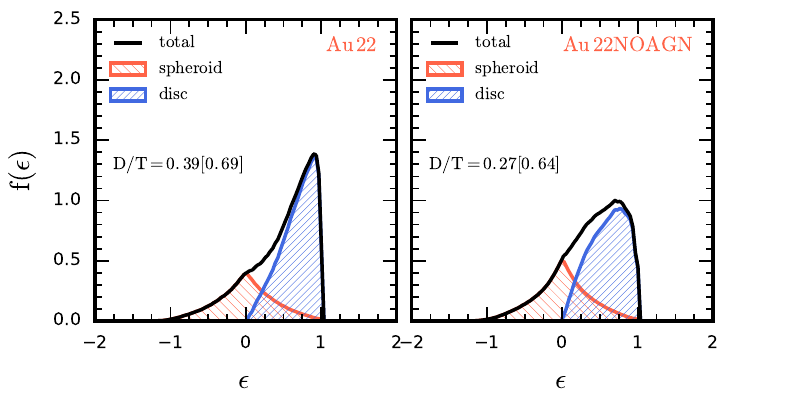}\\
\includegraphics[scale=1.1,trim={0 0 0 0},clip]{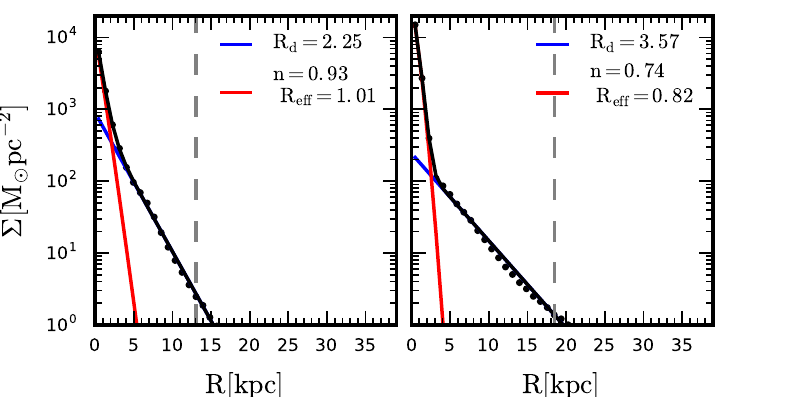} \\
\includegraphics[scale=1.1,trim={0 0 0 0},clip]{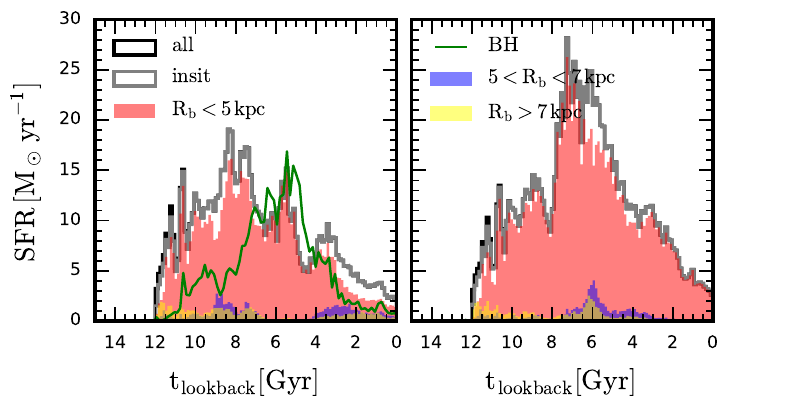} \\
\includegraphics[scale=0.93,trim={0.1cm 0 1.3cm 0},clip]{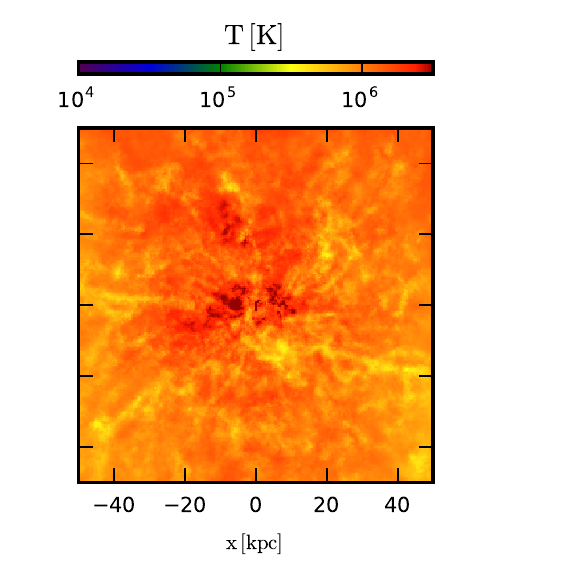}  \includegraphics[scale=0.93,trim={0.6cm 0 1.15cm 0},clip]{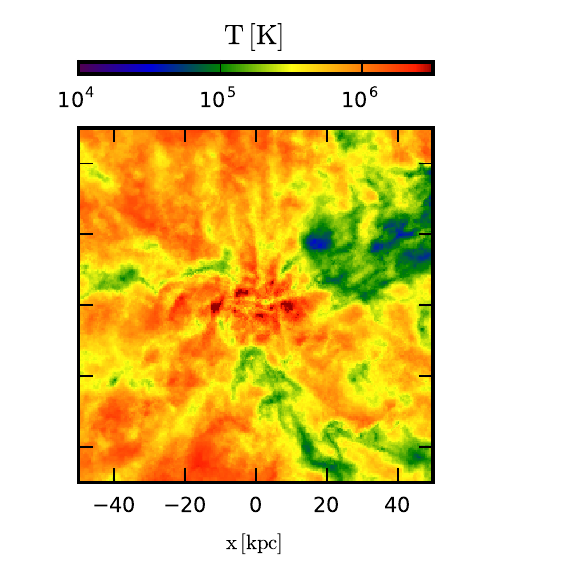} 
\caption{Comparison between the fiducial run of halo Au 22 (left column) and the resimulation with AGN feedback turned off for $z<1$ (right column). We show the orbital circularity distribution (top row) and the surface density profile of the star particles (second row) at $z=0$, the SFHs (third row) and the edge-on projected gas temperature at $t_{\rm lookback}$ = 5 Gyr (fourth row). We note that the kinematically defined disc is more prominent in the fiducial run of Au 22, whereas the bulge is larger in the run without AGN feedback. The surface density profile indicates, however, that the disc scale length is larger for the run without AGN feedback, in addition to the central surface density being larger. These effects are illustrated in the SFH of the simulations, in which SF is reduced in the outer disc regions during the time of high AGN activity, whereas the central star formation is greatly increased without the presence of AGN feedback.}
\label{noagn}
\end{figure}

As described in Section 2, the galaxy formation model includes black hole feedback in both a radio and a quasar mode. In this Section, we investigate how AGN feedback shapes the distribution and size of the stellar disc. In Fig.~\ref{dmbh}, we show the mass growth of the black hole (a good proxy for AGN feedback energy injection into the gas) since $z=1$ as a function of the present-day disc scale length. An anti-correlation between the two quantities is discernible, with some outliers such as Au 2, which lies far to the right of the other simulations. This trend {\bff might} suggest that black hole growth, and thus AGN feedback, is able to suppress the formation of disc stars at large radii, and hence lead to smaller discs. However, it is unclear from Fig.~\ref{dmbh} whether it is the degree of feedback that drives disc size, or whether black hole growth and disc size are both driven independently by another factor, for example, the angular momentum of halo gas: an important parameter that affects black hole growth is the gas density in the central regions (localised around the BH), which will be higher for more compact discs that arise when low angular momentum gas concentrates in the central regions. We note also that violent (major) mergers often {\bff cause star bursts and black hole growth and} result in small, compact stellar discs. For example, Au 4, 11, 28 and 29 all experience a major merger after $z=0.5$, and end up with present day accreted star fractions of about $30 \%$ and disc scale lengths of less than 4 kpc. 


In order to eliminate the effects of {\bff these factors}, we re-ran Au 22 unchanged until $z=1$, after which time we turned off all AGN feedback (while still allowing for the black hole to swallow gas). We term this new simulation Au 22NOAGN. We focus on Au 22 because it has a quiet merger history and a significant amount of AGN feedback. In the top row of Fig.~\ref{noagn}, we compare the circularity distributions of the two simulations. There are noticeable differences between them: the disc component is less well defined in Au 22NOAGN, resulting in a $D/T$ ratio 0.27, to be compared to 0.39 in Au 22 (in both cases we use the conservative definition, i.e., $\epsilon > 0.7$ to belong to the disc). The other disc fraction definition is smaller also in the simulation with no AGN feedback. Clearly, the central bulge is more dominant compared to Au 22. This is reflected in the surface density decomposition in the second row of Fig.~\ref{noagn}, in which the central density is visibly higher in the simulation without AGN feedback. Interestingly, the extent of the disc component is greater in the simulation without AGN feedback, which exhibits a scale length about 1.6 times larger than that of Au 22.

\begin{figure*} 
\centering
\includegraphics[scale=1.,trim={1.cm 2.5cm 1.5cm 0.5cm},clip]{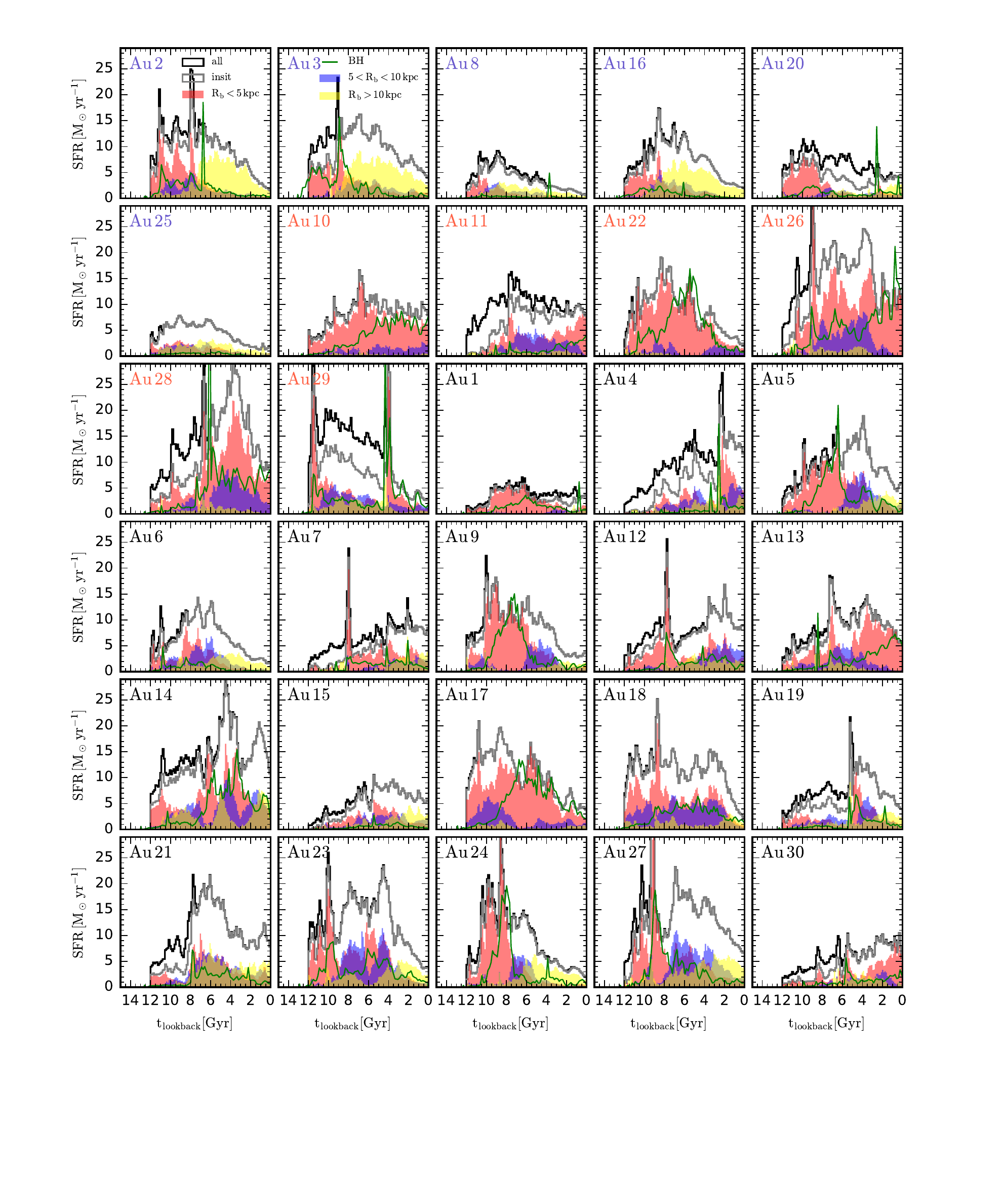}
\caption{Star formation histories for all simulations. The histories are shown for all star particles (black curves) and star particles born {\it in situ} (grey curves) that are present in the main halo at $z=0$. In addition, the star particles born {\it in situ} are further split according to their birth radius, as indicated in the top-left panel. The black hole mass growth rate (green curves) is shown multiplied by a factor $2.5 \times 10^3$.}
\label{sfr}
\end{figure*}


\begin{figure*} 
\centering
\includegraphics[scale=0.62,trim={0 0 0.2cm 0},clip]{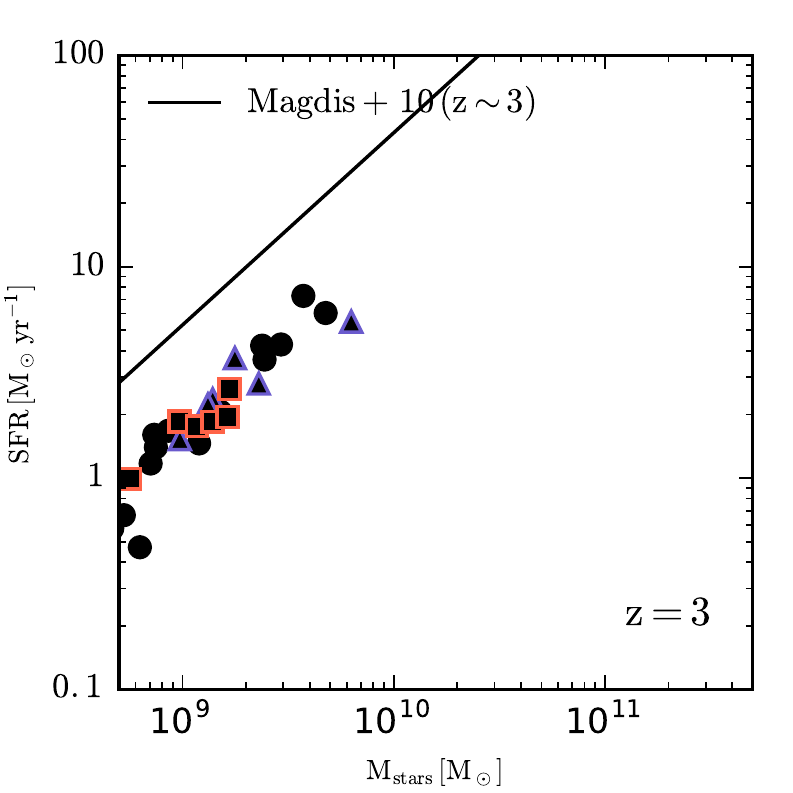}
\includegraphics[scale=0.62,trim={1.2cm 0 0.2cm 0},clip]{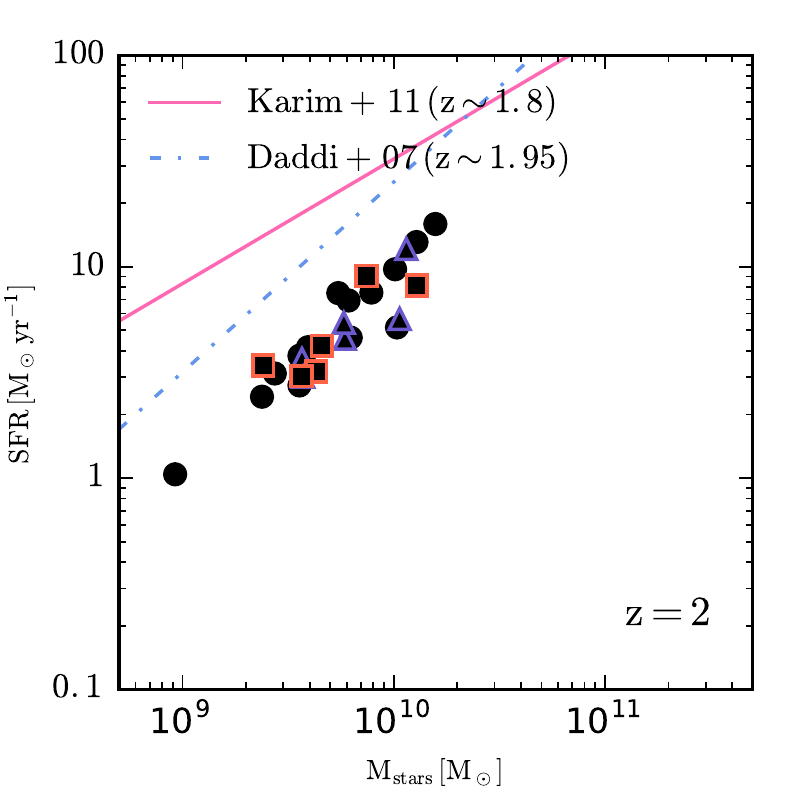}
\includegraphics[scale=0.62,trim={1.2cm 0 0.2cm 0},clip]{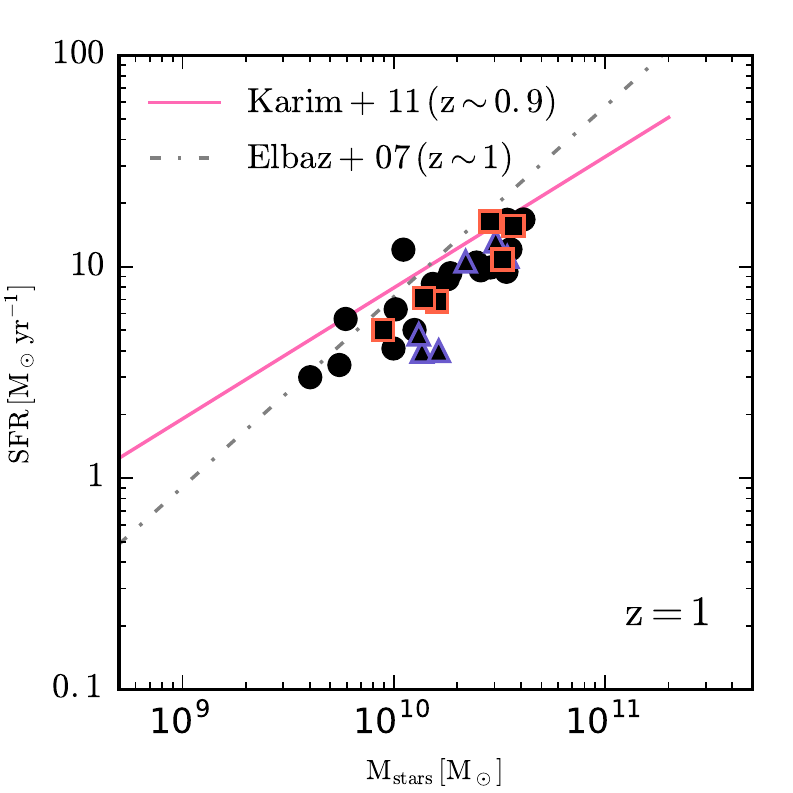}
\includegraphics[scale=0.62,trim={1.2cm 0 0.2cm 0},clip]{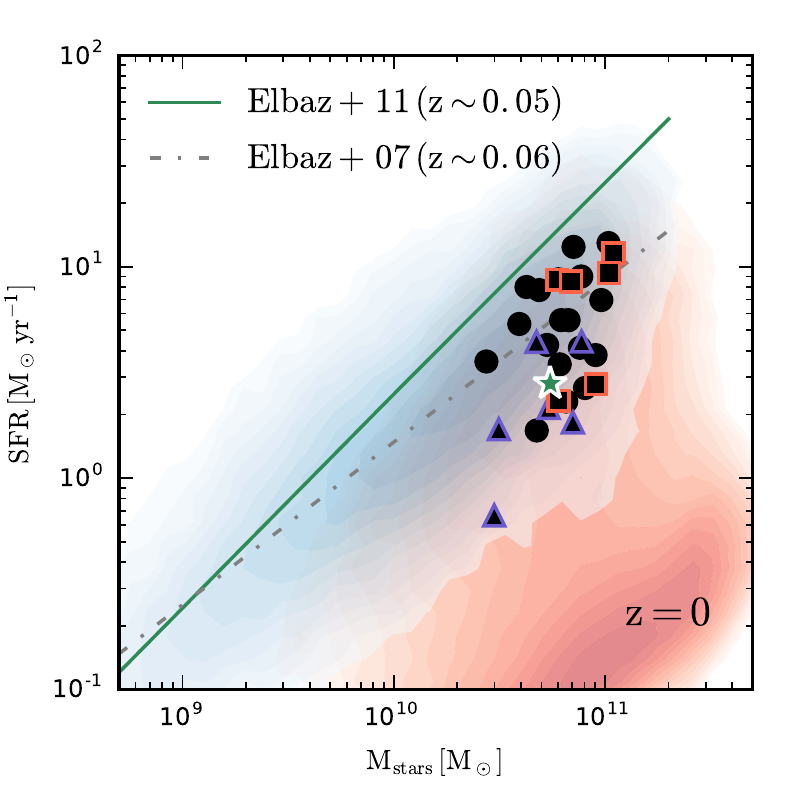} 
\caption{{\bff The star formation rate as a function stellar mass for all simulated galaxies at a series of redshifts. In each panel we also show observationally derived scalings for late-type galaxies \citep{MRH10,DMC07,KSM11,EDL07,EDH11}.} The Milky Way is represented by the green star in the right panel. Observations from the SDSS MPA-JHU DR7 catalogue are shown as the blue cloud and red sequence according to the classification described in \citet{MPS14}. }
\label{color}
\end{figure*}


\begin{figure*}
\centering
\includegraphics[scale=0.7,trim={0 0 0 0},clip]{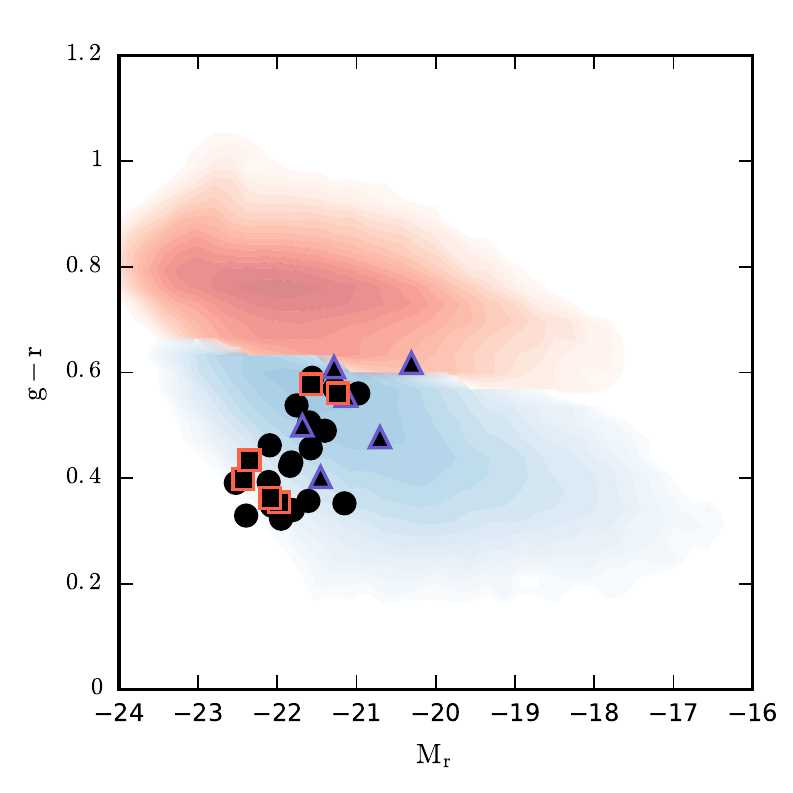}
\includegraphics[scale=0.7,trim={0 0 0 0},clip]{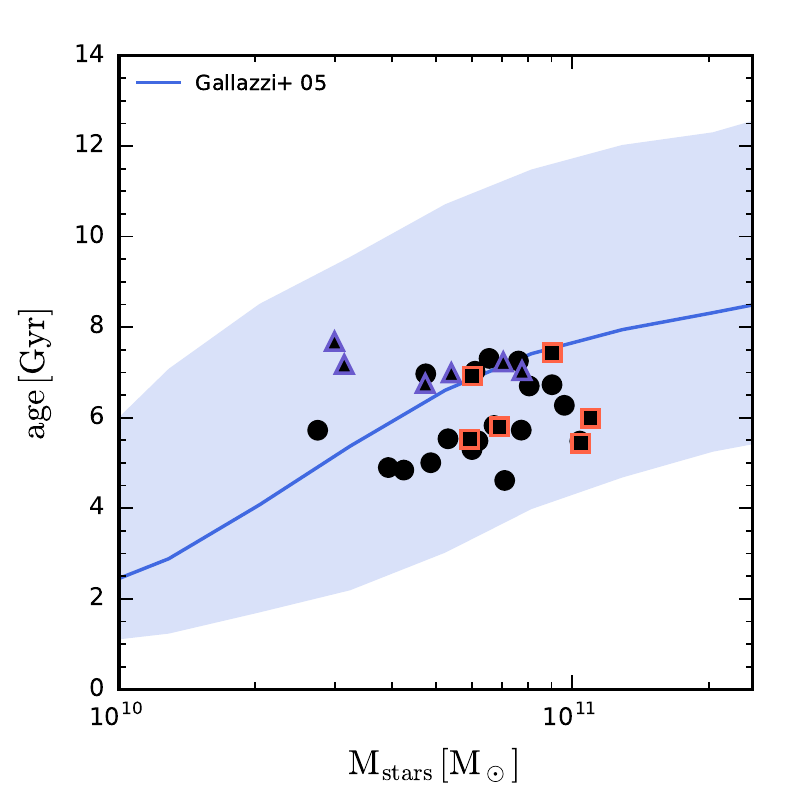} 
\includegraphics[scale=0.7,trim={0 0 0 0},clip]{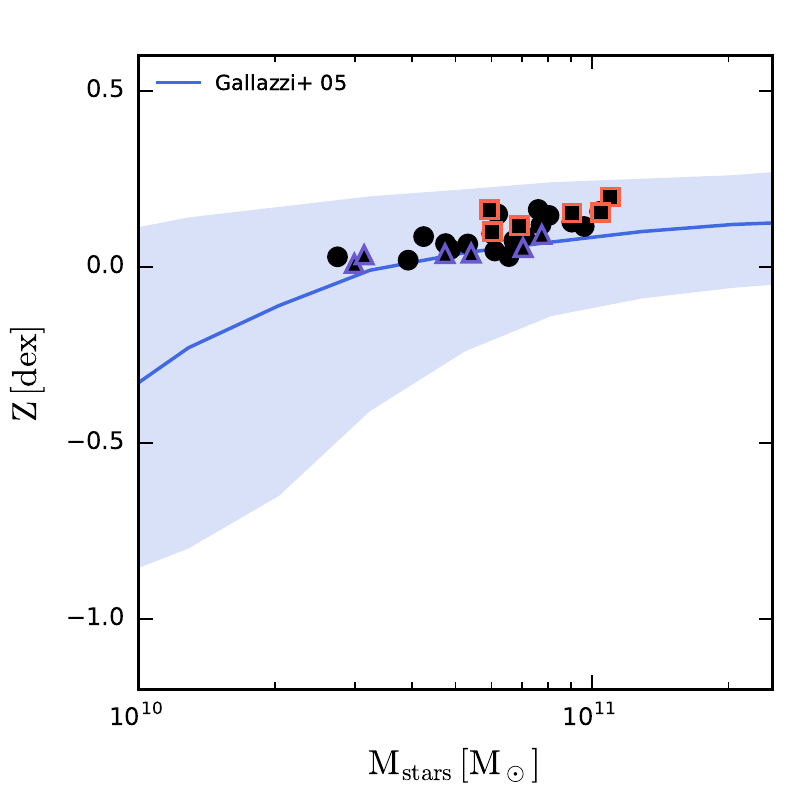}  
\caption{\emph{Left panel}: g-r color plotted as a function of r-band absolute magnitude plotted over the same blue/red sequence of SDSS MPA-JHU DR7 galaxies shown in Fig.~\ref{color}. \emph{Middle} and \emph{Right panels}: The mean stellar age plotted as a function of stellar mass (left), and the mass-metallicity relation (right) of all simulated galaxies. The observational relations presented in \citet{GCB05} are shown. The simulations are clustered about the mean (solid blue line) and lie within the scatter (blue shaded regions) for both relations. }
\label{agemet}
\end{figure*}

The SFHs shown in the third row of Fig.~\ref{noagn} clearly show the effects of AGN feedback on the evolution of both bulge and disc: without AGN feedback, the SFR in the inner disc ($R < 5$ kpc) peaks at a value of nearly $30$ $\rm M_{\odot} yr^{-1}$ at around $t_{\rm lookback} \sim 7$ Gyr, whereas this is reduced to less than $20$ $\rm M_{\odot} yr^{-1}$ in Au 22. In this case, AGN feedback helps to reduce star formation in the inner regions and thus reduces the mass of the central bulge. The other noticeable difference in the SFHs is that the amount of star formation at large radii ($R>5$ kpc) is enhanced in Au 22NOAGN compared to Au 22. In particular, at $t_{\rm lookback} \sim 5$ Gyr, when AGN activity is at its maximum, there is no star formation in the outer disc. However, during the same time period in Au 22NOAGN, a peak of about 5 $\rm M_{\odot} yr^{-1}$ is reached in the outer regions. It is this star formation that likely leads to the larger radial disc scale length reported in the panels above. The fourth row of Fig.~\ref{noagn} shows the edge-on projection of the gas temperature at $t_{\rm lookback} = 5$ Gyr. The temperature of the halo gas is significantly higher for Au 22 than for Au 22NOAGN: in the latter, there is evidence of gas condensing onto the central galaxy, whereas in the former, halo gas is heated by strong AGN feedback and unable to cool and become star-forming. {\bff We thus conclude that strong AGN feedback is able to stifle the growth of the stellar disc. However, this is likely one of several contributing factors to the correlation shown in Fig.~\ref{dmbh}. For example, the angular momentum of gas is linked both to black hole growth and to the radial extent of star formation (see Section~\ref{sec5}), and mergers can destroy, reduce or even increase the size of discs and create bursts of both star formation and AGN activity. A full quantitative explanation of the trend shown in Fig.~\ref{dmbh} is therefore far from trivial, and is beyond the scope of this paper. }


\section{Evolution and stellar populations}
\label{sec5}

In this Section, we analyse the stellar mass growth history of the simulations and the resulting observational properties of stellar populations.

\subsection{Star formation histories}

In Fig.~\ref{sfr} we present star formation histories for all the simulated galaxies. Here we consider all the star particles within $0.1$ $R_{200}$ of the main halo at $z=0$ and determine the star formation history by binning \emph{initial} stellar mass (before stellar mass loss occurs) by age into 140 Myr intervals, which is similar to the time resolution of the simulations. A common feature in most of the SFHs is that they peak at around a redshift of between $z=2$ and $1$ ($t_{\rm lookback} \sim$ 10 - 8 Gyr), and gradually decline before reaching a roughly constant SFR of order a few solar masses per year. However, there are a variety of SFHs: 1) The peaks vary from about 5 $\rm M_{\odot}$ yr$^{-1}$ (Au 25) to more than 30 $\rm M_{\odot}$ yr$^{-1}$ (Au 28); 2) Many simulated galaxies exhibit bursty star formation histories; for example, Au 23 exhibits several sharp peaks between 4 and 10 Gyr look back time. 

We subdivide the total star formation histories to obtain SFHs for {\it in situ} star particles only. From these curves, it is evident that in the majority of cases accreted star particles are a negligible fraction of the total in the halo at $z=0$, typically between $10\%$ and $30\%$, usually as a result of a major merger that occurred at a time indicated by a burst of star formation (e.g., Au 29). We further subdivide the star particles born {\it in situ} into three groups according to their birth radius, $R_b$, specifically: $R_b < 5$ kpc, $5 < R_b < 10$ kpc and $R_b > 10$ kpc, where $R_b$ is the birth radius. This information reveals two broad classes of SFH: inside-out formation that invariably leads to radially extended discs, particularly in haloes of high specific angular momentum, e.g., Au 2, 3 and 16, and star formation that remains centrally concentrated for most of the evolution, e.g., Au 10, 17 and 18. This leads to the more compact discs among the simulation suite (Fig.~\ref{sden}). In addition to the SFHs, Fig.~\ref{sfr} shows the black hole mass growth rate, multiplied by a factor of $2.5\times 10^3$ in order to make it visible on the same scale. The characteristics of the black hole growth histories vary widely among the haloes: for example, Au 17 and Au 22 exhibit significant black hole growth particularly during the period from $z\sim1$ to $z\sim0.1$, whereas Au 16 and Au 25 show almost no black hole growth at all. We note that short periods of black hole growth are correlated with bursts of star formation and are in many cases caused by merger related activity. Haloes that exhibit long periods of significant black hole growth tend to be those with more centrally concentrated SFHs (and more compact discs, see Section \ref{agn}), both of which are related to the gas density in the central regions. We emphasise that a success of our simulations is the coexistence of massive black holes and significant disc components, even after periods of strong black hole feedback.

{\bff In order to put the SFH of the Auriga galaxies into an observational context, in Fig.~\ref{color} we calculate the SFR averaged over a period of 0.5 Gyr at $z=3$, 2, 1 and 0, and compare these values with observed relations of SFR as a function of stellar mass. At all redshifts, particularly $z>0$, the slope of the simulated galaxies matches well at least one set of observations: at $z=3$, the slope is consistent with the slope derived from Lyman break galaxies obtained by \citet{MRH10}, and at $z=2$, the slope is consistent with the observations of \citet{DMC07}, although steeper than the observations of \citet{KSM11}. The peak of star formation for most galaxies occurs between $z=1$ and 2, consistent with Fig.~\ref{sfr}. The magnitude of star formation in the simulated galaxies agrees well with the observations from $z=1$ onwards, but at earlier times the SFR is about a factor of 2 too low compared with observations \citep[see also][]{AWN13}. In the right panel of Fig.~\ref{color} we compare the $z=0$ SFR of the simulated galaxies with nearby galaxies taken from SDSS MPA-JHU DR7\footnote{http://www.mpa-garching.mpg.de/SDSS/DR7/}, divided into blue and red sequences \citep[as in][]{SWP12,MPS14}. The simulations lie in the blue cloud region with present day SFRs between $\sim 1$ and $\sim 10$ $\rm M_{\odot} \, yr^{-1}$, and are clustered around the value obtained for the Milky Way by \citet{LK11}. } 

\begin{figure*} 
\centering
\includegraphics[scale=0.62,trim={0 0 0 0},clip]{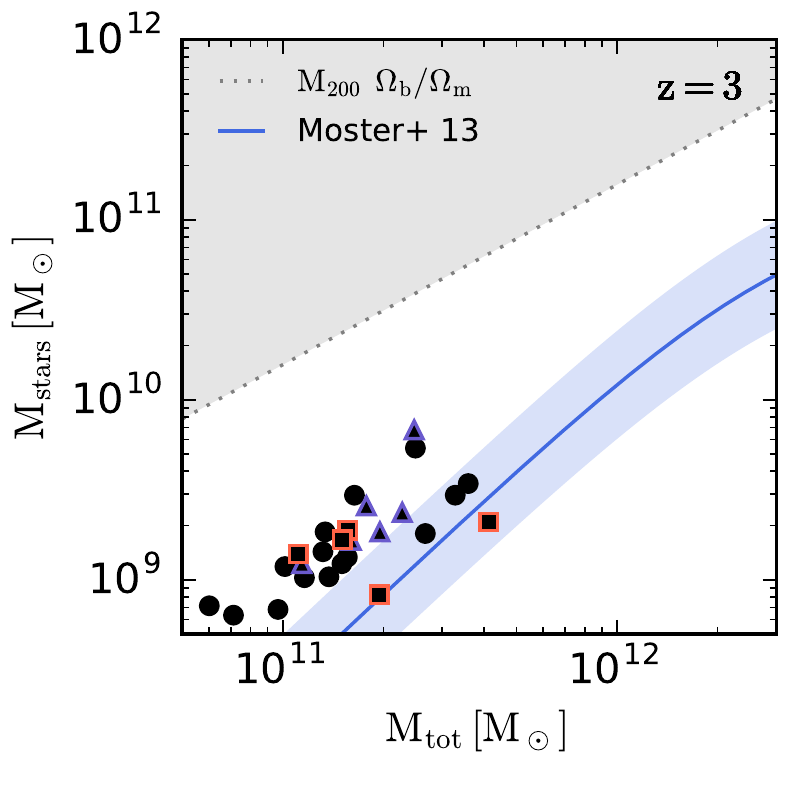}
\includegraphics[scale=0.62,trim={1.7cm 0 0 0},clip]{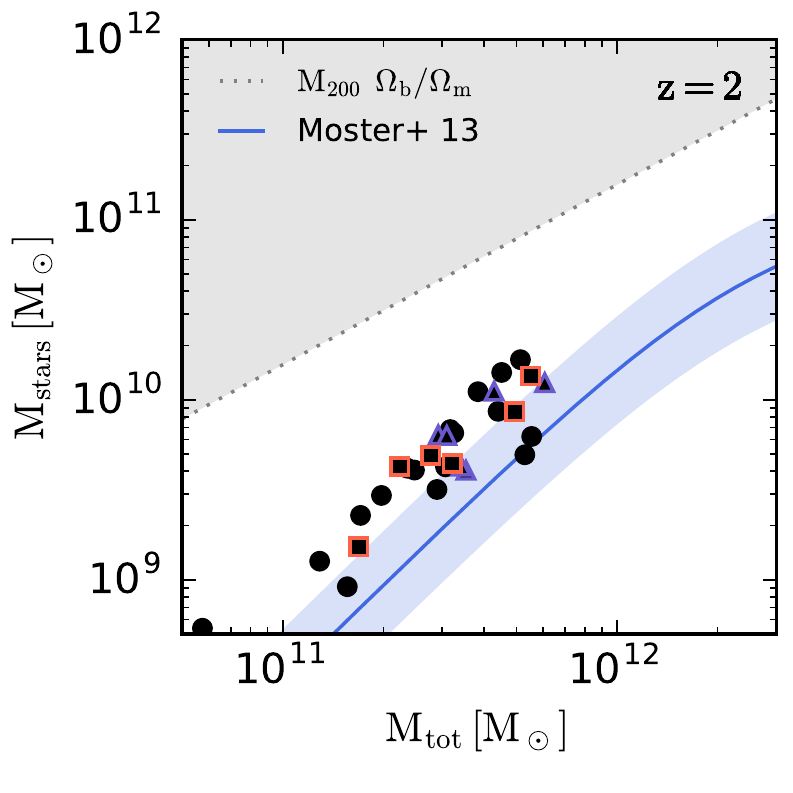}
\includegraphics[scale=0.62,trim={1.7cm 0 0 0},clip]{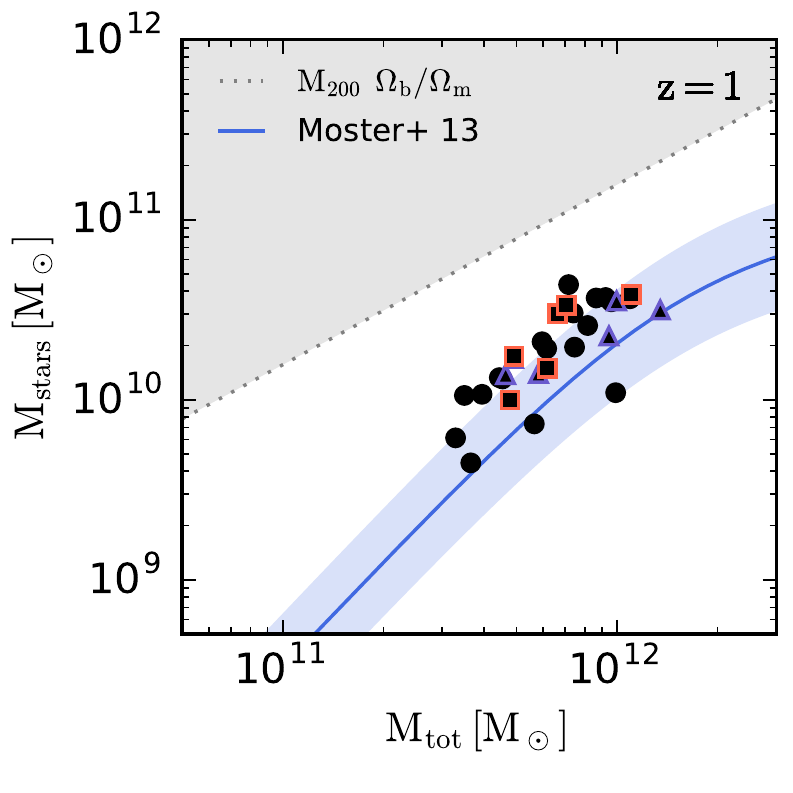}
\includegraphics[scale=0.62,trim={1.7cm 0 0 0},clip]{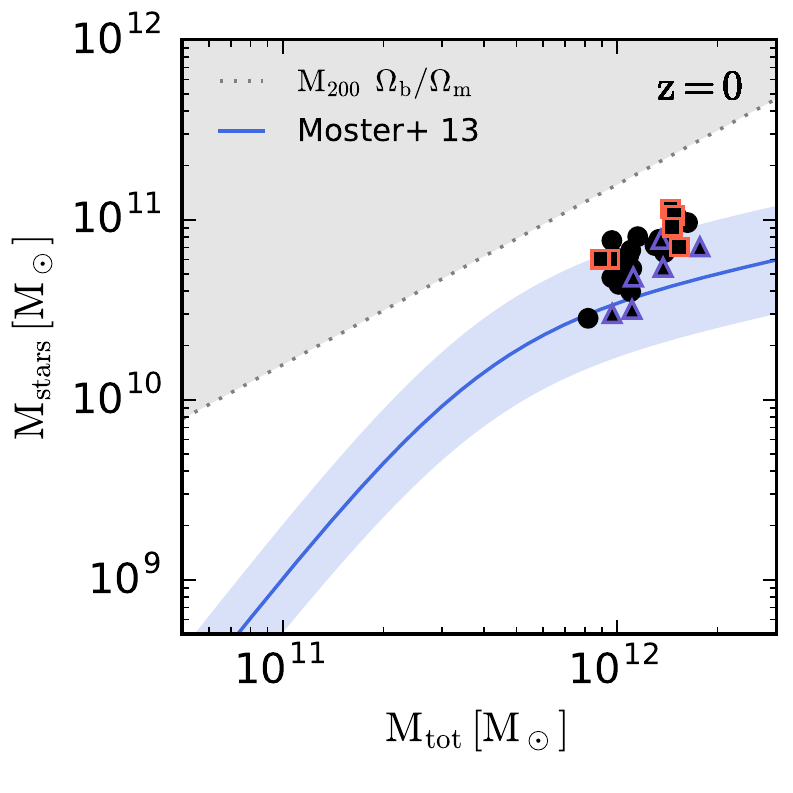}
\caption{The stellar mass-halo mass relation of the simulated galaxies at $z=3$ (first panel), {\bff $z=2$ (second panel), $z=1$ (third panel)} and $z=0$ (fourth panel). The mean and scatter of the semi-empirical relations derived in \citet{MNW13} are indicated by the solid blue line and the shaded region, respectively. At $z=3$, most haloes lie well above the relation, and tend to evolve in a direction slightly shallower than the relation such that they enter the region of 0.2 dex scatter (blue shaded regions) at around $z=1$ and lie mostly within the scatter at $z=0$. Some haloes have slightly too much stellar mass relative to the present-day abundance matching estimate, and it is interesting to note that these haloes tend to host the discs with the smallest scale lengths (red squares), whereas those that host the most radially extended discs scatter around the mean relation (blue triangles).  }
\label{srel}
\end{figure*}

\begin{figure} 
\centering
\includegraphics[scale=1.]{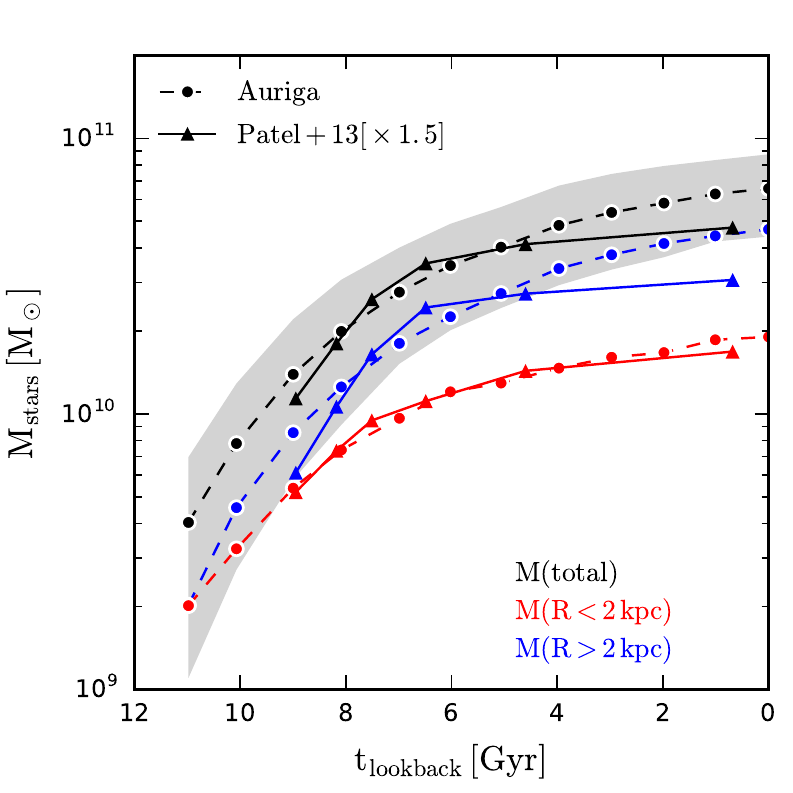}\\
\caption{The growth history of the total stellar mass (black dashed curve), and that of the inner (red dashed curve) and outer (blue dashed curve) regions of the mass distribution, averaged over the simulation sample. The 1-$\sigma$ dispersion is indicated by the grey shaded region. The solid curves show the observationally inferred growth histories from \citet{PFF13}, multiplied by a factor 1.5 in order to correspond to a {\bff final stellar mass similar to that of the Milky Way}. }
\label{mgrowth}
\end{figure}

\subsection{Stellar properties}

The SFH, together with the stellar evolution assumed by the simulations, determine the properties of the $z=0$ stellar population, which we here compare to a variety of observations. In the left panel of Fig.~\ref{agemet}, we show the simulated galaxies in $g-r$ colour vs. $r$-magnitude space, which are both shaped by the integrated SFH. In the background, we show contours of blue and red nearby galaxies taken from SDSS MPA-JHU DR7. {\bff This verifies that the simulated galaxies are blue in colour, i.e., they are not part of the red sequence shown in the right panel of Fig.~\ref{color}}.  

Further implications of the late-time star formation necessary to maintain the galaxies in the blue cloud region populated by late-type spiral galaxies can be found in the mass-weighted mean stellar age of the star particles, shown in the {\bff middle} panel of Fig.~\ref{agemet}. The distribution of simulated galaxies is centred around a mean stellar age of $\sim 5$-$7$ Gyr, which is roughly the age range expected {\bff for galaxies like the Milky Way} from the observations of \citet{GCB05}, although the scatter from measurement uncertainties is large and easily contains that of the simulations. The {\bff third} panel of Fig.~\ref{agemet} shows the metal content of the simulated galaxies, represented as the decimal logarithm of the mass-weighted mean iron abundance relative to the solar, compared to the observational results of \citet{GCB05}. It is clear that the simulations reproduce the observed metal content {\bff for galaxies like the Milky Way.} The SFHs of our simulated galaxies are thus able to broadly reproduce a variety of stellar properties.

\subsection{Mass evolution}

In order to see how our simulated galaxies grow in stellar mass, we show, in Fig.~\ref{srel}, the stellar mass-halo mass relation at redshifts 3, 2, 1 and 0. We define the stellar mass to be the sum over all star particles within 0.1 $R_{200}$ and plot this mass against the dark matter mass contained within $R_{200}$. The simulated galaxies are then compared with the semi-empirically derived stellar to halo mass relation of \citet{MNW13}, who used abundance matching from simulated dark matter-only haloes constrained to match the observed stellar mass function. At $z=0$, most of the Auriga galaxies lie above this relation but within the 0.15 dex scatter assumed by \citet{MNW13}. A few have have stellar masses above the $1 \sigma$ scatter boundary. At earlier redshifts ({\bff first and second} panels of Fig.~\ref{srel}), we note that the haloes evolve almost parallel to the \citet{MNW13} relation; however they lie noticeably above the relation at $z=3$, which indicates an excess of early star formation {\bff at even earlier epochs}. Although the AGN feedback in these simulations plays a role in suppressing early star formation, its effectiveness varies, as noted above, and in some cases is insufficient. This may indicate the need for additional sources of feedback such as early stellar feedback \citep[e.g.,][]{SB13,AWN13,HKO14,WDS15}. {\bff \citet{SB13} showed that early stellar feedback reduced the amount of early star formation in the central regions by heating gas and extending its distribution, which helped prevent the formation of compact discs. The inclusion of such feedback in our model may help to bring our simulations into better agreement with abundance matching predictions, which will be investigated in future work.}

Haloes that grow quiescently, including the largest discs (highlighted), appear to evolve smoothly along the relation, whereas those that experience major mergers can evolve more sporadically. Such discontinuous behaviour violates the assumptions made in abundance matching at constant galaxy comoving number density and the subsequent derivation of progenitor-descendent relations \citep[see][]{TWM16}. {\bff This may become especially relevant at higher redshifts, when mergers are more frequent and abundance matching results more uncertain than for $z=0$.}

For a more detailed analysis of the build up of the galaxy, we follow the growth of stellar mass in the inner and outer regions of the simulated galaxies in Fig.~\ref{mgrowth}. For comparison, we show also the evolution of the corresponding quantities presented in \citet{PFF13}, who analysed the structural parameters of a set of star-forming galaxies, observed with HST WFC3 (\emph{Hubble Space Telescope Wide Field Camera 3}) as part of the CANDELS survey \citep{VBH12}. They assume that every galaxy remains close to the ridge of the star-forming main sequence as it grows, so that star-forming galaxies of given present-day stellar mass formed from star-forming galaxies of a well-defined and smaller stellar mass at earlier times. This allows them to link a sample of low-redshift disc galaxies from SDSS to representative sets of ``progenitors" at a sequence of earlier redshifts back to $z \sim 1.3$. In order to facilitate a fair comparison, we define the inner and outer regions in the same way as in the observed sample: inside and outside $R=2$ kpc, which are proxies for the bulge and disc components. The simulation means of the total mass and of the mass in the two regions evolve as observed if the observations are renormalised by a factor 1.5, particularly the inner region. A renormalisation of this type is expected because the results presented in \citet{PFF13} are for galaxies selected to be around the stellar mass $10^{10.5} \sim 3 \times 10^{10}$ $\rm M_{\odot}$, which is about half the mass of our simulated galaxy average. After this renormalisation, the most notable difference between the simulated galaxies and the observations is found in the outer region, which begins growing in mass earlier and continues growing later than observationally inferred. These discrepancies may be because the haloes in the Auriga simulation sample are selected to be isolated, and therefore may constitute a sample biased toward prominent discs that experience prolonged disc formation. However, we note that the assumption that all galaxies grow smoothly along the ridge of the star-forming main sequence cannot be precisely correct and is violated by mergers, which are particularly frequent at early times, so that the observationally derived curves do not necessarily reflect the true mass growth of late-type star-forming galaxies.

\begin{figure*} 
\centering
\includegraphics[scale=1.5,trim={0 0 0 0},clip]{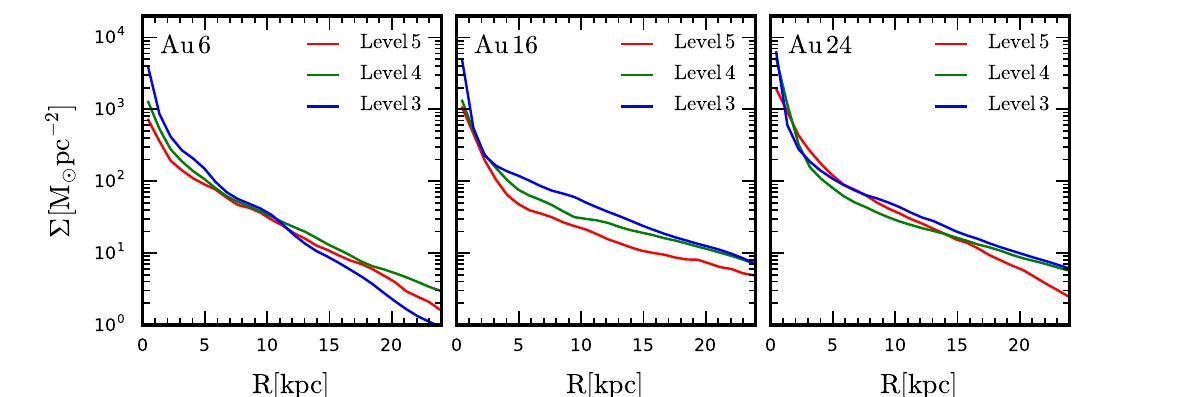} \\
\includegraphics[scale=1.5]{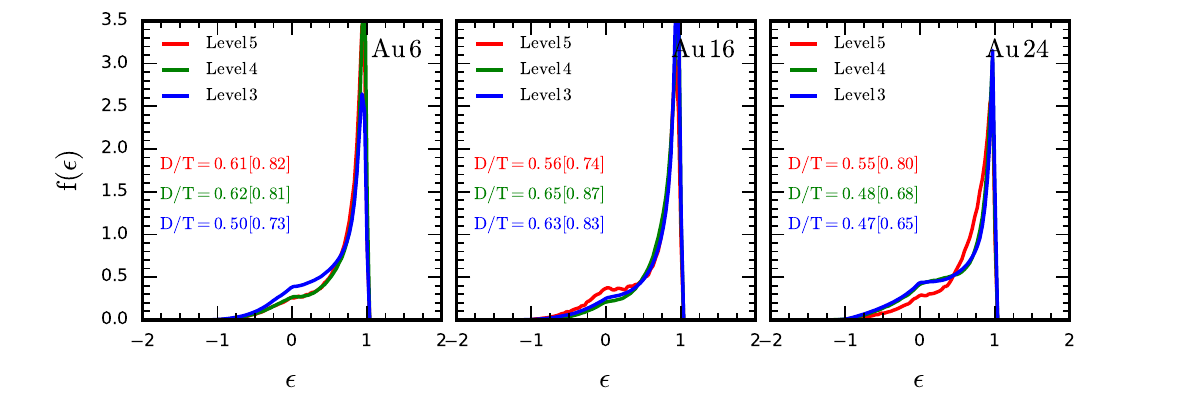} \\
\includegraphics[scale=1.54]{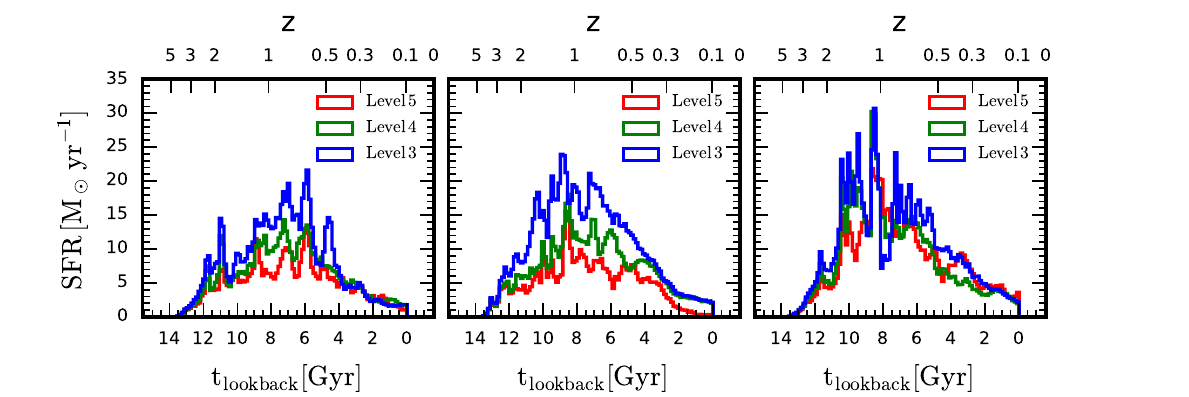} 
\caption{Stellar surface density profiles (top row), orbital circularity distribution (middle row) and the SFH of haloes Au 6 (first column), Au 16 (middle column) and Au 24 (third column) at three different resolution levels.}
\label{res1}
\end{figure*}

\begin{figure*} 
\centering
\includegraphics[scale=1.5]{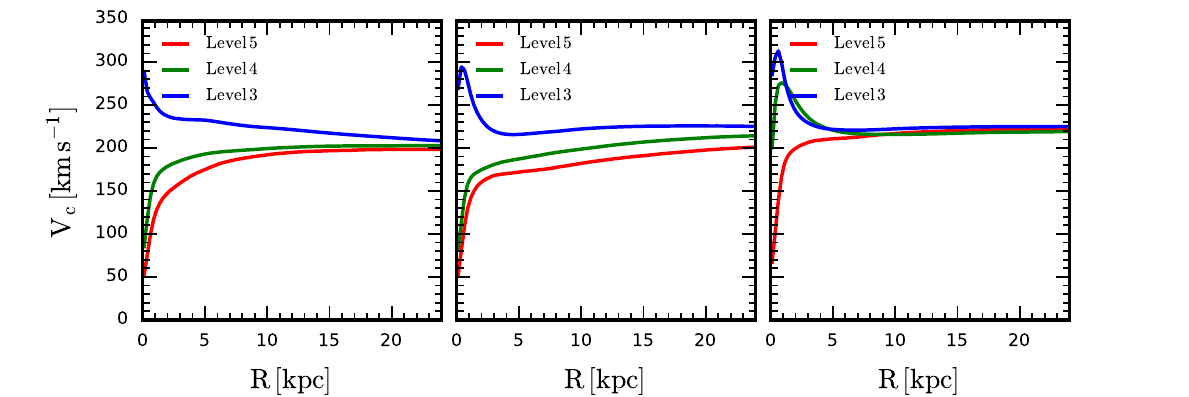} \\
\includegraphics[scale=1.54,trim={1.8cm 0 -0.1cm 0},clip]{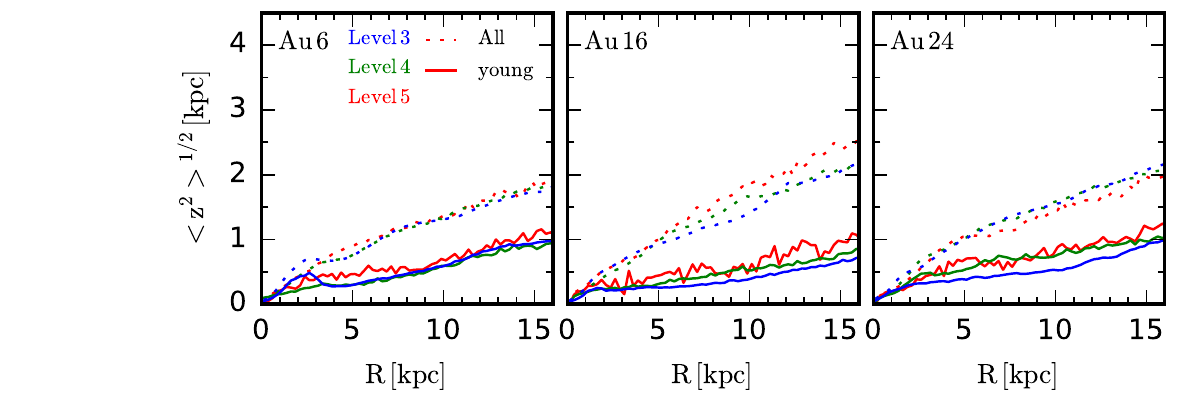} \\
\caption{As Fig.~\ref{res1}, but for the radial profiles of the circular velocity (top row) and the root mean square height (bottom row) for all star particles (dotted lines) and for star particles younger than 3 Gyr (solid lines).}
\label{res2}
\end{figure*}

\begin{figure} 
\centering
\includegraphics[scale=2.]{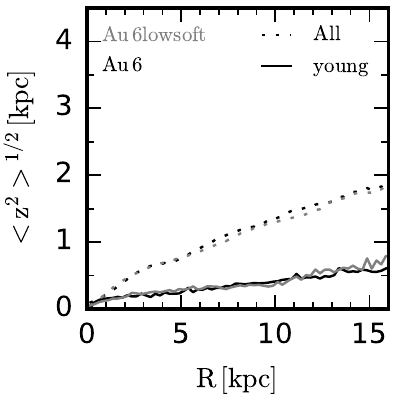} 
\caption{The radial profile of the root mean square height of simulation Au 6 and a re-run of the same halo with ten times lower softening lengths. }
\label{res3}
\end{figure}

\section{Resolution study}
\label{resstudy}

\begin{table}
\caption{Table of numerical resolution parameters at $z=0$. The columns are 1) resolution level; 2) dark matter particle mass; 3) typical baryonic element mass; 4) softening length of collisionless particles.}
\centering
\begin{tabular}{c c c c}
\\\toprule
Resolution level  & $\frac{m_{\rm DM}}{[\rm M_{\odot}]}$ & $\frac{m_{\rm b}}{[\rm M_{\odot}]}$ & $\frac{\epsilon}{[\rm pc]}$ \\
\hline                                                                     
4 & $3 \times 10 ^5$ & $5 \times 10^4$ & $369$ \\
5 & $2 \times 10 ^6$ & $4 \times 10^5$ & $738$ \\
3 & $4 \times 10 ^4$ & $6 \times 10^3$ & $184$ 
\\\bottomrule
\end{tabular}
\end{table}

In this section, we present a resolution study of our simulations. As discussed in the introduction, convergence of galaxy properties across different levels of resolution is difficult to achieve in simulations of galaxy formation. In \citet{MPS14}, one of the Aquarius simulations was tested across three levels of resolution and good convergence was obtained. The highest resolution level in that study is equivalent to that of the 30 simulations we have presented in this paper (known as level 4 in Aquarius nomenclature). This corresponds to a few million star particles in the main halo at $z=0$. In this paper, we attain a new resolution level (level 3) for three of our haloes, which is a factor of 8 and 2 better in mass and spatial resolution, respectively. In order to assess convergence, we here compare some basic quantities for these three simulations across three levels of resolution (see Table 2), which span a mass (spatial) resolution range of 64 (4), and evaluate their convergence.

In Fig.~\ref{res1}, we present the surface density profiles (top), the orbital circularity distribution (middle) and the SFHs (bottom) for simulations Au 6, Au 16 and Au 24. We find that each halo is generally well converged across the three levels of resolution. In particular, the circularity distributions and the inferred disc-total mass ratios are very similar, which indicates that the disc and bulge kinematic structure do not change significantly. The surface density profiles do show some shifts, however: in the case of Au 6, the slope of the profile becomes slightly steeper at the highest resolution level than in the others, although those of Au 16 and Au 24 appear relatively stable. The SFHs show a weak trend towards increased star formation at higher resolution, although the SFH of Au 24 is barely changed. Despite this tendency, the SFH retains its shape in all haloes.

The largest and most significant change with resolution is most easily seen in the circular velocity curves of the haloes, which we present in the top panels of \mbox{Fig.~\ref{res2}}. At the highest level of resolution, these curves become peaked near the centre, which reflects a more compact core. We investigated the origin of the relative excess of stellar density in the galaxy centre in the level 3 run: we examined the birth positions and ages of all star particles that are within 5 kpc of the galaxy centre at $z=0$. We find that most of the excess originates from star particles born in the central regions but over a range of times. The excess apparently reflects variations in the highly non-linear black hole accretion and feedback loops that help regulate central star formation \mbox{\citep[as postulated by][]{MPS14}}. It is notoriously difficult to obtain convergence for these processes.

To compare the vertical disc structure across resolution, we calculate the radial profile of the root mean square height for disc star particles, which in this resolution study is preferable to the vertical scale height owing to the poor particle sampling, and therefore poor density profile fitting, in the low resolution level 5 simulations. In the bottom panels of Fig.~\ref{res2}, we show the radial profile of the root mean square height of the stellar disc component for all star particles (dotted lines) and for star particles younger than 3 Gyr (solid lines). There appears to be tentative indications in the total stellar disc population in Au 16, and to a lesser extent the young disc stars, that discs become vertically thinner with increasing resolution, which one may na\"{i}vely interpret as the softening length setting a lower limit on the disc scale height. However, this trend is not seen in Au 6 nor Au 24, which both show very similar height profiles despite the reduction in softening length by a factor of 4 from level 5 to level 3, which indicates that overall, the vertical disc structure is well converged.

To investigate further the concern that the reduction in softening lengths leads to thinner discs, we re-run simulation Au 6 with ten times lower softening lengths (labelled Au 6lowsoft), and show the radial profile of the root mean square height of all stars and young stars in Fig.~\ref{res3} in addition to that of the fiducial Au 6 simulation. The difference in vertical disc thickness between the two simulations is negligible for both young stars and all stars at all radii, which clearly demonstrates that lower softening lengths do not lead to thinner discs. Based on these results, we caution that the softening length ought not to be referred to as a resolution limit, given that gravity is not simply turned off below the softening length scale.

We conclude from this study that our simulations are generally well converged across three resolution levels, the last of which is one level higher than probed in \mbox{\citet{MPS14}}. We stress that this level of convergence is extremely difficult to achieve in simulations of galaxy formation, and it suggests that our models are numerically well posed in the sense that the evolution of the simulated galaxies depends on physical parameters only, not on numerical resolution.

\section{Discussion and Conclusions}
\label{sec7}

In this paper, we have introduced a suite of cosmological simulations of the formation of galaxies in isolated Milky Way-mass haloes using the zoom-in technique. The simulation code follows many aspects of galaxy formation, including black hole accretion and feedback, stellar feedback, stellar evolution, chemical evolution, metallicity-dependent cooling, star formation and magnetic fields, and has been shown to produce realistic galaxy populations in large-scale cosmological boxes \citep{VGS13,GVS14}. The most novel aspects of the simulations with respect to previous studies are the large simulation sample, high resolution, the inclusion of magnetic fields that are seeded, grown and evolved in a fully cosmological context, and the black hole model, which has been used in only one simulation set previous to this study \citep{MPS14}. The suite comprises 30 simulations with typical baryonic mass resolution of around $5\times10^4$ $\rm M_{\odot}$, three of which have higher resolution counterparts with 8 times better mass resolution. The sample size, resolution, and galaxy formation model make these simulations some of the most advanced and comprehensive found in the literature. 

We analysed the properties of the stellar components of these simulations, and found that they lie on the expected stellar mass-halo mass relation, and are consistent with other key observables such as flat rotation curves, the mass-metallicity relation and present day star formation rates. The simulated galaxies are, on average, slightly too massive in the stellar component, particularly at earlier times such as $z=3$. This may indicate the need for {\bff early stellar feedback \citep[see e.g.,][]{SB13,AWN13}. Such feedback may bring  down the stellar masses of our simulations and reduce the number of compact galaxies (see Fig.~\ref{masssize}), which we will investigate in future work.}

The mass distributions of our simulated galaxies are realistic and produce rotationally supported discs: most of our simulations have flat rotation curves, with velocities that match the Tully-Fisher relation for Milky Way mass galaxies. We note that in the past is has proven difficult to reproduce flat circular velocity curves for galaxy discs in haloes in the virial mass range $1-2 \times 10^{12} $ $\rm M_{\odot}$ \citep[see for example][]{SBC10,WDS15,MNT15}, although a variety of recent models have achieved some success \citep{SB13,AWN13,MPS14,SFB15,CAR16}. The circularity distributions indicate that at least two thirds of the simulations have well defined, rotationally supported disc components, with disc-total mass ratios of up to 0.6-0.7 for the most disc dominated simulations. These are among the highest disc-total mass ratios in the literature, along with \citet{AWN13}. 

{\bff The vertical structure of our simulated discs appears to be somewhat thicker than expected from observations in the Milky Way, which may be related to energy from stellar feedback that acts to puff up the gas disc \citep[see][]{MGP16}. \citet[][]{RTA14} reported a similar conclusion, although their feedback recipe included radiation feedback that destroyed the morphology of their disc in addition to thickening it. However, the inner discs of our simulations show some consistency with observations of external galaxies.}

{\bff The SFR - stellar mass relation for the simulations is lower than observed by about a factor of 2 for $z=2$-3, but similar to observations for $z=0$-1. The simulations lie on a slope consistent with observed scaling relations for $z=1$-3. The SFRs tend to} peak at around redshift {\bff $1$ - 2}, followed by a steady decline until reaching a present-day star formation rate of order a few solar masses per year. The resulting present-day metallicities, mean stellar ages and colours are consistent with observations of late-type, star-forming galaxies. 

The radial distribution of star formation may be broadly categorised into either centrally concentrated or radially extended, inside-out formation. This variety of SFH in the suite translates into a wide range of {\bff disc sizes (with scale lengths that range from about 2 kpc to 11 kpc), a diversity that grows in time from a smaller range of early-time disc sizes that were consistent with observations.} We investigated the cause of this diversity and found that disc size correlates well with basic measures of halo spin, in agreement with theoretical expectations. We identified gas-rich quiescent mergers as a prominent mechanism by which the halo dark matter and gas acquire a high degree of specific angular momentum, which leads to high angular momentum star-forming material condensing around the disc, and therefore to inside-out star formation and large disc scale lengths. In addition, a fraction of the orbital angular momentum of the satellite is transferred to the pre-existing stellar disc. In contrast, compact discs typically have no such quiescent merger and a low degree of halo angular momentum.

We investigated the effect of AGN on disc size, and found that strong AGN feedback could reduce the size of the disc by injecting thermal energy into the halo and preventing the cooling and condensation of new star-forming gas around the disc. However, in terms of disc size this effect is mild, and we conclude that the dominant effect is the addition of angular momentum from infalling gas and stars. The most important effect of AGN feedback is to suppress central star formation, which is particularly important in haloes that have low angular momentum gas that leads to high gas densities in the galaxy centre. In these cases, the high gas density leads to strong AGN feedback that regulates star formation and prevents the formation of overly massive bulges. Thus, AGN feedback can play an important role in {\bff galaxies of Milky Way mass.}

Overall, we believe that our simulations represent one of the best currently available models for the formation {\bff of galaxies like the Milky Way}. The high fidelity of our model is supported by the strong convergence of many galaxy properties across three levels of resolution \citep[4 if one extrapolates from][]{MPS14}, without the need to retune parameters. As pointed out in \citet{MPS14}, this is a desirable trait that indicates that the results of the simulations reflect their physical assumptions rather than their numerical parameters. The simulations do not rely on resolution dependent star formation density thresholds, high star formation density thresholds \citep{GBM10,ATM11} or high resolution \citep{KMW07} to curtail star formation and obtain physically meaningful results.

In this paper, we have presented an overview of the Auriga simulation suite, with a focus on the present properties and formation history of the stellar disc. We note that there are other studies that have already made use of a subset of these simulations. In particular, an analysis of dynamical heating mechanisms of stellar discs is presented in \citet{GSG16}, and a study of the origin of disc warps is presented in \citet{GWG16} \citep[see also][]{GWM15}. There are also studies that provide predictions that are directly comparable to observations: chemo-dynamical features associated to spiral structure are studied in \citet{GSK16}, and an analysis of the stellar halo metal content is presented in \citet{MGG16}. Concerning the complete Auriga suite, the properties of the HI gas discs are presented in \citet{MGP16}. In the near future, we will present studies concerned with magnetic fields (Pakmor et al. in prep), satellites (Simpson et al. in prep), ex-situ stellar discs (G\'{o}mez et al. in prep), stellar halo properties (Monachesi et al. in prep), the circumgalactic medium (Nelson et al. in prep) and X-ray haloes (Campbell et al. in prep). Together, these papers will provide a comprehensive picture of the Auriga simulations.

\section*{acknowledgements}
We thank the referee for constructive and helpful comments that helped improve this paper. It is a pleasure to thank Antonela Monachesi, Dylan Nelson, Annalisa Pillepich, Christine Simpson, Martin Sparre, Rainer Weinberger and Jolanta Zjupa for helpful discussions and constructive comments. We thank Shannon Patel for providing his observationally derived multi-epoch stellar masses of star-forming galaxies, and Matthieu Schaller for providing the FOF group identifications for the Auriga sample in the Eagle Ref-L100N1504 volume and its corresponding dark matter only counterpart. RG is grateful for the hospitality of the Max Planck Institute for Astrophysics during the advanced stages of this work. RG and VS acknowledge support by the DFG Research Centre SFB-881 `The Milky Way System' through project A1. This work has also been supported by the European Research Council under ERC-StG grant EXAGAL- 308037. Part of the simulations of this paper used the SuperMUC system at the Leibniz Computing Centre, Garching, under the project PR85JE of the Gauss Centre for Supercomputing. This work used the DiRAC Data Centric system at Durham University, operated by the Institute for Computational Cosmology on behalf of the STFC DiRAC HPC Facility `www.dirac.ac.uk'. This equipment was funded by BIS National E-infrastructure capital grant ST/K00042X/1, STFC capital grant ST/H008519/1, and STFC DiRAC Operations grant ST/K003267/1 and Durham University. DiRAC is part of the UK National E-Infrastructure.

\bibliographystyle{mnras}
\bibliography{GG2R2.bbl}

\begin{thebibliography}{}
\makeatletter
\relax
\def\mn@urlcharsother{\let\do\@makeother \do\$\do\&\do\#\do\^\do\_\do\%\do\~}
\def\mn@doi{\begingroup\mn@urlcharsother \@ifnextchar [ {\mn@doi@}
  {\mn@doi@[]}}
\def\mn@doi@[#1]#2{\def\@tempa{#1}\ifx\@tempa\@empty \href
  {http://dx.doi.org/#2} {doi:#2}\else \href {http://dx.doi.org/#2} {#1}\fi
  \endgroup}
\def\mn@eprint#1#2{\mn@eprint@#1:#2::\@nil}
\def\mn@eprint@arXiv#1{\href {http://arxiv.org/abs/#1} {{\tt arXiv:#1}}}
\def\mn@eprint@dblp#1{\href {http://dblp.uni-trier.de/rec/bibtex/#1.xml}
  {dblp:#1}}
\def\mn@eprint@#1:#2:#3:#4\@nil{\def\@tempa {#1}\def\@tempb {#2}\def\@tempc
  {#3}\ifx \@tempc \@empty \let \@tempc \@tempb \let \@tempb \@tempa \fi \ifx
  \@tempb \@empty \def\@tempb {arXiv}\fi \@ifundefined
  {mn@eprint@\@tempb}{\@tempb:\@tempc}{\expandafter \expandafter \csname
  mn@eprint@\@tempb\endcsname \expandafter{\@tempc}}}

\bibitem[\protect\citeauthoryear{{Abadi}, {Navarro}, {Steinmetz}  \&
  {Eke}}{{Abadi} et~al.}{2003}]{ANS03}
{Abadi} M.~G.,  {Navarro} J.~F.,  {Steinmetz} M.,   {Eke} V.~R.,  2003, \mn@doi
  [\apj] {10.1086/375512}, \href
  {http://adsabs.harvard.edu/abs/2003ApJ...591..499A} {591, 499}

\bibitem[\protect\citeauthoryear{{Agertz}, {Teyssier}  \& {Moore}}{{Agertz}
  et~al.}{2011}]{ATM11}
{Agertz} O.,  {Teyssier} R.,   {Moore} B.,  2011, \mn@doi [\mnras]
  {10.1111/j.1365-2966.2010.17530.x}, \href
  {http://adsabs.harvard.edu/abs/2011MNRAS.410.1391A} {410, 1391}

\bibitem[\protect\citeauthoryear{{Aumer}, {White}, {Naab}  \&
  {Scannapieco}}{{Aumer} et~al.}{2013}]{AWN13}
{Aumer} M.,  {White} S.~D.~M.,  {Naab} T.,   {Scannapieco} C.,  2013, \mn@doi
  [\mnras] {10.1093/mnras/stt1230}, \href
  {http://adsabs.harvard.edu/abs/2013MNRAS.434.3142A} {434, 3142}

\bibitem[\protect\citeauthoryear{{Barnes} \& {Hut}}{{Barnes} \&
  {Hut}}{1986}]{BH86}
{Barnes} J.,  {Hut} P.,  1986, \mn@doi [\nat] {10.1038/324446a0}, \href
  {http://adsabs.harvard.edu/abs/1986Natur.324..446B} {324, 446}

\bibitem[\protect\citeauthoryear{{Bauer} \& {Springel}}{{Bauer} \&
  {Springel}}{2012}]{BS12}
{Bauer} A.,  {Springel} V.,  2012, \mn@doi [\mnras]
  {10.1111/j.1365-2966.2012.21058.x}, \href
  {http://adsabs.harvard.edu/abs/2012MNRAS.423.2558B} {423, 2558}

\bibitem[\protect\citeauthoryear{{Benson}, {Bower}, {Frenk}, {Lacey}, {Baugh}
  \& {Cole}}{{Benson} et~al.}{2003}]{BBF03}
{Benson} A.~J.,  {Bower} R.~G.,  {Frenk} C.~S.,  {Lacey} C.~G.,  {Baugh} C.~M.,
    {Cole} S.,  2003, \mn@doi [\apj] {10.1086/379160}, \href
  {http://adsabs.harvard.edu/abs/2003ApJ...599...38B} {599, 38}

\bibitem[\protect\citeauthoryear{{Bershady}, {Verheijen}, {Swaters},
  {Andersen}, {Westfall}  \& {Martinsson}}{{Bershady} et~al.}{2010}]{BVS10}
{Bershady} M.~A.,  {Verheijen} M.~A.~W.,  {Swaters} R.~A.,  {Andersen} D.~R.,
  {Westfall} K.~B.,   {Martinsson} T.,  2010, \mn@doi [\apj]
  {10.1088/0004-637X/716/1/198}, \href
  {http://adsabs.harvard.edu/abs/2010ApJ...716..198B} {716, 198}

\bibitem[\protect\citeauthoryear{{Bland-Hawthorn} \&
  {Gerhard}}{{Bland-Hawthorn} \& {Gerhard}}{2016}]{BHG16}
{Bland-Hawthorn} J.,  {Gerhard} O.,  2016, \mn@doi [\araa]
  {10.1146/annurev-astro-081915-023441}, \href
  {http://adsabs.harvard.edu/abs/2016ARA%26A..54..529B} {54, 529}

\bibitem[\protect\citeauthoryear{{Bondi}}{{Bondi}}{1952}]{B52}
{Bondi} H.,  1952, \mn@doi [\mnras] {10.1093/mnras/112.2.195}, \href
  {http://adsabs.harvard.edu/abs/1952MNRAS.112..195B} {112, 195}

\bibitem[\protect\citeauthoryear{{Bondi} \& {Hoyle}}{{Bondi} \&
  {Hoyle}}{1944}]{BH44}
{Bondi} H.,  {Hoyle} F.,  1944, \mn@doi [\mnras] {10.1093/mnras/104.5.273},
  \href {http://adsabs.harvard.edu/abs/1944MNRAS.104..273B} {104, 273}

\bibitem[\protect\citeauthoryear{{Brook} et~al.,}{{Brook} et~al.}{2011}]{BGR11}
{Brook} C.~B.,  et~al., 2011, \mn@doi [\mnras]
  {10.1111/j.1365-2966.2011.18545.x}, \href
  {http://adsabs.harvard.edu/abs/2011MNRAS.415.1051B} {415, 1051}

\bibitem[\protect\citeauthoryear{{Brooks} et~al.,}{{Brooks}
  et~al.}{2011}]{BSG11}
{Brooks} A.~M.,  et~al., 2011, \mn@doi [\apj] {10.1088/0004-637X/728/1/51},
  \href {http://adsabs.harvard.edu/abs/2011ApJ...728...51B} {728, 51}

\bibitem[\protect\citeauthoryear{{Bruzual} \& {Charlot}}{{Bruzual} \&
  {Charlot}}{2003}]{BC03}
{Bruzual} G.,  {Charlot} S.,  2003, \mn@doi [\mnras]
  {10.1046/j.1365-8711.2003.06897.x}, \href
  {http://adsabs.harvard.edu/abs/2003MNRAS.344.1000B} {344, 1000}

\bibitem[\protect\citeauthoryear{{Bryan} et~al.}{{Bryan} et~al.}{2014}]{BNO14}
{Bryan} G.~L.,  et~al., 2014, \mn@doi [\apjs] {10.1088/0067-0049/211/2/19},
  \href {http://adsabs.harvard.edu/abs/2014ApJS..211...19B} {211, 19}

\bibitem[\protect\citeauthoryear{{Bullock}, {Dekel}, {Kolatt}, {Kravtsov},
  {Klypin}, {Porciani}  \& {Primack}}{{Bullock} et~al.}{2001}]{BDK01}
{Bullock} J.~S.,  {Dekel} A.,  {Kolatt} T.~S.,  {Kravtsov} A.~V.,  {Klypin}
  A.~A.,  {Porciani} C.,   {Primack} J.~R.,  2001, \mn@doi [\apj]
  {10.1086/321477}, \href {http://adsabs.harvard.edu/abs/2001ApJ...555..240B}
  {555, 240}

\bibitem[\protect\citeauthoryear{{Chabrier}}{{Chabrier}}{2003}]{C03}
{Chabrier} G.,  2003, \mn@doi [\pasp] {10.1086/376392}, \href
  {http://adsabs.harvard.edu/abs/2003PASP..115..763C} {115, 763}

\bibitem[\protect\citeauthoryear{{Col{\'{\i}}n}, {Avila-Reese},
  {Roca-F{\`a}brega}  \& {Valenzuela}}{{Col{\'{\i}}n} et~al.}{2016}]{CAR16}
{Col{\'{\i}}n} P.,  {Avila-Reese} V.,  {Roca-F{\`a}brega} S.,   {Valenzuela}
  O.,  2016, \mn@doi [\apj] {10.3847/0004-637X/829/2/98}, \href
  {http://adsabs.harvard.edu/abs/2016ApJ...829...98C} {829, 98}

\bibitem[\protect\citeauthoryear{{Courteau}, {Dutton}, {van den Bosch},
  {MacArthur}, {Dekel}, {McIntosh}  \& {Dale}}{{Courteau} et~al.}{2007}]{CDV07}
{Courteau} S.,  {Dutton} A.~A.,  {van den Bosch} F.~C.,  {MacArthur} L.~A.,
  {Dekel} A.,  {McIntosh} D.~H.,   {Dale} D.~A.,  2007, \mn@doi [\apj]
  {10.1086/522193}, \href {http://adsabs.harvard.edu/abs/2007ApJ...671..203C}
  {671, 203}

\bibitem[\protect\citeauthoryear{{Creasey}, {Theuns}  \& {Bower}}{{Creasey}
  et~al.}{2013}]{CTB13}
{Creasey} P.,  {Theuns} T.,   {Bower} R.~G.,  2013, \mn@doi [\mnras]
  {10.1093/mnras/sts439}, \href
  {http://adsabs.harvard.edu/abs/2013MNRAS.429.1922C} {429, 1922}

\bibitem[\protect\citeauthoryear{{D'Onghia} \& {Navarro}}{{D'Onghia} \&
  {Navarro}}{2007}]{DN07}
{D'Onghia} E.,  {Navarro} J.~F.,  2007, \mn@doi [\mnras]
  {10.1111/j.1745-3933.2007.00348.x}, \href
  {http://adsabs.harvard.edu/abs/2007MNRAS.380L..58D} {380, L58}

\bibitem[\protect\citeauthoryear{{Daddi} et~al.}{{Daddi} et~al.}{2007}]{DMC07}
{Daddi} E.,  et~al., 2007, \mn@doi [\apj] {10.1086/521818}, \href
  {http://adsabs.harvard.edu/abs/2007ApJ...670..156D} {670, 156}

\bibitem[\protect\citeauthoryear{{Dalla Vecchia} \& {Schaye}}{{Dalla Vecchia}
  \& {Schaye}}{2008}]{DS08}
{Dalla Vecchia} C.,  {Schaye} J.,  2008, \mn@doi [\mnras]
  {10.1111/j.1365-2966.2008.13322.x}, \href
  {http://adsabs.harvard.edu/abs/2008MNRAS.387.1431D} {387, 1431}

\bibitem[\protect\citeauthoryear{{Dalla Vecchia} \& {Schaye}}{{Dalla Vecchia}
  \& {Schaye}}{2012}]{DVS12}
{Dalla Vecchia} C.,  {Schaye} J.,  2012, \mn@doi [\mnras]
  {10.1111/j.1365-2966.2012.21704.x}, \href
  {http://adsabs.harvard.edu/abs/2012MNRAS.426..140D} {426, 140}

\bibitem[\protect\citeauthoryear{{Davis}, {Efstathiou}, {Frenk}  \&
  {White}}{{Davis} et~al.}{1985}]{DEF95}
{Davis} M.,  {Efstathiou} G.,  {Frenk} C.~S.,   {White} S.~D.~M.,  1985,
  \mn@doi [\apj] {10.1086/163168}, \href
  {http://adsabs.harvard.edu/abs/1985ApJ...292..371D} {292, 371}

\bibitem[\protect\citeauthoryear{{Diemand}, {Kuhlen}  \& {Madau}}{{Diemand}
  et~al.}{2007}]{DKM07}
{Diemand} J.,  {Kuhlen} M.,   {Madau} P.,  2007, \mn@doi [\apj]
  {10.1086/510736}, \href {http://adsabs.harvard.edu/abs/2007ApJ...657..262D}
  {657, 262}

\bibitem[\protect\citeauthoryear{{Dubois}, {Gavazzi}, {Peirani}  \&
  {Silk}}{{Dubois} et~al.}{2013}]{DGP13}
{Dubois} Y.,  {Gavazzi} R.,  {Peirani} S.,   {Silk} J.,  2013, \mn@doi [\mnras]
  {10.1093/mnras/stt997}, \href
  {http://adsabs.harvard.edu/abs/2013MNRAS.433.3297D} {433, 3297}

\bibitem[\protect\citeauthoryear{{Dutton} et~al.,}{{Dutton}
  et~al.}{2011}]{DBF11}
{Dutton} A.~A.,  et~al., 2011, \mn@doi [\mnras]
  {10.1111/j.1365-2966.2010.17555.x}, \href
  {http://adsabs.harvard.edu/abs/2011MNRAS.410.1660D} {410, 1660}

\bibitem[\protect\citeauthoryear{{Elbaz} et~al.}{{Elbaz} et~al.}{2007}]{EDL07}
{Elbaz} D.,  et~al., 2007, \mn@doi [\aap] {10.1051/0004-6361:20077525}, \href
  {http://adsabs.harvard.edu/abs/2007A%26A...468...33E} {468, 33}

\bibitem[\protect\citeauthoryear{{Elbaz} et~al.}{{Elbaz} et~al.}{2011}]{EDH11}
{Elbaz} D.,  et~al., 2011, \mn@doi [\aap] {10.1051/0004-6361/201117239}, \href
  {http://adsabs.harvard.edu/abs/2011A%26A...533A.119E} {533, A119}

\bibitem[\protect\citeauthoryear{{Fall} \& {Efstathiou}}{{Fall} \&
  {Efstathiou}}{1980}]{FE80}
{Fall} S.~M.,  {Efstathiou} G.,  1980, \mn@doi [\mnras]
  {10.1093/mnras/193.2.189}, \href
  {http://adsabs.harvard.edu/abs/1980MNRAS.193..189F} {193, 189}

\bibitem[\protect\citeauthoryear{{Faucher-Gigu{\`e}re}, {Lidz}, {Zaldarriaga}
  \& {Hernquist}}{{Faucher-Gigu{\`e}re} et~al.}{2009}]{FG09}
{Faucher-Gigu{\`e}re} C.-A.,  {Lidz} A.,  {Zaldarriaga} M.,   {Hernquist} L.,
  2009, \mn@doi [\apj] {10.1088/0004-637X/703/2/1416}, \href
  {http://adsabs.harvard.edu/abs/2009ApJ...703.1416F} {703, 1416}

\bibitem[\protect\citeauthoryear{{Few}, {Dobbs}, {Pettitt}  \&
  {Konstandin}}{{Few} et~al.}{2016}]{FDP16}
{Few} C.~G.,  {Dobbs} C.,  {Pettitt} A.,   {Konstandin} L.,  2016, \mn@doi
  [\mnras] {10.1093/mnras/stw1226}, \href
  {http://adsabs.harvard.edu/abs/2016MNRAS.460.4382F} {460, 4382}

\bibitem[\protect\citeauthoryear{{Flynn}, {Holmberg}, {Portinari}, {Fuchs}  \&
  {Jahrei{\ss}}}{{Flynn} et~al.}{2006}]{FHP06}
{Flynn} C.,  {Holmberg} J.,  {Portinari} L.,  {Fuchs} B.,   {Jahrei{\ss}} H.,
  2006, \mn@doi [\mnras] {10.1111/j.1365-2966.2006.10911.x}, \href
  {http://adsabs.harvard.edu/abs/2006MNRAS.372.1149F} {372, 1149}

\bibitem[\protect\citeauthoryear{{Gadotti}}{{Gadotti}}{2009}]{G09}
{Gadotti} D.~A.,  2009, \mn@doi [\mnras] {10.1111/j.1365-2966.2008.14257.x},
  \href {http://adsabs.harvard.edu/abs/2009MNRAS.393.1531G} {393, 1531}

\bibitem[\protect\citeauthoryear{{Gallazzi}, {Charlot}, {Brinchmann}, {White}
  \& {Tremonti}}{{Gallazzi} et~al.}{2005}]{GCB05}
{Gallazzi} A.,  {Charlot} S.,  {Brinchmann} J.,  {White} S.~D.~M.,   {Tremonti}
  C.~A.,  2005, \mn@doi [\mnras] {10.1111/j.1365-2966.2005.09321.x}, \href
  {http://adsabs.harvard.edu/abs/2005MNRAS.362...41G} {362, 41}

\bibitem[\protect\citeauthoryear{{Genel} et~al.,}{{Genel} et~al.}{2014}]{GVS14}
{Genel} S.,  et~al., 2014, \mn@doi [\mnras] {10.1093/mnras/stu1654}, \href
  {http://adsabs.harvard.edu/abs/2014MNRAS.445..175G} {445, 175}

\bibitem[\protect\citeauthoryear{{G{\'o}mez}, {White}, {Marinacci}, {Slater},
  {Grand}, {Springel}  \& {Pakmor}}{{G{\'o}mez} et~al.}{2016}]{GWM15}
{G{\'o}mez} F.~A.,  {White} S.~D.~M.,  {Marinacci} F.,  {Slater} C.~T.,
  {Grand} R.~J.~J.,  {Springel} V.,   {Pakmor} R.,  2016, \mn@doi [\mnras]
  {10.1093/mnras/stv2786}, \href
  {http://adsabs.harvard.edu/abs/2016MNRAS.456.2779G} {456, 2779}

\bibitem[\protect\citeauthoryear{{G{\'o}mez}, {White}, {Grand}, {Marinacci},
  {Springel}  \& {Pakmor}}{{G{\'o}mez} et~al.}{2017}]{GWG16}
{G{\'o}mez} F.~A.,  {White} S.~D.~M.,  {Grand} R.~J.~J.,  {Marinacci} F.,
  {Springel} V.,   {Pakmor} R.,  2017, \mn@doi [\mnras]
  {10.1093/mnras/stw2957}, \href
  {http://adsabs.harvard.edu/abs/2017MNRAS.465.3446G} {465, 3446}

\bibitem[\protect\citeauthoryear{{Governato} et~al.,}{{Governato}
  et~al.}{2010}]{GBM10}
{Governato} F.,  et~al., 2010, \mn@doi [\nat] {10.1038/nature08640}, \href
  {http://adsabs.harvard.edu/abs/2010Natur.463..203G} {463, 203}

\bibitem[\protect\citeauthoryear{{Grand}, {Springel}, {G{\'o}mez}, {Marinacci},
  {Pakmor}, {Campbell}  \& {Jenkins}}{{Grand} et~al.}{2016a}]{GSG16}
{Grand} R.~J.~J.,  {Springel} V.,  {G{\'o}mez} F.~A.,  {Marinacci} F.,
  {Pakmor} R.,  {Campbell} D.~J.~R.,   {Jenkins} A.,  2016a, \mn@doi [\mnras]
  {10.1093/mnras/stw601}, \href
  {http://adsabs.harvard.edu/abs/2016MNRAS.tmp..395G} {}

\bibitem[\protect\citeauthoryear{{Grand} et~al.,}{{Grand}
  et~al.}{2016b}]{GSK16}
{Grand} R.~J.~J.,  et~al., 2016b, \mn@doi [\mnras] {10.1093/mnrasl/slw086},
  \href {http://adsabs.harvard.edu/abs/2016MNRAS.460L..94G} {460, L94}

\bibitem[\protect\citeauthoryear{{Guedes}, {Callegari}, {Madau}  \&
  {Mayer}}{{Guedes} et~al.}{2011}]{GC11}
{Guedes} J.,  {Callegari} S.,  {Madau} P.,   {Mayer} L.,  2011, \mn@doi [\apj]
  {10.1088/0004-637X/742/2/76}, \href
  {http://adsabs.harvard.edu/abs/2011ApJ...742...76G} {742, 76}

\bibitem[\protect\citeauthoryear{{Guo}, {White}, {Li}  \&
  {Boylan-Kolchin}}{{Guo} et~al.}{2010}]{GWB10}
{Guo} Q.,  {White} S.,  {Li} C.,   {Boylan-Kolchin} M.,  2010, \mn@doi [\mnras]
  {10.1111/j.1365-2966.2010.16341.x}, \href
  {http://adsabs.harvard.edu/abs/2010MNRAS.404.1111G} {404, 1111}

\bibitem[\protect\citeauthoryear{{Herpich} et~al.,}{{Herpich}
  et~al.}{2015}]{HSD15}
{Herpich} J.,  et~al., 2015, \mn@doi [\mnras] {10.1093/mnrasl/slv006}, \href
  {http://adsabs.harvard.edu/abs/2015MNRAS.448L..99H} {448, L99}

\bibitem[\protect\citeauthoryear{{Hopkins}, {Quataert}  \& {Murray}}{{Hopkins}
  et~al.}{2012}]{HQM12}
{Hopkins} P.~F.,  {Quataert} E.,   {Murray} N.,  2012, \mn@doi [\mnras]
  {10.1111/j.1365-2966.2012.20578.x}, \href
  {http://adsabs.harvard.edu/abs/2012MNRAS.421.3488H} {421, 3488}

\bibitem[\protect\citeauthoryear{{Hopkins}, {Kere{\v s}}, {O{\~n}orbe},
  {Faucher-Gigu{\`e}re}, {Quataert}, {Murray}  \& {Bullock}}{{Hopkins}
  et~al.}{2014}]{HKO14}
{Hopkins} P.~F.,  {Kere{\v s}} D.,  {O{\~n}orbe} J.,  {Faucher-Gigu{\`e}re}
  C.-A.,  {Quataert} E.,  {Murray} N.,   {Bullock} J.~S.,  2014, \mn@doi
  [\mnras] {10.1093/mnras/stu1738}, \href
  {http://adsabs.harvard.edu/abs/2014MNRAS.445..581H} {445, 581}

\bibitem[\protect\citeauthoryear{{Huang} et~al.}{{Huang} et~al.}{2016}]{HLY16}
{Huang} Y.,  et~al., 2016, \mn@doi [\mnras] {10.1093/mnras/stw2096}, \href
  {http://adsabs.harvard.edu/abs/2016MNRAS.463.2623H} {463, 2623}

\bibitem[\protect\citeauthoryear{{Jenkins}}{{Jenkins}}{2010}]{J10}
{Jenkins} A.,  2010, \mn@doi [\mnras] {10.1111/j.1365-2966.2010.16259.x}, \href
  {http://adsabs.harvard.edu/abs/2010MNRAS.403.1859J} {403, 1859}

\bibitem[\protect\citeauthoryear{{Jenkins}}{{Jenkins}}{2013}]{J13}
{Jenkins} A.,  2013, \mn@doi [\mnras] {10.1093/mnras/stt1154}, \href
  {http://adsabs.harvard.edu/abs/2013MNRAS.434.2094J} {434, 2094}

\bibitem[\protect\citeauthoryear{{Juri{\'c}}, {Ivezi{\'c}}, {Brooks}
  et~al.}{{Juri{\'c}} et~al.}{2008}]{JIB08}
{Juri{\'c}} M.,  {Ivezi{\'c}} {\v Z}.,  {Brooks} A.,   et~al., 2008, \mn@doi
  [\apj] {10.1086/523619}, \href
  {http://adsabs.harvard.edu/abs/2008ApJ...673..864J} {673, 864}

\bibitem[\protect\citeauthoryear{{Karakas}}{{Karakas}}{2010}]{K10}
{Karakas} A.~I.,  2010, \mn@doi [\mnras] {10.1111/j.1365-2966.2009.16198.x},
  \href {http://adsabs.harvard.edu/abs/2010MNRAS.403.1413K} {403, 1413}

\bibitem[\protect\citeauthoryear{{Karim} et~al.}{{Karim} et~al.}{2011}]{KSM11}
{Karim} A.,  et~al., 2011, \mn@doi [\apj] {10.1088/0004-637X/730/2/61}, \href
  {http://adsabs.harvard.edu/abs/2011ApJ...730...61K} {730, 61}

\bibitem[\protect\citeauthoryear{{Katz}}{{Katz}}{1992}]{K92}
{Katz} N.,  1992, \mn@doi [\apj] {10.1086/171366}, \href
  {http://adsabs.harvard.edu/abs/1992ApJ...391..502K} {391, 502}

\bibitem[\protect\citeauthoryear{{Katz} \& {Gunn}}{{Katz} \&
  {Gunn}}{1991}]{KG91}
{Katz} N.,  {Gunn} J.~E.,  1991, \mn@doi [\apj] {10.1086/170367}, \href
  {http://adsabs.harvard.edu/abs/1991ApJ...377..365K} {377, 365}

\bibitem[\protect\citeauthoryear{{Kauffmann} et~al.,}{{Kauffmann}
  et~al.}{2003}]{KHW03}
{Kauffmann} G.,  et~al., 2003, \mn@doi [\mnras]
  {10.1046/j.1365-8711.2003.06291.x}, \href
  {http://adsabs.harvard.edu/abs/2003MNRAS.341...33K} {341, 33}

\bibitem[\protect\citeauthoryear{{Kaufmann}, {Mayer}, {Wadsley}, {Stadel}  \&
  {Moore}}{{Kaufmann} et~al.}{2007}]{KMW07}
{Kaufmann} T.,  {Mayer} L.,  {Wadsley} J.,  {Stadel} J.,   {Moore} B.,  2007,
  \mn@doi [\mnras] {10.1111/j.1365-2966.2006.11314.x}, \href
  {http://adsabs.harvard.edu/abs/2007MNRAS.375...53K} {375, 53}

\bibitem[\protect\citeauthoryear{{Kawata}}{{Kawata}}{2001}]{K01}
{Kawata} D.,  2001, \mn@doi [\apj] {10.1086/322309}, \href
  {http://adsabs.harvard.edu/abs/2001ApJ...558..598K} {558, 598}

\bibitem[\protect\citeauthoryear{{Kawata} \& {Gibson}}{{Kawata} \&
  {Gibson}}{2003}]{KG03}
{Kawata} D.,  {Gibson} B.~K.,  2003, \mn@doi [\mnras]
  {10.1046/j.1365-8711.2003.06356.x}, \href
  {http://adsabs.harvard.edu/abs/2003MNRAS.340..908K} {340, 908}

\bibitem[\protect\citeauthoryear{{Keller}, {Wadsley}, {Benincasa}  \&
  {Couchman}}{{Keller} et~al.}{2014}]{KWB14}
{Keller} B.~W.,  {Wadsley} J.,  {Benincasa} S.~M.,   {Couchman} H.~M.~P.,
  2014, \mn@doi [\mnras] {10.1093/mnras/stu1058}, \href
  {http://adsabs.harvard.edu/abs/2014MNRAS.442.3013K} {442, 3013}

\bibitem[\protect\citeauthoryear{{Krumholz} \& {Thompson}}{{Krumholz} \&
  {Thompson}}{2013}]{KT13}
{Krumholz} M.~R.,  {Thompson} T.~A.,  2013, \mn@doi [\mnras]
  {10.1093/mnras/stt1174}, \href
  {http://adsabs.harvard.edu/abs/2013MNRAS.434.2329K} {434, 2329}

\bibitem[\protect\citeauthoryear{{Lange} et~al.}{{Lange} et~al.}{2016}]{LMD16}
{Lange} R.,  et~al., 2016, \mn@doi [\mnras] {10.1093/mnras/stw1495}, \href
  {http://adsabs.harvard.edu/abs/2016MNRAS.462.1470L} {462, 1470}

\bibitem[\protect\citeauthoryear{{Leitner} \& {Kravtsov}}{{Leitner} \&
  {Kravtsov}}{2011}]{LK11}
{Leitner} S.~N.,  {Kravtsov} A.~V.,  2011, \mn@doi [\apj]
  {10.1088/0004-637X/734/1/48}, \href
  {http://adsabs.harvard.edu/abs/2011ApJ...734...48L} {734, 48}

\bibitem[\protect\citeauthoryear{{Magdis}, {Rigopoulou}, {Huang}  \&
  {Fazio}}{{Magdis} et~al.}{2010}]{MRH10}
{Magdis} G.~E.,  {Rigopoulou} D.,  {Huang} J.-S.,   {Fazio} G.~G.,  2010,
  \mn@doi [\mnras] {10.1111/j.1365-2966.2009.15779.x}, \href
  {http://adsabs.harvard.edu/abs/2010MNRAS.401.1521M} {401, 1521}

\bibitem[\protect\citeauthoryear{{Maoz}, {Mannucci}  \& {Brandt}}{{Maoz}
  et~al.}{2012}]{MMB12}
{Maoz} D.,  {Mannucci} F.,   {Brandt} T.~D.,  2012, \mn@doi [\mnras]
  {10.1111/j.1365-2966.2012.21871.x}, \href
  {http://adsabs.harvard.edu/abs/2012MNRAS.426.3282M} {426, 3282}

\bibitem[\protect\citeauthoryear{{Marinacci}, {Pakmor}  \&
  {Springel}}{{Marinacci} et~al.}{2014a}]{MPS14}
{Marinacci} F.,  {Pakmor} R.,   {Springel} V.,  2014a, \mn@doi [\mnras]
  {10.1093/mnras/stt2003}, \href
  {http://adsabs.harvard.edu/abs/2014MNRAS.437.1750M} {437, 1750}

\bibitem[\protect\citeauthoryear{{Marinacci}, {Pakmor}, {Springel}  \&
  {Simpson}}{{Marinacci} et~al.}{2014b}]{MPS14b}
{Marinacci} F.,  {Pakmor} R.,  {Springel} V.,   {Simpson} C.~M.,  2014b,
  \mn@doi [\mnras] {10.1093/mnras/stu1136}, \href
  {http://adsabs.harvard.edu/abs/2014MNRAS.442.3745M} {442, 3745}

\bibitem[\protect\citeauthoryear{{Marinacci}, {Vogelsberger}, {Mocz}  \&
  {Pakmor}}{{Marinacci} et~al.}{2015}]{MVM15}
{Marinacci} F.,  {Vogelsberger} M.,  {Mocz} P.,   {Pakmor} R.,  2015, \mn@doi
  [\mnras] {10.1093/mnras/stv1692}, \href
  {http://adsabs.harvard.edu/abs/2015MNRAS.453.3999M} {453, 3999}

\bibitem[\protect\citeauthoryear{{Marinacci}, {Grand}, {Pakmor}, {Springel},
  {G{\'o}mez}, {Frenk}  \& {White}}{{Marinacci} et~al.}{2016}]{MGP16}
{Marinacci} F.,  {Grand} R.,  {Pakmor} R.,  {Springel} V.,  {G{\'o}mez} F.,
  {Frenk} C.,   {White} S.,  2016, preprint, \href
  {http://adsabs.harvard.edu/abs/2016arXiv161001594M} {} (\mn@eprint {arXiv}
  {1610.01594})

\bibitem[\protect\citeauthoryear{{Martinsson}, {Verheijen}, {Westfall},
  {Bershady}, {Schechtman-Rook}, {Andersen}  \& {Swaters}}{{Martinsson}
  et~al.}{2013}]{MVW13}
{Martinsson} T.~P.~K.,  {Verheijen} M.~A.~W.,  {Westfall} K.~B.,  {Bershady}
  M.~A.,  {Schechtman-Rook} A.,  {Andersen} D.~R.,   {Swaters} R.~A.,  2013,
  \mn@doi [\aap] {10.1051/0004-6361/201220515}, \href
  {http://adsabs.harvard.edu/abs/2013A%26A...557A.130M} {557, A130}

\bibitem[\protect\citeauthoryear{{McAlpine} et~al.}{{McAlpine}
  et~al.}{2016}]{MHS16}
{McAlpine} S.,  et~al., 2016, \mn@doi [Astronomy and Computing]
  {10.1016/j.ascom.2016.02.004}, \href
  {http://adsabs.harvard.edu/abs/2016A%26C....15...72M} {15, 72}

\bibitem[\protect\citeauthoryear{{McGaugh} \& {Schombert}}{{McGaugh} \&
  {Schombert}}{2015}]{MS15}
{McGaugh} S.~S.,  {Schombert} J.~M.,  2015, \mn@doi [\apj]
  {10.1088/0004-637X/802/1/18}, \href
  {http://adsabs.harvard.edu/abs/2015ApJ...802...18M} {802, 18}

\bibitem[\protect\citeauthoryear{{Mo}, {Mao}  \& {White}}{{Mo}
  et~al.}{1998}]{MMW98}
{Mo} H.~J.,  {Mao} S.,   {White} S.~D.~M.,  1998, \mn@doi [\mnras]
  {10.1046/j.1365-8711.1998.01227.x}, \href
  {http://adsabs.harvard.edu/abs/1998MNRAS.295..319M} {295, 319}

\bibitem[\protect\citeauthoryear{{Mollitor}, {Nezri}  \& {Teyssier}}{{Mollitor}
  et~al.}{2015}]{MNT15}
{Mollitor} P.,  {Nezri} E.,   {Teyssier} R.,  2015, \mn@doi [\mnras]
  {10.1093/mnras/stu2466}, \href
  {http://adsabs.harvard.edu/abs/2015MNRAS.447.1353M} {447, 1353}

\bibitem[\protect\citeauthoryear{{Monachesi}, {G{\'o}mez}, {Grand},
  {Kauffmann}, {Marinacci}, {Pakmor}, {Springel}  \& {Frenk}}{{Monachesi}
  et~al.}{2016}]{MGG16}
{Monachesi} A.,  {G{\'o}mez} F.~A.,  {Grand} R.~J.~J.,  {Kauffmann} G.,
  {Marinacci} F.,  {Pakmor} R.,  {Springel} V.,   {Frenk} C.~S.,  2016, \mn@doi
  [\mnras] {10.1093/mnrasl/slw052}, \href
  {http://adsabs.harvard.edu/abs/2016MNRAS.459L..46M} {459, L46}

\bibitem[\protect\citeauthoryear{{Moster}, {Naab}  \& {White}}{{Moster}
  et~al.}{2013}]{MNW13}
{Moster} B.~P.,  {Naab} T.,   {White} S.~D.~M.,  2013, \mn@doi [\mnras]
  {10.1093/mnras/sts261}, \href
  {http://adsabs.harvard.edu/abs/2013MNRAS.428.3121M} {428, 3121}

\bibitem[\protect\citeauthoryear{{Navarro} \& {Steinmetz}}{{Navarro} \&
  {Steinmetz}}{2000}]{NS00}
{Navarro} J.~F.,  {Steinmetz} M.,  2000, \mn@doi [\apj] {10.1086/309175}, \href
  {http://adsabs.harvard.edu/abs/2000ApJ...538..477N} {538, 477}

\bibitem[\protect\citeauthoryear{{Navarro} \& {White}}{{Navarro} \&
  {White}}{1994}]{NW94}
{Navarro} J.~F.,  {White} S.~D.~M.,  1994, \mn@doi [\mnras]
  {10.1093/mnras/267.2.401}, \href
  {http://adsabs.harvard.edu/abs/1994MNRAS.267..401N} {267, 401}

\bibitem[\protect\citeauthoryear{{Nulsen} \& {Fabian}}{{Nulsen} \&
  {Fabian}}{2000}]{NF00}
{Nulsen} P.~E.~J.,  {Fabian} A.~C.,  2000, \mn@doi [\mnras]
  {10.1046/j.1365-8711.2000.03038.x}, \href
  {http://adsabs.harvard.edu/abs/2000MNRAS.311..346N} {311, 346}

\bibitem[\protect\citeauthoryear{{Obreja}, {Stinson}, {Dutton}, {Macci{\`o}},
  {Wang}  \& {Kang}}{{Obreja} et~al.}{2016}]{OSD16}
{Obreja} A.,  {Stinson} G.~S.,  {Dutton} A.~A.,  {Macci{\`o}} A.~V.,  {Wang}
  L.,   {Kang} X.,  2016, \mn@doi [\mnras] {10.1093/mnras/stw690}, \href
  {http://adsabs.harvard.edu/abs/2016MNRAS.459..467O} {459, 467}

\bibitem[\protect\citeauthoryear{{Okamoto}}{{Okamoto}}{2013}]{O13}
{Okamoto} T.,  2013, \mn@doi [\mnras] {10.1093/mnras/sts067}, \href
  {http://adsabs.harvard.edu/abs/2013MNRAS.428..718O} {428, 718}

\bibitem[\protect\citeauthoryear{{Okamoto}, {Eke}, {Frenk}  \&
  {Jenkins}}{{Okamoto} et~al.}{2005}]{OEF05}
{Okamoto} T.,  {Eke} V.~R.,  {Frenk} C.~S.,   {Jenkins} A.,  2005, \mn@doi
  [\mnras] {10.1111/j.1365-2966.2005.09525.x}, \href
  {http://adsabs.harvard.edu/abs/2005MNRAS.363.1299O} {363, 1299}

\bibitem[\protect\citeauthoryear{{Okamoto}, {Frenk}, {Jenkins}  \&
  {Theuns}}{{Okamoto} et~al.}{2010}]{OFJ10}
{Okamoto} T.,  {Frenk} C.~S.,  {Jenkins} A.,   {Theuns} T.,  2010, \mn@doi
  [\mnras] {10.1111/j.1365-2966.2010.16690.x}, \href
  {http://adsabs.harvard.edu/abs/2010MNRAS.406..208O} {406, 208}

\bibitem[\protect\citeauthoryear{{Oppenheimer} \& {Dav{\'e}}}{{Oppenheimer} \&
  {Dav{\'e}}}{2006}]{OD06}
{Oppenheimer} B.~D.,  {Dav{\'e}} R.,  2006, \mn@doi [\mnras]
  {10.1111/j.1365-2966.2006.10989.x}, \href
  {http://adsabs.harvard.edu/abs/2006MNRAS.373.1265O} {373, 1265}

\bibitem[\protect\citeauthoryear{{Pakmor} \& {Springel}}{{Pakmor} \&
  {Springel}}{2013}]{PakS13}
{Pakmor} R.,  {Springel} V.,  2013, \mn@doi [\mnras] {10.1093/mnras/stt428},
  \href {http://adsabs.harvard.edu/abs/2013MNRAS.432..176P} {432, 176}

\bibitem[\protect\citeauthoryear{{Pakmor}, {Marinacci}  \& {Springel}}{{Pakmor}
  et~al.}{2014}]{PMS14}
{Pakmor} R.,  {Marinacci} F.,   {Springel} V.,  2014, \mn@doi [\apjl]
  {10.1088/2041-8205/783/1/L20}, \href
  {http://adsabs.harvard.edu/abs/2014ApJ...783L..20P} {783, L20}

\bibitem[\protect\citeauthoryear{{Pakmor}, {Springel}, {Bauer}, {Mocz},
  {Munoz}, {Ohlmann}, {Schaal}  \& {Zhu}}{{Pakmor} et~al.}{2016a}]{PSB15}
{Pakmor} R.,  {Springel} V.,  {Bauer} A.,  {Mocz} P.,  {Munoz} D.~J.,
  {Ohlmann} S.~T.,  {Schaal} K.,   {Zhu} C.,  2016a, \mn@doi [\mnras]
  {10.1093/mnras/stv2380}, \href
  {http://adsabs.harvard.edu/abs/2016MNRAS.455.1134P} {455, 1134}

\bibitem[\protect\citeauthoryear{{Pakmor}, {Pfrommer}, {Simpson}  \&
  {Springel}}{{Pakmor} et~al.}{2016b}]{PPS16b}
{Pakmor} R.,  {Pfrommer} C.,  {Simpson} C.~M.,   {Springel} V.,  2016b, \mn@doi
  [\apjl] {10.3847/2041-8205/824/2/L30}, \href
  {http://adsabs.harvard.edu/abs/2016ApJ...824L..30P} {824, L30}

\bibitem[\protect\citeauthoryear{{Patel} et~al.}{{Patel} et~al.}{2013}]{PFF13}
{Patel} S.~G.,  et~al., 2013, \mn@doi [\apj] {10.1088/0004-637X/778/2/115},
  \href {http://adsabs.harvard.edu/abs/2013ApJ...778..115P} {778, 115}

\bibitem[\protect\citeauthoryear{{Pearce} et~al.,}{{Pearce}
  et~al.}{1999}]{PJF99}
{Pearce} F.~R.,  et~al., 1999, \mn@doi [\apjl] {10.1086/312196}, \href
  {http://adsabs.harvard.edu/abs/1999ApJ...521L..99P} {521, L99}

\bibitem[\protect\citeauthoryear{{Peebles}}{{Peebles}}{1969}]{P69}
{Peebles} P.~J.~E.,  1969, \mn@doi [\apj] {10.1086/149876}, \href
  {http://adsabs.harvard.edu/abs/1969ApJ...155..393P} {155, 393}

\bibitem[\protect\citeauthoryear{{Pfrommer}, {Pakmor}, {Schaal}, {Simpson}  \&
  {Springel}}{{Pfrommer} et~al.}{2017}]{PPS16}
{Pfrommer} C.,  {Pakmor} R.,  {Schaal} K.,  {Simpson} C.~M.,   {Springel} V.,
  2017, \mn@doi [\mnras] {10.1093/mnras/stw2941}, \href
  {http://adsabs.harvard.edu/abs/2017MNRAS.465.4500P} {465, 4500}

\bibitem[\protect\citeauthoryear{{Pizagno} et~al.,}{{Pizagno}
  et~al.}{2007}]{PPW07}
{Pizagno} J.,  et~al., 2007, \mn@doi [\aj] {10.1086/519522}, \href
  {http://adsabs.harvard.edu/abs/2007AJ....134..945P} {134, 945}

\bibitem[\protect\citeauthoryear{{Planck Collaboration} et~al.,}{{Planck
  Collaboration} et~al.}{2014}]{PC13}
{Planck Collaboration} et~al., 2014, \mn@doi [\aap]
  {10.1051/0004-6361/201321591}, \href
  {http://adsabs.harvard.edu/abs/2014A%26A...571A..16P} {571, A16}

\bibitem[\protect\citeauthoryear{{Portinari}, {Chiosi}  \&
  {Bressan}}{{Portinari} et~al.}{1998}]{PCB98}
{Portinari} L.,  {Chiosi} C.,   {Bressan} A.,  1998, \aap, \href
  {http://adsabs.harvard.edu/abs/1998A%26A...334..505P} {334, 505}

\bibitem[\protect\citeauthoryear{{Powell}, {Roe}, {Linde}, {Gombosi}  \& {De
  Zeeuw}}{{Powell} et~al.}{1999}]{PR99}
{Powell} K.~G.,  {Roe} P.~L.,  {Linde} T.~J.,  {Gombosi} T.~I.,   {De Zeeuw}
  D.~L.,  1999, \mn@doi [Journal of Computational Physics]
  {10.1006/jcph.1999.6299}, \href
  {http://adsabs.harvard.edu/abs/1999JCoPh.154..284P} {154, 284}

\bibitem[\protect\citeauthoryear{{Power}, {Navarro}, {Jenkins}, {Frenk},
  {White}, {Springel}, {Stadel}  \& {Quinn}}{{Power} et~al.}{2003}]{PNJ03}
{Power} C.,  {Navarro} J.~F.,  {Jenkins} A.,  {Frenk} C.~S.,  {White} S.~D.~M.,
   {Springel} V.,  {Stadel} J.,   {Quinn} T.,  2003, \mn@doi [\mnras]
  {10.1046/j.1365-8711.2003.05925.x}, \href
  {http://adsabs.harvard.edu/abs/2003MNRAS.338...14P} {338, 14}

\bibitem[\protect\citeauthoryear{{Pratt}, {Croston}, {Arnaud}  \&
  {B{\"o}hringer}}{{Pratt} et~al.}{2009}]{PCA09}
{Pratt} G.~W.,  {Croston} J.~H.,  {Arnaud} M.,   {B{\"o}hringer} H.,  2009,
  \mn@doi [\aap] {10.1051/0004-6361/200810994}, \href
  {http://adsabs.harvard.edu/abs/2009A%26A...498..361P} {498, 361}

\bibitem[\protect\citeauthoryear{{Puchwein} \& {Springel}}{{Puchwein} \&
  {Springel}}{2013}]{PS13}
{Puchwein} E.,  {Springel} V.,  2013, \mn@doi [\mnras] {10.1093/mnras/sts243},
  \href {http://adsabs.harvard.edu/abs/2013MNRAS.428.2966P} {428, 2966}

\bibitem[\protect\citeauthoryear{{Read}, {Hayfield}  \& {Agertz}}{{Read}
  et~al.}{2010}]{RHA10}
{Read} J.~I.,  {Hayfield} T.,   {Agertz} O.,  2010, \mn@doi [\mnras]
  {10.1111/j.1365-2966.2010.16577.x}, \href
  {http://adsabs.harvard.edu/abs/2010MNRAS.405.1513R} {405, 1513}

\bibitem[\protect\citeauthoryear{{Ro{\v s}kar}, {Teyssier}, {Agertz},
  {Wetzstein}  \& {Moore}}{{Ro{\v s}kar} et~al.}{2014}]{RTA14}
{Ro{\v s}kar} R.,  {Teyssier} R.,  {Agertz} O.,  {Wetzstein} M.,   {Moore} B.,
  2014, \mn@doi [\mnras] {10.1093/mnras/stu1548}, \href
  {http://adsabs.harvard.edu/abs/2014MNRAS.444.2837R} {444, 2837}

\bibitem[\protect\citeauthoryear{{Saitoh} \& {Makino}}{{Saitoh} \&
  {Makino}}{2013}]{SM13}
{Saitoh} T.~R.,  {Makino} J.,  2013, \mn@doi [\apj]
  {10.1088/0004-637X/768/1/44}, \href
  {http://adsabs.harvard.edu/abs/2013ApJ...768...44S} {768, 44}

\bibitem[\protect\citeauthoryear{{Sales}, {Navarro}, {Schaye}, {Dalla Vecchia},
  {Springel}  \& {Booth}}{{Sales} et~al.}{2010}]{SNS10}
{Sales} L.~V.,  {Navarro} J.~F.,  {Schaye} J.,  {Dalla Vecchia} C.,  {Springel}
  V.,   {Booth} C.~M.,  2010, \mn@doi [\mnras]
  {10.1111/j.1365-2966.2010.17391.x}, \href
  {http://adsabs.harvard.edu/abs/2010MNRAS.409.1541S} {409, 1541}

\bibitem[\protect\citeauthoryear{{Sawala} et~al.,}{{Sawala}
  et~al.}{2015}]{SFF15}
{Sawala} T.,  et~al., 2015, \mn@doi [\mnras] {10.1093/mnras/stu2753}, \href
  {http://adsabs.harvard.edu/abs/2015MNRAS.448.2941S} {448, 2941}

\bibitem[\protect\citeauthoryear{{Scannapieco}, {Tissera}, {White}  \&
  {Springel}}{{Scannapieco} et~al.}{2008}]{Sc08}
{Scannapieco} C.,  {Tissera} P.~B.,  {White} S.~D.~M.,   {Springel} V.,  2008,
  \mn@doi [\mnras] {10.1111/j.1365-2966.2008.13678.x}, \href
  {http://adsabs.harvard.edu/abs/2008MNRAS.389.1137S} {389, 1137}

\bibitem[\protect\citeauthoryear{{Scannapieco}, {White}, {Springel}  \&
  {Tissera}}{{Scannapieco} et~al.}{2009}]{SWS09}
{Scannapieco} C.,  {White} S.~D.~M.,  {Springel} V.,   {Tissera} P.~B.,  2009,
  \mn@doi [\mnras] {10.1111/j.1365-2966.2009.14764.x}, \href
  {http://adsabs.harvard.edu/abs/2009MNRAS.396..696S} {396, 696}

\bibitem[\protect\citeauthoryear{{Scannapieco} et~al.,}{{Scannapieco}
  et~al.}{2012}]{SWP12}
{Scannapieco} C.,  et~al., 2012, \mn@doi [\mnras]
  {10.1111/j.1365-2966.2012.20993.x}, \href
  {http://adsabs.harvard.edu/abs/2012MNRAS.423.1726S} {423, 1726}

\bibitem[\protect\citeauthoryear{{Schaller} et~al.,}{{Schaller}
  et~al.}{2015a}]{SFB15}
{Schaller} M.,  et~al., 2015a, \mn@doi [\mnras] {10.1093/mnras/stv1067}, \href
  {http://adsabs.harvard.edu/abs/2015MNRAS.451.1247S} {451, 1247}

\bibitem[\protect\citeauthoryear{{Schaller}, {Dalla Vecchia}, {Schaye},
  {Bower}, {Theuns}, {Crain}, {Furlong}  \& {McCarthy}}{{Schaller}
  et~al.}{2015b}]{SDS15}
{Schaller} M.,  {Dalla Vecchia} C.,  {Schaye} J.,  {Bower} R.~G.,  {Theuns} T.,
   {Crain} R.~A.,  {Furlong} M.,   {McCarthy} I.~G.,  2015b, \mn@doi [\mnras]
  {10.1093/mnras/stv2169}, \href
  {http://adsabs.harvard.edu/abs/2015MNRAS.454.2277S} {454, 2277}

\bibitem[\protect\citeauthoryear{{Schaye} et~al.,}{{Schaye}
  et~al.}{2015}]{SCB15}
{Schaye} J.,  et~al., 2015, \mn@doi [\mnras] {10.1093/mnras/stu2058}, \href
  {http://adsabs.harvard.edu/abs/2015MNRAS.446..521S} {446, 521}

\bibitem[\protect\citeauthoryear{{S{\'e}rsic}}{{S{\'e}rsic}}{1963}]{S63}
{S{\'e}rsic} J.~L.,  1963, Boletin de la Asociacion Argentina de Astronomia La
  Plata Argentina, \href {http://adsabs.harvard.edu/abs/1963BAAA....6...41S}
  {6, 41}

\bibitem[\protect\citeauthoryear{{Sharma}, {Steinmetz}  \&
  {Bland-Hawthorn}}{{Sharma} et~al.}{2012}]{SSB12}
{Sharma} S.,  {Steinmetz} M.,   {Bland-Hawthorn} J.,  2012, \mn@doi [\apj]
  {10.1088/0004-637X/750/2/107}, \href
  {http://adsabs.harvard.edu/abs/2012ApJ...750..107S} {750, 107}

\bibitem[\protect\citeauthoryear{{Sijacki}, {Springel}, {Di Matteo}  \&
  {Hernquist}}{{Sijacki} et~al.}{2007}]{SSD07}
{Sijacki} D.,  {Springel} V.,  {Di Matteo} T.,   {Hernquist} L.,  2007, \mn@doi
  [\mnras] {10.1111/j.1365-2966.2007.12153.x}, \href
  {http://adsabs.harvard.edu/abs/2007MNRAS.380..877S} {380, 877}

\bibitem[\protect\citeauthoryear{{Simpson}, {Pakmor}, {Marinacci}, {Pfrommer},
  {Springel}, {Glover}, {Clark}  \& {Smith}}{{Simpson} et~al.}{2016}]{SPM16}
{Simpson} C.~M.,  {Pakmor} R.,  {Marinacci} F.,  {Pfrommer} C.,  {Springel} V.,
   {Glover} S.~C.~O.,  {Clark} P.~C.,   {Smith} R.~J.,  2016, \mn@doi [\apjl]
  {10.3847/2041-8205/827/2/L29}, \href
  {http://adsabs.harvard.edu/abs/2016ApJ...827L..29S} {827, L29}

\bibitem[\protect\citeauthoryear{{Sparre} \& {Springel}}{{Sparre} \&
  {Springel}}{2016}]{SS16}
{Sparre} M.,  {Springel} V.,  2016, \mn@doi [\mnras] {10.1093/mnras/stw1793},
  \href {http://adsabs.harvard.edu/abs/2016MNRAS.462.2418S} {462, 2418}

\bibitem[\protect\citeauthoryear{{Sparre}, {Hayward}, {Feldmann},
  {Faucher-Gigu{\`e}re}, {Muratov}, {Kere{\v s}}  \& {Hopkins}}{{Sparre}
  et~al.}{2015}]{SHF16}
{Sparre} M.,  {Hayward} C.~C.,  {Feldmann} R.,  {Faucher-Gigu{\`e}re} C.-A.,
  {Muratov} A.~L.,  {Kere{\v s}} D.,   {Hopkins} P.~F.,  2015, preprint, \href
  {http://adsabs.harvard.edu/abs/2015arXiv151003869S} {} (\mn@eprint {arXiv}
  {1510.03869})

\bibitem[\protect\citeauthoryear{{Springel}}{{Springel}}{2005}]{SP05}
{Springel} V.,  2005, \mn@doi [\mnras] {10.1111/j.1365-2966.2005.09655.x},
  \href {http://adsabs.harvard.edu/abs/2005MNRAS.364.1105S} {364, 1105}

\bibitem[\protect\citeauthoryear{{Springel}}{{Springel}}{2010}]{Sp10}
{Springel} V.,  2010, \mn@doi [\mnras] {10.1111/j.1365-2966.2009.15715.x},
  \href {http://adsabs.harvard.edu/abs/2010MNRAS.401..791S} {401, 791}

\bibitem[\protect\citeauthoryear{{Springel} \& {Hernquist}}{{Springel} \&
  {Hernquist}}{2003}]{SH03}
{Springel} V.,  {Hernquist} L.,  2003, \mn@doi [\mnras]
  {10.1046/j.1365-8711.2003.06206.x}, \href
  {http://adsabs.harvard.edu/abs/2003MNRAS.339..289S} {339, 289}

\bibitem[\protect\citeauthoryear{{Springel}, {Di Matteo}  \&
  {Hernquist}}{{Springel} et~al.}{2005}]{SMH05}
{Springel} V.,  {Di Matteo} T.,   {Hernquist} L.,  2005, \mn@doi [\mnras]
  {10.1111/j.1365-2966.2005.09238.x}, \href
  {http://adsabs.harvard.edu/abs/2005MNRAS.361..776S} {361, 776}

\bibitem[\protect\citeauthoryear{{Springel} et~al.,}{{Springel}
  et~al.}{2008}]{SWV08}
{Springel} V.,  et~al., 2008, \mn@doi [\mnras]
  {10.1111/j.1365-2966.2008.14066.x}, \href
  {http://adsabs.harvard.edu/abs/2008MNRAS.391.1685S} {391, 1685}

\bibitem[\protect\citeauthoryear{{Stinson}, {Seth}, {Katz}, {Wadsley},
  {Governato}  \& {Quinn}}{{Stinson} et~al.}{2006}]{SSK06}
{Stinson} G.,  {Seth} A.,  {Katz} N.,  {Wadsley} J.,  {Governato} F.,   {Quinn}
  T.,  2006, \mn@doi [\mnras] {10.1111/j.1365-2966.2006.11097.x}, \href
  {http://adsabs.harvard.edu/abs/2006MNRAS.373.1074S} {373, 1074}

\bibitem[\protect\citeauthoryear{{Stinson}, {Bailin}, {Couchman}, {Wadsley},
  {Shen}, {Nickerson}, {Brook}  \& {Quinn}}{{Stinson} et~al.}{2010}]{SBC10}
{Stinson} G.~S.,  {Bailin} J.,  {Couchman} H.,  {Wadsley} J.,  {Shen} S.,
  {Nickerson} S.,  {Brook} C.,   {Quinn} T.,  2010, \mn@doi [\mnras]
  {10.1111/j.1365-2966.2010.17187.x}, \href
  {http://adsabs.harvard.edu/abs/2010MNRAS.408..812S} {408, 812}

\bibitem[\protect\citeauthoryear{{Stinson}, {Brook}, {Macci{\`o}}, {Wadsley},
  {Quinn}  \& {Couchman}}{{Stinson} et~al.}{2013a}]{SB13}
{Stinson} G.~S.,  {Brook} C.,  {Macci{\`o}} A.~V.,  {Wadsley} J.,  {Quinn}
  T.~R.,   {Couchman} H.~M.~P.,  2013a, \mn@doi [\mnras]
  {10.1093/mnras/sts028}, \href
  {http://adsabs.harvard.edu/abs/2013MNRAS.428..129S} {428, 129}

\bibitem[\protect\citeauthoryear{{Stinson} et~al.,}{{Stinson}
  et~al.}{2013b}]{SBR13}
{Stinson} G.~S.,  et~al., 2013b, \mn@doi [\mnras] {10.1093/mnras/stt1600},
  \href {http://adsabs.harvard.edu/abs/2013MNRAS.436..625S} {436, 625}

\bibitem[\protect\citeauthoryear{{Sur}, {Scannapieco}  \& {Ostriker}}{{Sur}
  et~al.}{2016}]{SSO16}
{Sur} S.,  {Scannapieco} E.,   {Ostriker} E.~C.,  2016, \mn@doi [\apj]
  {10.3847/0004-637X/818/1/28}, \href
  {http://adsabs.harvard.edu/abs/2016ApJ...818...28S} {818, 28}

\bibitem[\protect\citeauthoryear{{Teyssier}}{{Teyssier}}{2002}]{T02}
{Teyssier} R.,  2002, \mn@doi [\aap] {10.1051/0004-6361:20011817}, \href
  {http://adsabs.harvard.edu/abs/2002A%26A...385..337T} {385, 337}

\bibitem[\protect\citeauthoryear{{Thielemann} et~al.,}{{Thielemann}
  et~al.}{2003}]{TAB03}
{Thielemann} F.-K.,  et~al., 2003, \mn@doi [Nuclear Physics A]
  {10.1016/S0375-9474(03)00704-8}, \href
  {http://adsabs.harvard.edu/abs/2003NuPhA.718..139T} {718, 139}

\bibitem[\protect\citeauthoryear{{Torrey}, {Wellons}, {Ma}, {Hopkins}  \&
  {Vogelsberger}}{{Torrey} et~al.}{2016}]{TWM16}
{Torrey} P.,  {Wellons} S.,  {Ma} C.-P.,  {Hopkins} P.~F.,   {Vogelsberger} M.,
   2016, preprint, \href {http://adsabs.harvard.edu/abs/2016arXiv160607271T} {}
  (\mn@eprint {arXiv} {1606.07271})

\bibitem[\protect\citeauthoryear{{Travaglio}, {Hillebrandt}, {Reinecke}  \&
  {Thielemann}}{{Travaglio} et~al.}{2004}]{THR04}
{Travaglio} C.,  {Hillebrandt} W.,  {Reinecke} M.,   {Thielemann} F.-K.,  2004,
  \mn@doi [\aap] {10.1051/0004-6361:20041108}, \href
  {http://adsabs.harvard.edu/abs/2004A%26A...425.1029T} {425, 1029}

\bibitem[\protect\citeauthoryear{{Verheijen}}{{Verheijen}}{2001}]{V01}
{Verheijen} M.~A.~W.,  2001, \mn@doi [\apj] {10.1086/323887}, \href
  {http://adsabs.harvard.edu/abs/2001ApJ...563..694V} {563, 694}

\bibitem[\protect\citeauthoryear{{Vitvitska}, {Klypin}, {Kravtsov}, {Wechsler},
  {Primack}  \& {Bullock}}{{Vitvitska} et~al.}{2002}]{VKK02}
{Vitvitska} M.,  {Klypin} A.~A.,  {Kravtsov} A.~V.,  {Wechsler} R.~H.,
  {Primack} J.~R.,   {Bullock} J.~S.,  2002, \mn@doi [\apj] {10.1086/344361},
  \href {http://adsabs.harvard.edu/abs/2002ApJ...581..799V} {581, 799}

\bibitem[\protect\citeauthoryear{{Vogelsberger}, {Genel}, {Sijacki}, {Torrey},
  {Springel}  \& {Hernquist}}{{Vogelsberger} et~al.}{2013}]{VGS13}
{Vogelsberger} M.,  {Genel} S.,  {Sijacki} D.,  {Torrey} P.,  {Springel} V.,
  {Hernquist} L.,  2013, \mn@doi [\mnras] {10.1093/mnras/stt1789}, \href
  {http://adsabs.harvard.edu/abs/2013MNRAS.436.3031V} {436, 3031}

\bibitem[\protect\citeauthoryear{{Walch} \& {Naab}}{{Walch} \&
  {Naab}}{2015}]{WN15}
{Walch} S.,  {Naab} T.,  2015, \mn@doi [\mnras] {10.1093/mnras/stv1155}, \href
  {http://adsabs.harvard.edu/abs/2015MNRAS.451.2757W} {451, 2757}

\bibitem[\protect\citeauthoryear{{Wang}, {Dutton}, {Stinson}, {Macci{\`o}},
  {Penzo}, {Kang}, {Keller}  \& {Wadsley}}{{Wang} et~al.}{2015}]{WDS15}
{Wang} L.,  {Dutton} A.~A.,  {Stinson} G.~S.,  {Macci{\`o}} A.~V.,  {Penzo} C.,
   {Kang} X.,  {Keller} B.~W.,   {Wadsley} J.,  2015, \mn@doi [\mnras]
  {10.1093/mnras/stv1937}, \href
  {http://adsabs.harvard.edu/abs/2015MNRAS.454...83W} {454, 83}

\bibitem[\protect\citeauthoryear{{Weinberger} et~al.,}{{Weinberger}
  et~al.}{2017}]{WSH16}
{Weinberger} R.,  et~al., 2017, \mn@doi [\mnras] {10.1093/mnras/stw2944}, \href
  {http://adsabs.harvard.edu/abs/2017MNRAS.465.3291W} {465, 3291}

\bibitem[\protect\citeauthoryear{{Zel'dovich}}{{Zel'dovich}}{1970}]{Z70}
{Zel'dovich} Y.~B.,  1970, \aap, \href
  {http://adsabs.harvard.edu/abs/1970A%26A.....5...84Z} {5, 84}

\bibitem[\protect\citeauthoryear{{Zjupa} \& {Springel}}{{Zjupa} \&
  {Springel}}{2016}]{ZS16}
{Zjupa} J.,  {Springel} V.,  2016, preprint, \href
  {http://adsabs.harvard.edu/abs/2016arXiv160801323Z} {} (\mn@eprint {arXiv}
  {1608.01323})

\bibitem[\protect\citeauthoryear{{van der Wel} et~al.}{{van der Wel}
  et~al.}{2012}]{VBH12}
{van der Wel} A.,  et~al., 2012, \mn@doi [\apjs] {10.1088/0067-0049/203/2/24},
  \href {http://adsabs.harvard.edu/abs/2012ApJS..203...24V} {203, 24}

\bibitem[\protect\citeauthoryear{{van der Wel} et~al.}{{van der Wel}
  et~al.}{2014}]{vdW14}
{van der Wel} A.,  et~al., 2014, \mn@doi [\apj] {10.1088/0004-637X/788/1/28},
  \href {http://adsabs.harvard.edu/abs/2014ApJ...788...28V} {788, 28}

\makeatother
\end{thebibliography}

\appendix
\section{Parent box halo Identification}
\label{appa}


The Auriga sample were selected from the dark matter only version of the EAGLE Ref-L100N1504 cosmological volume \citet{SCB15}. In Table A1, we list the ID numbers of the Auriga simulations and provide a corresponding identification number for these haloes in the parent dark matter only and in the full hydrodynamic EAGLE cosmological box simulations.

The positions and properties of the corresponding central galaxies in the EAGLE Ref-L100N1504 simulation can be obtained by making an SQL query to the public Eagle Database, which is described in \citet{MHS16}. For example, to find the EAGLE counterpart to Au 3 in the Eagle database, we would take the group number, 1449, from Table A1, and execute the following SQL query, to obtain the location and properties of the central galaxy associated with the FOF group at $z=0$:

\begin{Verbatim}
	select * from refl0100n1504_subhalo
	where snapnum=28 and subgroupnumber=0 
	and groupnumber = 1449
\end{Verbatim}
Amongst the properties this query returns are the cartesian coordinates for the gravitational potential minimum of the central galaxy, which is located for Au 3 at $(x,y,z)=(10.204818,98.30205,72.66644)$~~Mpc.

\begin{table}
\caption{Table of halo ID numbers at redshift zero. The left column lists the Auriga IDs referred to in this paper. The middle and right columns list the corresponding FOF halo IDs in the parent EAGLE dark matter only (DMO) cosmological volume and its hydrodynamical counterpart.}
\centering
\begin{tabular}{c c c}
\\\toprule
Auriga ID & EAGLE DMO  & EAGLE REF   \\
 & group number & group number \\
\hline                                                                     
1 & 931 & 949  \\
2 & 942 & 897  \\
3 & 1266 & 1449 \\ 
4 & 1018 & 954  \\
5 & 1468 & 1446  \\
6 & 1574 & 1619 \\ 
7 & 1257 & 1359  \\
8 & 1606 & 1522  \\
9 & 1495 & 1763 \\ 
10 & 1679 & 1649  \\
11 & 1101 & 1101  \\
12 & 1538 & 1547 \\ 
13 & 1355 & 1379  \\
14 & 1212 & 1173  \\
15 & 1436 & 1457 \\ 
16 & 1231 & 1143  \\
17 & 1689 & 1566  \\
18 & 844 & 848 \\ 
19 & 1432 & 1467  \\
20 & 1164 & 1152  \\
21 & 1110 & 983 \\ 
22 & 1632 & 1692  \\
23 & 1168 & 1285  \\
24 & 938 & 1065 \\ 
25 & 1440 & 1470  \\
26 & 1026 & 1099  \\
27 & 998 & 1071 \\ 
28 & 1173 & 1224  \\
29 & 997 & 1019  \\
30 & 1429 & 1481 \\ 
\\\bottomrule
\end{tabular}
\end{table}

\label{lastpage}

\end{document}